\pdfoutput=1

\documentclass[12pt,a4paper]{article}

\usepackage{ifthen} 
\newboolean{pdflatex}
\setboolean{pdflatex}{true} 

\newboolean{articletitles}
\setboolean{articletitles}{true} 

\newboolean{uprightparticles}
\setboolean{uprightparticles}{false} 

\newcommand{\tip}{\ensuremath{\text{TIP}}\xspace}

\newcommand{\deltaACP}{\ensuremath{\Delta A_{\CP}}\xspace}
\newcommand{\DACP}{\deltaACP}

\newcommand{\Araw}{\ensuremath{A_{\rm raw}}\xspace}

\newcommand{\AP}{\ensuremath{A_{\rm P}}\xspace}

\newcommand{\APB}{\ensuremath{A_{\rm P}(\PB)}\xspace}
\newcommand{\AD}{\ensuremath{A_{\rm D}}\xspace}

\def\dkk        {\decay{\Dz}{\Km\Kp}}
\def\dpipi      {\decay{\Dz}{\pim\pip}}

\def\dkpi       {\decay{\Dz}{\Km\pip}}
\def\dkpiDCS       {\decay{\Dz}{\Kp\pim}}

\newcommand{\mypaperversion}{}
\newcommand{\mydate}{May 31, 2019}
\newcommand{\mylhcbpapernumber}{LHCb-PAPER-2019-006}
\newcommand{\mycernpapernumber}{CERN-EP-2019-042}

\def\paperauthors{LHCb collaboration} 
\def\paperasciititle{Observation of CP violation in neutral charm decays} 
\def\papertitle{Observation of \CP violation in charm decays} 
\def\paperkeywords{{High Energy Physics}, {LHCb}} 
\def\papercopyright{\the\year\ CERN for the benefit of the LHCb collaboration} 
\def\paperlicence{CC-BY-4.0 licence}
\def\paperlicenceurl{https://creativecommons.org/licenses/by/4.0/}

\usepackage[top=1in, bottom=1.25in, left=1in, right=1in]{geometry}

\usepackage{microtype}
\usepackage{lineno}  
\usepackage{xspace} 
\usepackage{caption} 

\usepackage{graphicx}  
\usepackage{color}
\usepackage{colortbl}
\graphicspath{{./figs/}{./figs/paper/}{./figs/supplementary/}} 
\DeclareGraphicsExtensions{.pdf,.PDF,png,.PNG}

\usepackage{amsmath} 
\usepackage{amssymb}
\usepackage{amsfonts}
\usepackage{upgreek} 
\usepackage{overpic}
\newcommand*\patchAmsMathEnvironmentForLineno[1]{%
\expandafter\let\csname old#1\expandafter\endcsname\csname #1\endcsname
\expandafter\let\csname oldend#1\expandafter\endcsname\csname
end#1\endcsname
 \renewenvironment{#1}%
   {\linenomath\csname old#1\endcsname}%
   {\csname oldend#1\endcsname\endlinenomath}%
}
\newcommand*\patchBothAmsMathEnvironmentsForLineno[1]{%
  \patchAmsMathEnvironmentForLineno{#1}%
  \patchAmsMathEnvironmentForLineno{#1*}%
}
\AtBeginDocument{%
\patchBothAmsMathEnvironmentsForLineno{equation}%
\patchBothAmsMathEnvironmentsForLineno{align}%
\patchBothAmsMathEnvironmentsForLineno{flalign}%
\patchBothAmsMathEnvironmentsForLineno{alignat}%
\patchBothAmsMathEnvironmentsForLineno{gather}%
\patchBothAmsMathEnvironmentsForLineno{multline}%
\patchBothAmsMathEnvironmentsForLineno{eqnarray}%
}

\usepackage{hyperxmp}

\usepackage[pdftex,
            pdfauthor={\paperauthors},
            pdftitle={\paperasciititle},
            pdfkeywords={\paperkeywords},
            pdfcopyright={Copyright (C) \papercopyright},
            pdflicenseurl={\paperlicenceurl}]{hyperref}

\usepackage[all]{hypcap} 

\usepackage{xspace} 
\usepackage{upgreek}

\newcommand{\offsetoverline}[2][0.1em]{\kern #1\overline{\kern -#1 #2}}%

\def\MagUp {\mbox{\em Mag\kern -0.05em Up}\xspace}

\ifthenelse{\boolean{uprightparticles}}%
{

 \def\Pmu         {\ensuremath{\upmu}\xspace}                 
 \def\Pnu         {\ensuremath{\upnu}\xspace}                 
                  
 \def\Ppi         {\ensuremath{\uppi}\xspace}

 \def\Ppsi        {\ensuremath{\uppsi}\xspace}

 \def\PDelta      {\ensuremath{\Delta}\xspace}                 
 \def\PXi         {\ensuremath{\Xi}\xspace}                 
 \def\PLambda     {\ensuremath{\Lambda}\xspace}                 
 \def\PSigma      {\ensuremath{\Sigma}\xspace}                 
 \def\POmega      {\ensuremath{\Omega}\xspace}                 
 \def\PUpsilon    {\ensuremath{\Upsilon}\xspace}

 \def\PB      {\ensuremath{\mathrm{B}}\xspace}                 
                  
 \def\PD      {\ensuremath{\mathrm{D}}\xspace}

 \def\PJ      {\ensuremath{\mathrm{J}}\xspace}                 
 \def\PK      {\ensuremath{\mathrm{K}}\xspace}

 \def\Pb      {\ensuremath{\mathrm{b}}\xspace}                 
 \def\Pc      {\ensuremath{\mathrm{c}}\xspace}                 
                  
 \def\Pe      {\ensuremath{\mathrm{e}}\xspace}

 \def\Pi      {\ensuremath{\mathrm{i}}\xspace}

 \def\Pp      {\ensuremath{\mathrm{p}}\xspace}

}
{

 \def\Pmu         {\ensuremath{\mu}\xspace}                 
 \def\Pnu         {\ensuremath{\nu}\xspace}                 
                  
 \def\Ppi         {\ensuremath{\pi}\xspace}

 \def\Ppsi        {\ensuremath{\psi}\xspace}                 
                  
 \mathchardef\PDelta="7101
 \mathchardef\PXi="7104
 \mathchardef\PLambda="7103
 \mathchardef\PSigma="7106
 \mathchardef\POmega="710A
 \mathchardef\PUpsilon="7107
                  
 \def\PB      {\ensuremath{B}\xspace}                 
                  
 \def\PD      {\ensuremath{D}\xspace}

 \def\PJ      {\ensuremath{J}\xspace}                 
 \def\PK      {\ensuremath{K}\xspace}

 \def\Pb      {\ensuremath{b}\xspace}                 
 \def\Pc      {\ensuremath{c}\xspace}                 
                  
 \def\Pe      {\ensuremath{e}\xspace}

 \def\Pi      {\ensuremath{i}\xspace}

 \def\Pp      {\ensuremath{p}\xspace}

}

\makeatletter
\ifcase \@ptsize \relax
  \newcommand{\miniscule}{\@setfontsize\miniscule{4}{5}}
\or
  \newcommand{\miniscule}{\@setfontsize\miniscule{5}{6}}
\or
  \newcommand{\miniscule}{\@setfontsize\miniscule{5}{6}}
\fi
\makeatother

\DeclareRobustCommand{\optbar}[1]{\shortstack{{\miniscule (\rule[.5ex]{1.25em}{.18mm})}
  \\ [-.7ex] $#1$}}

\def\ep         {{\ensuremath{\Pe^+}}\xspace}

\def\mup        {{\ensuremath{\Pmu^+}}\xspace}

\def\neu        {{\ensuremath{\Pnu}}\xspace}

\def\neue       {{\ensuremath{\neu_e}}\xspace}

\def\neum       {{\ensuremath{\neu_\mu}}\xspace}

\def\cquark    {{\ensuremath{\Pc}}\xspace}
\def\cquarkbar {{\ensuremath{\overline \cquark}}\xspace}

\def\bquark    {{\ensuremath{\Pb}}\xspace}

\def\pion   {{\ensuremath{\Ppi}}\xspace}
\def\piz    {{\ensuremath{\pion^0}}\xspace}
\def\pip    {{\ensuremath{\pion^+}}\xspace}
\def\pim    {{\ensuremath{\pion^-}}\xspace}

\def\kaon    {{\ensuremath{\PK}}\xspace}
  \def\Kbar    {{\kern 0.2em\overline{\kern -0.2em \PK}{}}\xspace}

\def\KorKbar {\kern 0.18em\optbar{\kern -0.18em K}{}\xspace}

\def\Kp      {{\ensuremath{\kaon^+}}\xspace}
\def\Km      {{\ensuremath{\kaon^-}}\xspace}

  \def\Dbar    {{\kern 0.2em\overline{\kern -0.2em \PD}{}}\xspace}
\def\D       {{\ensuremath{\PD}}\xspace}

\def\DorDbar {\kern 0.18em\optbar{\kern -0.18em D}{}\xspace}
\def\Dz      {{\ensuremath{\D^0}}\xspace}
\def\Dzb     {{\ensuremath{\Dbar{}^0}}\xspace}

\def\Dstar   {{\ensuremath{\D^*}}\xspace}

\def\Dstarp  {{\ensuremath{\D^{*+}}}\xspace}
\def\Dstarm  {{\ensuremath{\D^{*-}}}\xspace}

\def\B       {{\ensuremath{\PB}}\xspace}
\def\Bbar    {{\ensuremath{\kern 0.18em\overline{\kern -0.18em \PB}{}}}\xspace}
\def\Bb      {{\ensuremath{\Bbar}}\xspace}
\def\BorBbar    {\kern 0.18em\optbar{\kern -0.18em B}{}\xspace}

\def\jpsi     {{\ensuremath{{\PJ\mskip -3mu/\mskip -2mu\Ppsi\mskip 2mu}}}\xspace}
\def\psitwos  {{\ensuremath{\Ppsi{(2S)}}}\xspace}

\def\Y#1S{\ensuremath{\PUpsilon{(#1S)}}\xspace}

\def\proton      {{\ensuremath{\Pp}}\xspace}

\def\LorLbar     {\kern 0.18em\optbar{\kern -0.18em \PLambda}{}\xspace}

\newcommand{\decay}[2]{\mbox{\ensuremath{#1\!\to #2}}\xspace}         

\def\to                 {\ensuremath{\rightarrow}\xspace}

\def\grpsuthree {{\ensuremath{\mathrm{SU}(3)}}\xspace}

\def\order   {{\ensuremath{\mathcal{O}}}\xspace}

\def\CP                {{\ensuremath{C\!P}}\xspace}

\def\AT#1     {\ensuremath{A_{\mathrm{T}}^{#1}}\xspace}           

\def\C#1      {\ensuremath{\mathcal{C}_{#1}}\xspace}                       
\def\Cp#1     {\ensuremath{\mathcal{C}_{#1}^{'}}\xspace}                    
\def\Ceff#1   {\ensuremath{\mathcal{C}_{#1}^{\mathrm{(eff)}}}\xspace}        
\def\Cpeff#1  {\ensuremath{\mathcal{C}_{#1}^{'\mathrm{(eff)}}}\xspace}       
\def\Ope#1    {\ensuremath{\mathcal{O}_{#1}}\xspace}                       
\def\Opep#1   {\ensuremath{\mathcal{O}_{#1}^{'}}\xspace}                    

\def\agamma     {\ensuremath{A_{\Gamma}}\xspace}

\newcommand{\aunit}[1]{\ensuremath{\text{\,#1}}}       

\newcommand{\tev}{\aunit{Te\kern -0.1em V}\xspace}
\newcommand{\gev}{\aunit{Ge\kern -0.1em V}\xspace}
\newcommand{\mev}{\aunit{Me\kern -0.1em V}\xspace}
\newcommand{\kev}{\aunit{ke\kern -0.1em V}\xspace}
\newcommand{\ev}{\aunit{e\kern -0.1em V}\xspace}
\newcommand{\mevc}{\ensuremath{\aunit{Me\kern -0.1em V\!/}c}\xspace}
\newcommand{\gevc}{\ensuremath{\aunit{Ge\kern -0.1em V\!/}c}\xspace}
\newcommand{\mevcc}{\ensuremath{\aunit{Me\kern -0.1em V\!/}c^2}\xspace}
\newcommand{\gevcc}{\ensuremath{\aunit{Ge\kern -0.1em V\!/}c^2}\xspace}

\def\fb   {\ensuremath{\aunit{fb}}\xspace}
\def\invfb   {\ensuremath{\fb^{-1}}\xspace}

\def\order{{\ensuremath{\mathcal{O}}}\xspace}
\newcommand{\chisq}{\ensuremath{\chi^2}\xspace}

\newcommand{\chisqip}{\ensuremath{\chi^2_{\text{IP}}}\xspace}

\def\gsim{{~\raise.15em\hbox{$>$}\kern-.85em
          \lower.35em\hbox{$\sim$}~}\xspace}
\def\lsim{{~\raise.15em\hbox{$<$}\kern-.85em
          \lower.35em\hbox{$\sim$}~}\xspace}

\newcommand{\mean}[1]{\ensuremath{\left\langle #1 \right\rangle}} 

\def\pt         {\ensuremath{p_{\mathrm{T}}}\xspace}

\xspace

\def\tell1  {TELL1\xspace}
\def\ukl1   {UKL1\xspace}

\newcommand{\ie}{\mbox{\itshape i.e.}\xspace}

\usepackage{cite} 
\usepackage{mciteplus}

\usepackage[capitalise]{cleveref}
\Crefname{figure}{Figure}{Figures}

\newcommand{\overbar}[1]{\mkern 1.5mu\overline{\mkern-1.5mu#1\mkern-1.5mu}\mkern 1.5mu}

\begin{document}
\renewcommand{\thefootnote}{\fnsymbol{footnote}}
\setcounter{footnote}{1}

\begin{titlepage}
\pagenumbering{roman}

\vspace*{-1.5cm}
\centerline{\large EUROPEAN ORGANIZATION FOR NUCLEAR RESEARCH (CERN)}
\vspace*{1.5cm}
\noindent
\begin{tabular*}{\linewidth}{lc@{\extracolsep{\fill}}r@{\extracolsep{0pt}}}
\vspace*{-1.5cm}\mbox{\!\!\!\includegraphics[width=.14\textwidth]{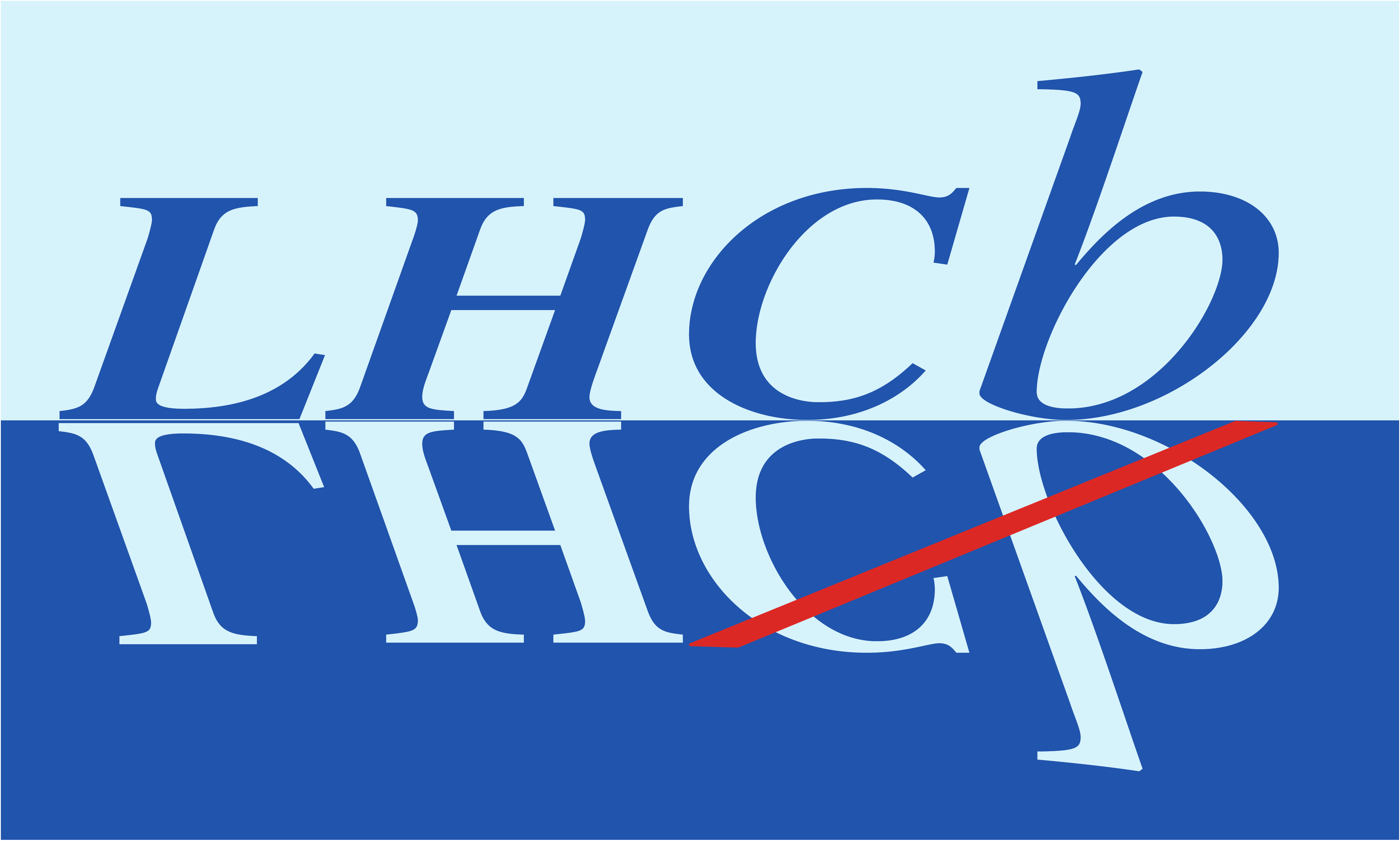}} & & \\
 & & \mycernpapernumber \\
 & & \mylhcbpapernumber \\
 & & \mydate \\ 
 & & \mypaperversion\\
\end{tabular*}

\vspace*{4.0cm}

{\normalfont\bfseries\boldmath\huge
\begin{center}
  \papertitle 
\end{center}
}

\vspace*{1.5cm}

\begin{center}
\paperauthors\footnote{Authors are listed at the end of this Letter.}
\end{center}

\vspace{\fill}

\begin{abstract}
\noindent A search for charge-parity~(\CP) violation in \dkk and \dpipi decays is reported, using $pp$ collision data corresponding to an integrated luminosity of 5.9\invfb collected at a center-of-mass energy of 13\tev with the LHCb detector. The flavor of the charm meson is inferred from the charge of the pion in \decay{\Dstar(2010)^{+}}{\Dz \pip} decays or from the charge of the muon in \decay{\Bb}{\Dz \mu^- \bar{\nu}_\mu X} decays. The difference between the $\CP$ asymmetries in $\dkk$ and $\dpipi$ decays is measured to be \mbox{$\DACP = [ -18.2 \pm 3.2\,(\rm stat.) \pm 0.9\,(\rm syst.) ] \times 10^{-4}$} for $\pi$-tagged and \mbox{$\DACP = [ -9 \pm 8\,(\rm stat.) \pm 5\,(\rm syst.) ] \times 10^{-4} $} for $\mu$-tagged \Dz mesons. Combining these with previous LHCb results leads to $$\DACP = ( -15.4 \pm 2.9) \times 10^{-4},$$ where the uncertainty includes both statistical and systematic contributions. The measured value differs from zero by more than five standard deviations. This is the first observation of \CP violation in the decay of charm hadrons.
\end{abstract}

\vspace*{1.5cm}

\begin{center}
  Published in Phys. Rev. Lett. \textbf{122} (2019) 211803
\end{center}

\vspace{\fill}

{\footnotesize 
\centerline{\copyright~\papercopyright. \href{\paperlicenceurl}{\paperlicence}.}}
\vspace*{2mm}

\end{titlepage}


\newpage
\setcounter{page}{2}
\mbox{~}

\cleardoublepage

\renewcommand{\thefootnote}{\arabic{footnote}}
\setcounter{footnote}{0}


\pagestyle{plain} 
\setcounter{page}{1}
\pagenumbering{arabic}


The noninvariance of fundamental interactions under the combined action of charge conjugation~($C$) and parity~($P$) transformations, so-called \CP violation, is a necessary condition for the dynamical generation of the baryon asymmetry of the universe~\cite{Sakharov:1967dj}. The Standard Model~(SM) of particle physics includes \CP violation through an irreducible complex phase in the Cabibbo-Kobayashi-Maskawa~(CKM) quark-mixing matrix~\cite{Cabibbo:1963yz,Kobayashi:1973fv}. The realization of \CP violation in weak interactions has been established in the $K$- and $B$-meson systems by several experiments~\cite{Christenson:1964fg,AlaviHarati:1999xp,Lai:2001ki,Aubert:2001nu,Abe:2001xe,Aubert:2004qm,Chao:2004mn,LHCb-PAPER-2013-018,LHCb-PAPER-2012-001}, and all results are well interpreted within the CKM formalism. However, the size of \CP violation in the SM appears to be too small to account for the observed matter-antimatter asymmetry~\cite{Cohen:1993nk,Riotto:1999yt,Hou:2008xd}, suggesting the existence of sources of \CP violation beyond the SM.

The observation of \CP violation in the charm sector has not been achieved yet, despite decades of experimental searches. Charm hadrons provide a unique opportunity to measure \CP violation with particles containing only up-type quarks. The size of \CP violation in charm decays is expected to be tiny in the SM, with asymmetries typically of the order of $10^{-4}$--$10^{-3}$, but due to the presence of low-energy strong-interaction effects, theoretical predictions are difficult to compute reliably~\cite{Golden:1989qx,Buccella:1994nf,Bianco:2003vb,Grossman:2006jg,Artuso:2008vf,Brod:2011re,Cheng:2012wr,Cheng:2012xb,Li:2012cfa,Franco:2012ck,Pirtskhalava:2011va,Feldmann:2012js,Brod:2012ud,Hiller:2012xm,Grossman:2012ry,Bhattacharya:2012ah,Muller:2015rna,Khodjamirian:2017zdu,Buccella:2019kpn}. Motivated by the fact that contributions of beyond-the-SM virtual particles may alter the size of \CP violation with respect to the SM expectation, a number of theoretical analyses have been performed~\cite{Grossman:2006jg,Feldmann:2012js,Atwood:2012ac,Muller:2015rna}.

Unprecedented experimental precision can be reached at LHCb in the measurement of \CP-violating asymmetries in \dkk and \dpipi decays. The inclusion of charge-conjugate decay modes is implied throughout except in asymmetry definitions. Searches for \CP violation in these decay modes have been performed by the BaBar~\cite{bib:babarpaper2008}, Belle~\cite{bib:bellepaper2008}, CDF~\cite{bib:cdfpaper,CDF:2012qw} and LHCb~\cite{LHCB-PAPER-2011-023,LHCB-PAPER-2013-003,LHCB-PAPER-2014-013,LHCB-PAPER-2015-055,LHCB-PAPER-2016-035} collaborations. The corresponding \CP asymmetries have been found to be consistent with zero within a precision of a few per mille.

This Letter presents a measurement of the difference of the time-integrated \CP asymmetries in \dkk and \dpipi decays, performed using \proton\proton collision data collected with the LHCb detector at a center-of-mass energy of 13\tev, and corresponding to an integrated luminosity of 5.9\invfb. 

The time-dependent \CP asymmetry, $A_{\CP}(f;\,t)$, between states produced as \Dz or \Dzb mesons decaying to a \CP eigenstate $f$ at time $t$ is defined as
\begin{equation}
A_{\CP}(f;\,t) \equiv \frac{\Gamma(\Dz(t) \to f)-\Gamma(\Dzb(t) \to f)}{\Gamma(\Dz(t) \to f)+\Gamma(\Dzb(t) \to f)}, \label{eq:acpf}
\end{equation}
where $\Gamma$ denotes the time-dependent rate of a given decay. For $f= \Km\Kp$ or $f= \pim\pip$, $A_{\CP}(f;\,t)$ can be expressed in terms of a direct component associated to \CP violation in the decay amplitude and another component associated to \CP violation in \Dz--\Dzb mixing or in the interference between mixing and decay. 

A time-integrated asymmetry, $A_{\CP}(f)$, can be determined, and its value will exhibit a dependence on the variation of the reconstruction efficiency as a function of the decay time. To first order in the \Dz--\Dzb mixing parameters, it can be written as~\cite{bib:cdfpaper,Gersabeck:2011xj}
\begin{equation}
A_{\CP}(f)  \approx a_{\CP}^{\rm dir}(f) - \frac{\langle t (f)\rangle}{\tau(\Dz)}\,A_\Gamma(f),\label{eq:acpphysics}
\end{equation}
where $\langle t (f)\rangle$ denotes the mean decay time of \Dz\to $f$ decays in the reconstructed sample, incorporating the effects of the time-dependent experimental efficiency, $a_{\CP}^{\rm dir}(f)$ is the direct \CP asymmetry, $\tau(\Dz)$ the \Dz lifetime and $A_\Gamma(f)$ the asymmetry between the $\Dz \to f$ and $\Dzb \to f$ effective decay widths~\cite{LHCb-PAPER-2014-069,LHCb-PAPER-2016-063}. In the limit of U-spin symmetry, the direct \CP asymmetry is equal in magnitude and opposite in sign for \Km\Kp and \pim\pip, though the size of U-spin-breaking effects at play is uncertain~\cite{Grossman:2006jg}. Taking $A_\Gamma$ to be independent of final state~\cite{Grossman:2006jg,bib:kagansokoloff,Du:2006jc}, the difference in \CP asymmetries between \decay{\Dz}{\Km\Kp} and \decay{\Dz}{\pim\pip} decays is
\begin{eqnarray}
\DACP  & \equiv &  A_{\CP}(\Km\Kp) - A_{\CP}(\pim\pip) \nonumber \label{eq:dacpdef1} \\
& \approx & \Delta a_{\CP}^{\rm dir} - \frac{\Delta \langle t \rangle}{\tau(\Dz)}\,A_\Gamma ,
\label{eq:dacpdef}
\end{eqnarray}
where $\Delta a_{\CP}^{\rm dir} \equiv a_{\CP}^{\rm dir} (K^-K^+) - a_{\CP}^{\rm dir} (\pi^-\pi^+)$ and $\Delta \langle t \rangle$ is the difference of the mean decay times $\langle t (\Km\Kp)\rangle$ and $\langle t (\pim\pip)\rangle$.

The \Dz mesons considered in this analysis are produced either promptly at a \proton\proton collision point (primary vertex, PV) in the strong decay of $\Dstar(2010)^+$ mesons~(hereafter referred to as $\Dstarp$) to a $\Dz \pi^+$ pair or at a vertex displaced from any PV in semileptonic $\Bb \to \Dz \mu^- \bar{\nu}_\mu X$ decays, where $\Bb$ denotes a hadron containing a $b$ quark and $X$ stands for potential additional particles. The flavor at production of \Dz mesons from $\Dstarp$ decays is determined from the charge of the accompanying pion~($\pi$-tagged), whereas that of \Dz mesons from semileptonic $b$-hadron decays is obtained from the charge of the accompanying muon~($\mu$-tagged). The raw asymmetries measured for $\pi$-tagged and $\mu$-tagged \Dz decays are defined as
\begin{align}
\begin{split}
\Araw^{\pi{\mbox{-}}\rm{tagged}}(f)& \equiv \frac {N\left(\Dstarp \to \Dz(f)\pi^{+}\right) - N\left(\Dstarm \to \Dzb(f)\pi^{-}\right)} {N\left(\Dstarp \to \Dz(f)\pi^{+}\right) + N\left(\Dstarm \to \Dzb(f)\pi^{-}\right)}, \\
\Araw^{\mu{\mbox{-}}\rm{tagged}}(f)& \equiv \frac {N(\Bb  \to \Dz(f)\, \mu^- \bar{\nu}_\mu X) - N(\B \to \Dzb(f)\, \mu^+ \nu_\mu X)} { N(\Bb \to \Dz(f)\, \mu^- \bar{\nu}_\mu X) + N(\B \to \Dzb(f)\, \mu^+ \nu_\mu X)},
\end{split}
\label{araw}
\end{align}
where $N$ is the measured signal yield for the given decay. These can be approximated as
\begin{align}
\begin{split}
\Araw^{\pi{\mbox{-}}\rm{tagged}}(f)& \approx A_{\CP}(f) + \AD(\pi) + \AP(\Dstar),\\
\Araw^{\mu{\mbox{-}}\rm{tagged}}(f)& \approx A_{\CP}(f) + \AD(\mu) + \AP(\B),
\end{split}
\label{def:arawstarcomponents}
\end{align}
where $\AD(\pi)$ and $\AD(\mu)$ are detection asymmetries due to different reconstruction efficiencies between positive and negative tagging pions and muons, whereas $\AP(\Dstar)$ and $\AP(\B)$ are the production asymmetries of \Dstar mesons and $b$ hadrons, arising from the hadronization of charm and beauty quarks in \proton\proton collisions~\cite{LHCb-PAPER-2016-062}. Owing to the smallness of the involved terms, which averaged over phase space for selected events are $\order(10^{-2})$ or less~\cite{LHCb-PAPER-2016-062,LHCb-PAPER-2013-033,LHCb-PAPER-2012-026,LHCb-PAPER-2012-009}, the approximations in Eqs.~\eqref{def:arawstarcomponents} are valid up to corrections of $\order(10^{-6})$. The values of $\AD(\pi)$ and $\AP(\Dstar)$, as well as those of $\AD(\mu)$ and $\AP(\B)$, are independent of the final state $f$, and thus cancel in the difference, resulting in
\begin{align}
\begin{split}
\DACP = \Araw(\Km\Kp) - \Araw(\pim\pip).
\label{DACP1}
\end{split}
\end{align}
This simple relation between $\DACP$ and the measurable raw asymmetries in $\Km\Kp$ and $\pim\pip$ makes the determination of $\DACP$ largely insensitive to systematic uncertainties.   

The LHCb detector is a single-arm forward spectrometer designed for the study of particles containing \bquark or \cquark quarks, as described in detail in Refs.~\cite{Alves:2008zz,LHCb-DP-2014-002}. The LHCb tracking system exploits a dipole magnet to measure the momentum of charged particles. Although the analysis presented in this Letter is expected to be insensitive to such effects, the magnetic-field polarity is reversed periodically during data taking to mitigate the differences of reconstruction efficiencies of particles with opposite charges. Data sets corresponding to about one half of the total integrated luminosity are recorded with each magnetic-field configuration.

The online event selection is performed by a trigger, which consists of a hardware stage based on information from the calorimeter and muon systems, followed by two software stages. In the first software stage, events used in this analysis are selected if at least one track has large transverse momentum and is incompatible with originating from any PV, or if any two-track combination forming a secondary vertex, consistent with that of a \Dz decay, is found in the event by a multivariate algorithm~\cite{BBDT,Likhomanenko:2015aba}. In between the first and second software stages, detector alignment and calibration are performed and updated constants are made available to the software trigger~\cite{LHCb-PROC-2015-011}. In the second stage, \Dz candidates are fully reconstructed using kinematic, topological and particle-identification~(PID) criteria. Requirements are placed on: the \Dz decay vertex, which must be well separated from all PVs in the event; the quality of reconstructed tracks; the \Dz transverse momentum; the angle between the \Dz momentum and its flight direction; PID information; and the impact-parameter significances~(\chisqip) of the \Dz decay products with respect to all PVs in the event, where the \chisqip is defined as the difference between the $\chi^2$ of the PV reconstructed with and without the considered particle. In the analysis of the $\mu$-tagged sample, \B candidates are formed by combining a \Dz candidate with a muon under the requirement that they are consistent with originating from a common vertex. In addition, requirements on the invariant mass of the $\Dz\mu$ system, $m(\Dz\mu)$, and on the corrected mass ($m_{\rm corr}$) are applied. The corrected mass partially recovers the missing energy of the unreconstructed particles and is defined as $m_{\rm corr} \equiv \sqrt{m(\Dz\mu)^2 + p_\perp(\Dz\mu)^2} + p_\perp(\Dz\mu)$~\cite{Kodama:1991ij}, where $p_\perp(\Dz\mu)$ is the momentum of the $\Dz\mu$ system transverse to the flight direction of the $b$ hadron, determined from the primary and $\Dz\mu$ vertices.

In the offline selection, trigger signals are associated to reconstructed particles. Selection requirements are applied on the trigger decision, taking into account the information on whether the decision was taken due to the signal decay products or to other particles produced in the event. Fiducial requirements are imposed to exclude kinematic regions characterized by large detection asymmetries for the tagging pion or muon. Very large raw asymmetries, up to 100\%, occur in certain kinematic regions because, for a given magnet polarity, low-momentum particles of one charge at large or small polar angles in the horizontal plane may be deflected out of the detector or into the (uninstrumented)~LHC beam pipe, whereas particles with the other charge are more likely to remain within the acceptance~\cite{PRL:supplemental}. About 35\% and 10\% of the selected candidates are rejected by these fiducial requirements for the $\pi$-tagged and $\mu$-tagged samples, respectively. In the retained samples, raw asymmetries are typically at the percent level or below. For $\pi$-tagged \Dz mesons, a requirement on the \Dz \chisqip is applied to suppress the background of \Dz mesons from $B$ decays, and PID requirements on the \Dz decay products are further tightened. Then the \Dz and pion candidates are combined to form \Dstarp candidates by requiring a good fit quality of the $D^{*+}$ vertex and the invariant mass of \Dz candidates to lie within a range of about $\pm3$ standard deviations around the known $\Dz$ mass. The $D^{*+}$ vertex is determined as a common vertex of $D^0$ and tagging $\pip$ candidates, and is constrained to coincide with the nearest PV~\cite{Hulsbergen2005566}.

For $\mu$-tagged mesons, the \B candidates are further filtered using a dedicated boosted decision tree~(BDT) to suppress the combinatorial background due to random combinations of charged kaon or pion pairs not originating from a \Dz decay. The variables used in the BDT to discriminate signal from combinatorial background are: the fit quality of the \Dz and the \B decay vertices; the \Dz flight distance; the \Dz impact parameter, \ie, the minimum distance of its trajectory to the nearest PV; the transverse momenta of the \Dz decay products, the significance of the distance between the \Dz and \B decay vertices; the invariant mass $m(\Dz\mu)$ and the corrected mass $m_{\rm corr}$. To suppress background from $b$-hadron decays to $\cquark\cquarkbar \pi^\pm X$ ($\cquark\cquarkbar K^\pm X$), where the $\cquark\cquarkbar$ resonance decays to a pair of muons, \Dz candidates are vetoed if the invariant mass of the $\mu^\mp \pi^\pm$~($\mu^\mp K^\pm$) pair, where the pion~(kaon) is given the muon mass hypothesis, lies within a window of about $\pm 50\mevcc$ around the \jpsi or \psitwos known masses.

The data sample includes events with multiple \Dstarp and \B candidates. The majority of these events contain the same reconstructed \Dz meson combined with different tagging pions or muons. When multiple candidates are present in the event, only one is kept randomly. The fractions of events with multiple candidates are about 10\% and 0.4\% in the $\pi$-tagged and $\mu$-tagged samples, respectively. A small fraction of events, of the order of per mille, belong to both the selected $\pi$-tagged and $\mu$-tagged samples.

As the detection and production asymmetries are expected to depend on the kinematics of the reconstructed particles, the cancellation in the difference between the raw asymmetries in Eq.~\eqref{DACP1} may be incomplete if the kinematic distributions of reconstructed \Dstarp or \B candidates and of the tagging pions or muons differ between the $K^-K^+$ and $\pi^-\pi^+$ decay modes. For this reason, a small correction to the $K^-K^+$ sample is applied by means of a weighting procedure~\cite{PRL:supplemental}. For the $\pi$-tagged sample, candidate-by-candidate weights are calculated by taking the ratio between the three-dimensional background-subtracted distributions of transverse momentum, azimuthal angle and pseudorapidity of the \Dstarp meson in the $K^-K^+$ and $\pi^-\pi^+$ modes. An analogous procedure is followed for the $\mu$-tagged sample, where \Dz distributions are used instead of those of the \Dstarp meson. It is then checked \emph{a posteriori} that the distributions of the same variables for tagging pions and muons are also equalized by the weighting. The application of the weights leads to a small variation of \DACP, below $10^{-4}$ for both the $\pi$-tagged and $\mu$-tagged samples.   

\begin{figure}[t]
\centering
\includegraphics[width=0.48\textwidth]{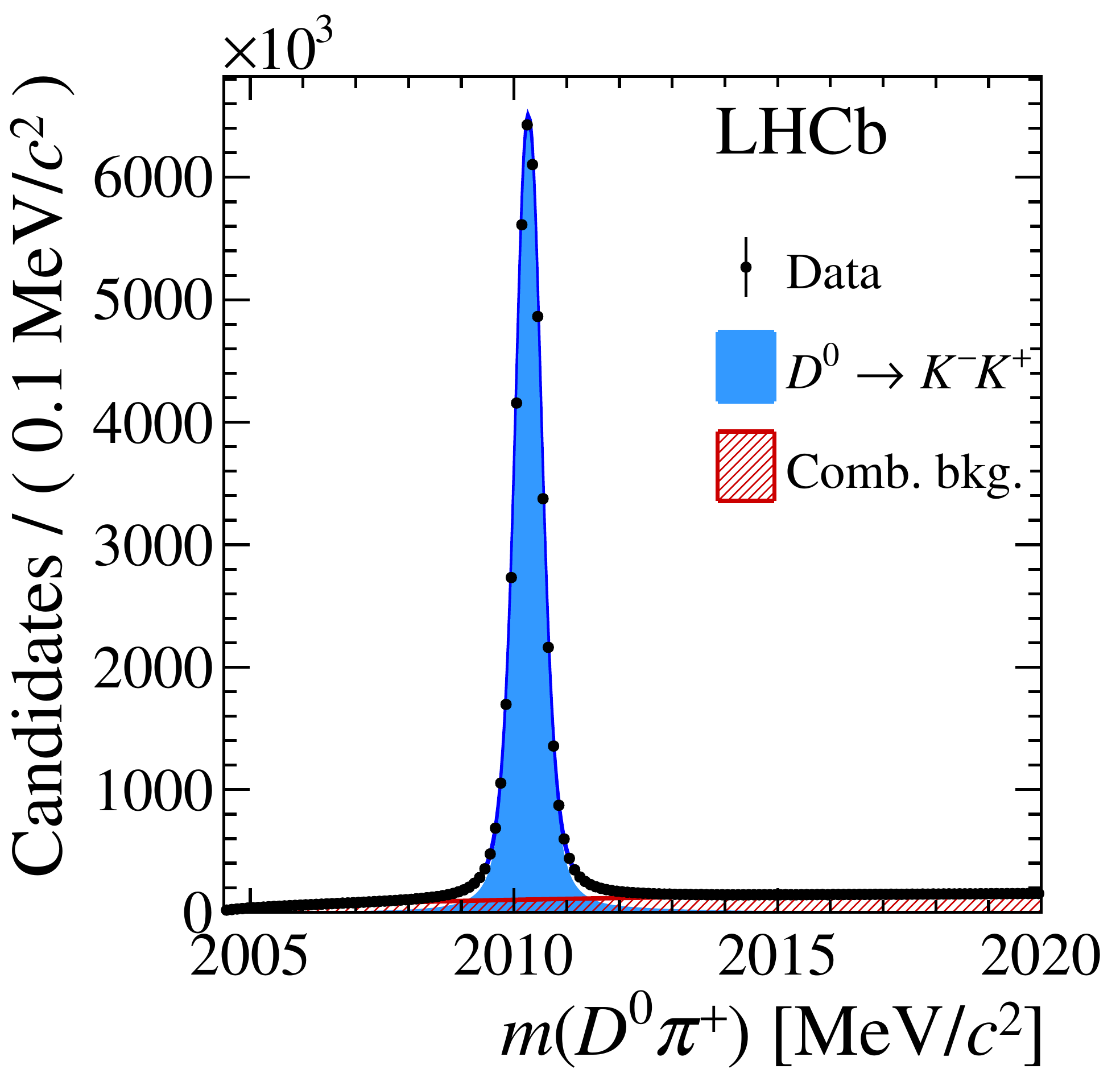}
\includegraphics[width=0.48\textwidth]{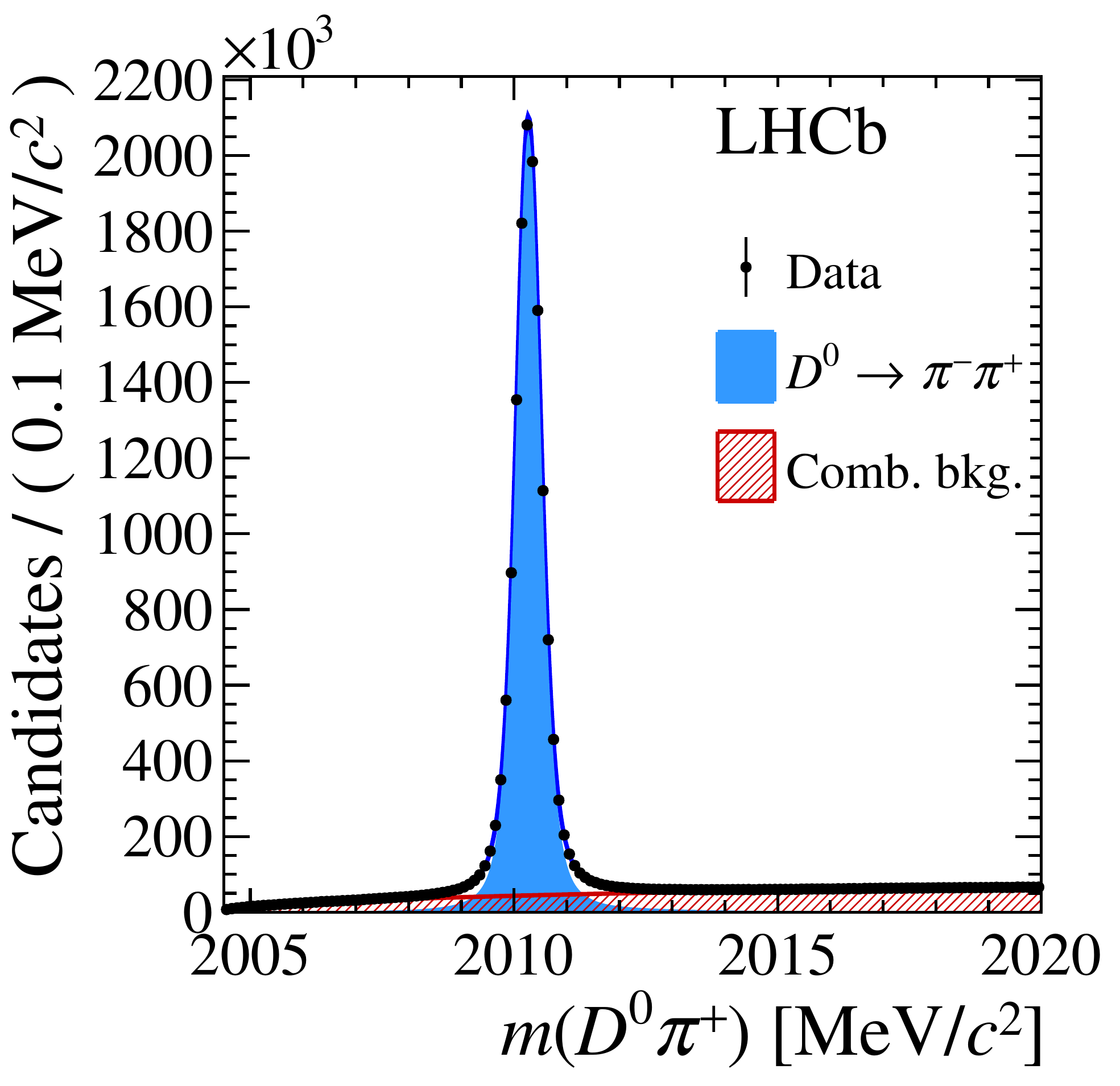}
\includegraphics[width=0.48\textwidth]{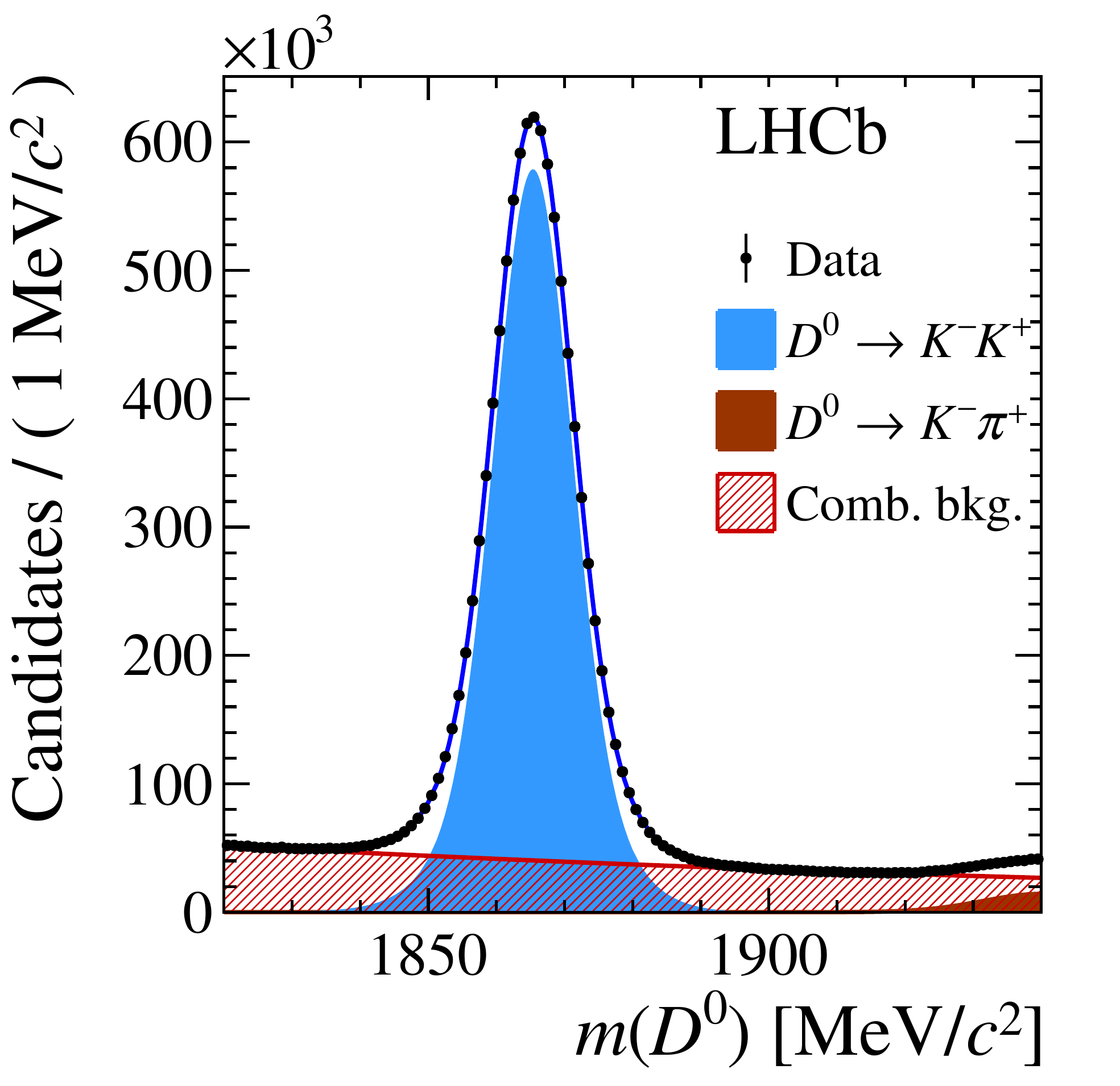}
\includegraphics[width=0.48\textwidth]{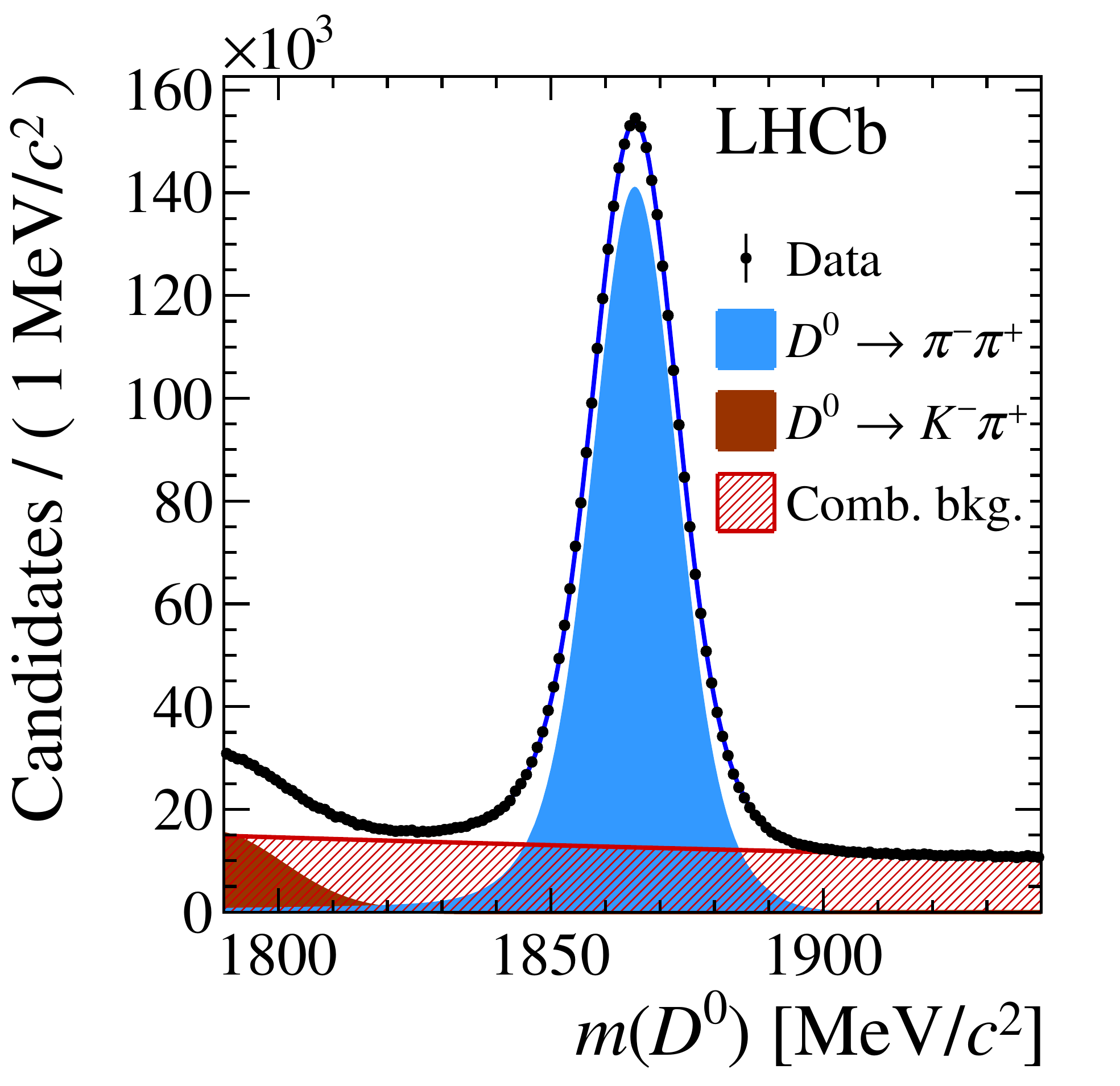}
\caption{Mass distributions of selected (top)~$\pi^\pm$-tagged and (bottom)~$\mu^\pm$-tagged candidates for (left)~$\Km\Kp$ and (right)~$\pim\pip$ final states of the \Dz-meson decays, with fit projections overlaid.}
\label{fig:fits}
\end{figure}

The raw asymmetries of signal and background components for each decay mode are free parameters determined by means of simultaneous least-square fits to the binned mass distributions of \Dstarp and \Dstarm candidates for the $\pi$-tagged sample, or \Dz and \Dzb candidates for the $\mu$-tagged sample. In particular, in the analysis of the $\pi$-tagged sample the fits are performed to the $m(\Dz\pip)$ and $m(\Dzb\pim)$ distributions. As outlined in Ref.~\cite{bib:cdfpaper}, using these distributions has the advantage that they are the same for both \dkk and \dpipi decay modes.

The signal mass model, which is obtained from simulation, consists of the sum of three Gaussian functions and a Johnson $S_{{U}}$ function~\cite{Johnson:1949zj}, whose parameters are free to be adjusted by the fit to the data. The mean values of the Gaussian functions are distinct for positive and negative tags, whereas widths and fractions are shared. The parameters of the Johnson $S_{{U}}$ function, which accounts for the slight asymmetric shape of the signal distribution due to the proximity of the $m(\Dz)+m(\pip)$ threshold, are also shared. The combinatorial background is described by an empirical function of the form \mbox{$[m(\Dz\pip) - m(\Dz) - m(\pip)]^\alpha e^{\beta\, m(\Dz\pip)}$}, where $\alpha$ and $\beta$ are two free parameters which are shared among positive and negative tags. In the analysis of the $\mu$-tagged sample, the fits are performed to the $m(\Dz)$ distributions. The signal is described by the sum of two Gaussian functions convolved with a truncated power-law function that accounts for final-state photon radiation effects, whereas the combinatorial background is described by an exponential function. A small contribution from $\decay{\Dz}{\Km\pip}$ decays with a misidentified kaon or pion is also visible, which is modeled as the tail of a Gaussian function. Separate fits are performed to subsamples of data collected with different magnet polarities and in different years. All partial \DACP values corresponding to each subsample are found to be in good agreement and then averaged to obtain the final results. If single fits are performed to the overall $\pi$-tagged and $\mu$-tagged samples, small differences of the order of a few $10^{-5}$ are found. The $m(\Dz\pip)$ and $m(\Dz)$ distributions corresponding to the entire samples are displayed in Fig.~\ref{fig:fits}~(see also Ref.~\cite{PRL:supplemental} for the corresponding asymmetries as a function of mass). The $\pi$-tagged~($\mu$-tagged) signal yields are approximately $44$~($9$) million  $\dkk$ decays and $14$~($3$) million $\D^0 \rightarrow \pi^-\pi^+$ decays. In the case of $\pi$-tagged decays, the fits to the $m(\Dz\pip)$ distributions do not distinguish between background that produces peaks in $m(\Dz\pip)$, which can arise from \Dstarp decays where the correct tagging pion is found but the \Dz meson is misreconstructed, and signal. The effect on \DACP of residual peaking backgrounds, suppressed by selection requirements to less than 1\% of the number of signal candidates, is evaluated as a systematic uncertainty.

Studies of systematic uncertainties on \DACP are carried out independently for the $\pi$-tagged and $\mu$-tagged samples. Several sources affecting the measurement are considered. In the case of $\pi$-tagged decays, the dominant systematic uncertainty is related to the knowledge of the signal and background mass models. It is evaluated by generating pseudoexperiments according to the baseline fit model, then fitting alternative models to those data. A value of $0.6\times 10^{-4}$ is assigned as a systematic uncertainty, corresponding to the largest variation observed using the alternative functions. Possible differences between $\Dz\pip$ and $\Dzb\pim$ invariant-mass shapes are investigated by studying a sample of 232 million $\decay{\Dstarp}{\Dz(\Km\pip)\pip}$ and $\decay{\Dstarm}{\Dzb(\Kp\pim)\pim}$ decays. The effect on \DACP is estimated to be order of $10^{-5}$ at most, hence negligible. A similar study with pseudoexperiments is also performed with the $\mu$-tagged sample and a value of $2\times 10^{-4}$ is found.

In the case of $\mu$-tagged decays, the main systematic uncertainty is due to the possibility that the \Dz flavor is not tagged correctly by the muon charge because of misreconstruction. The probability of wrongly assigning the \Dz flavor~(mistag) is studied with a large sample of $\mu$-tagged \dkpi decays by comparing the charges of kaon and muon candidates. Mistag rates are found to be at the percent level and compatible for positively and negatively tagged decays. The corresponding systematic uncertainty is estimated to be $4\times 10^{-4}$, also taking into account the fact that wrongly tagged decays include a fraction of doubly Cabibbo-suppressed \dkpiDCS and mixed $\Dz \to \Dzb \to \Kp\pim$ decays, calculated to be $0.39\%$ with negligible uncertainty for both the \Kp\pim and \Km\pip final states using input from Ref.~\cite{LHCb-PAPER-2017-046}. 

Systematic uncertainties of $0.2\times 10^{-4}$ and $1\times 10^{-4}$ accounting for the knowledge of the weights used in the kinematic weighting procedure are assessed for $\pi$-tagged and $\mu$-tagged decays, respectively. Although suppressed by the requirement that the \Dz trajectory points back to the PV, a fraction of \Dz mesons from $B$ decays is still present in the final $\pi$-tagged sample. As \dkk and \dpipi decays may have different levels of contamination, the value of \DACP may be biased because of an incomplete cancellation of the production asymmetries of $b$ hadrons. The fractions of \Dz mesons from $B$ decays are estimated by performing a fit to the distribution of the \Dz-candidate impact parameter in the plane transverse to the beam direction~\cite{PRL:supplemental}. The corresponding systematic uncertainty is estimated to be $0.3\times 10^{-4}$. A systematic uncertainty associated to the presence of background components peaking in $m(\Dz\pi)$ and not in $m(\Dz)$ is determined by fits to the $m(\Dz)$ distributions~\cite{PRL:supplemental}, where these components are modeled using fast simulation~\cite{Cowan:2016tnm}. The main sources are the $\decay{\Dz}{\Km\pip\piz}$ decay for the $\Kp\Km$ final state, and the $\Dz\to\pim\mup\neum$ and $\Dz\to\pim\ep\neue$ decays for the $\pip\pim$ final state. Yields and raw asymmetries of the peaking-background components measured from the fits are then used as inputs to pseudoexperiments designed to evaluate the corresponding effects on the determination of \DACP. A value of $0.5\times 10^{-4}$ is assigned as a systematic uncertainty.

In the case of $\mu$-tagged decays, the fractions of reconstructed \Bb decays can be slightly different between the $K^-K^+$ and $\pi^-\pi^+$ decay modes, which could lead to a small bias in \DACP. Using the LHCb measurements of the $b$-hadron production asymmetries~\cite{LHCb-PAPER-2016-062}, the systematic uncertainty on \DACP is estimated to be $1\times 10^{-4}$. The combination of a difference in the \B reconstruction efficiency as a function of the decay time between the \dkk and \dpipi modes and the presence of neutral $B$-meson oscillations may also cause an imperfect cancellation of \APB in \DACP. The associated systematic uncertainty is estimated to be $2\times 10^{-4}$.

All individual contributions are summed in quadrature to give total systematic uncertainties on \DACP of $0.9\times 10^{-4}$ and $5\times 10^{-4}$ for the $\pi$-tagged and $\mu$-tagged samples, respectively. A summary of all systematic uncertainties is reported in Table~\ref{tab:syst}. Other possible systematic uncertainties are investigated and found to be negligible. 

\begin{table}[t]
\centering
\caption{Systematic uncertainties on \DACP for $\pi$- and $\mu$-tagged decays (in $10^{-4}$). The total uncertainties are obtained as the sums in quadrature of the individual contributions.}
\label{tab:syst}
        \begin{tabular}{lcc}
            \hline\hline    
                Source & $\pi$-tagged & $\mu$-tagged \\
                \hline
                Fit model                        & 0.6         & 2  \\
                Mistag                           & --          & 4\\
                Weighting                        & 0.2         & 1 \\
                Secondary decays                 & 0.3         & -- \\
                Peaking background               & 0.5         & --\\
                \B fractions                     & --          & 1\\
                \B reco. efficiency              & --          & 2 \\
                \hline
                Total                            & 0.9         & 5\\
                \hline\hline
        \end{tabular}
\end{table}

Numerous additional robustness checks are carried out~\cite{PRL:supplemental}. The measured value of $\Delta A_{\CP}$ is studied as a function of several variables, notably including: the azimuthal angle, \chisqip, transverse momentum and pseudorapidity of $\pi$-tagged and $\mu$-tagged \Dz mesons as well as of the tagging pions or muons; the $\chisq$ of the \Dstarp and \B vertex fits; the track quality of the tagging pion and the charged-particle multiplicity in the event. Furthermore, the total sample is split into subsamples taken in different run periods within the years of data taking, also distinguishing different magnet polarities. No evidence for unexpected dependences of \DACP is found in any of these tests. A check using more stringent PID requirements is performed, and all variations of \DACP are found to be compatible within statistical uncertainties. An additional check concerns the measurement of $\Delta A_{\rm bkg}$, that is the difference of the background raw asymmetries in $\Km\Kp$ and $\pim\pip$ final states. As the prompt background is mainly composed of genuine \Dz candidates paired with unrelated pions originating from the PV, $\Delta A_{\rm bkg}$ is expected to be compatible with zero. A value of $\Delta A_{\rm bkg} = (-2 \pm 4) \times 10^{-4}$ is obtained.

The difference of time-integrated \CP asymmetries of \decay{\Dz}{\Km\Kp} and \Dz\to\pim\pip decays is measured using 13\tev $pp$ collision data collected with the LHCb detector and corresponding to an integrated luminosity of 5.9\invfb. The results are
\begin{align*}
\DACP^{\pi{\mbox{-}}\rm{tagged}} = \left[-18.2 \pm 3.2\,\rm{(stat.)} \pm 0.9\,\rm{(syst.)} \right]\times 10^{-4},\\ 
\DACP^{\mu{\mbox{-}}\rm{tagged}} = \left[-9 \pm 8\,\rm{(stat.)}  \pm 5\,\rm{(syst.)} \right] \times 10^{-4}.    
\end{align*}
Both measurements are in good agreement with world averages~\cite{HFLAV16} and previous LHCb results~\cite{LHCB-PAPER-2015-055,LHCB-PAPER-2014-013}. 

By making a full combination with previous LHCb measurements~\cite{LHCB-PAPER-2015-055,LHCB-PAPER-2014-013}, the following value of \DACP is obtained  
\begin{equation}
\DACP = \left(-15.4 \pm 2.9 \right)\times 10^{-4}, \nonumber
\end{equation}
where the uncertainty includes statistical and systematic contributions.
The significance of the deviation from zero corresponds to 5.3 standard deviations. This is the first observation of \CP violation in the decay of charm hadrons.

The interpretation of \DACP in terms of direct \CP violation and $A_\Gamma$ requires knowledge of the difference of reconstructed mean decay times for \dkk and \dpipi decays normalized to the \Dz lifetime, as shown in Eq.~\eqref{eq:dacpdef}. The values corresponding to the present measurements are $\Delta\!\mean{t}^{\pi{\mbox{-}}\rm{tagged}}/\tau(\Dz)  =  0.135 \pm 0.002$ and $\Delta\!\mean{t}^{\mu{\mbox{-}}\rm{tagged}}/\tau(\Dz)  =  -0.003 \pm 0.001$, 
whereas that corresponding to the full combination is $\Delta\mean{t}/\tau(\Dz)  =  0.115 \pm 0.002$. The uncertainties include statistical and systematic contributions, and the world average of the \Dz lifetime is used~\cite{PDG2018}.

By using in addition the LHCb average \mbox{$\agamma=(-2.8 \pm 2.8)\times 10^{-4}$}~\cite{LHCb-PAPER-2014-069,LHCb-PAPER-2016-063}, from Eq.~\eqref{eq:dacpdef} it is possible to derive
\begin{equation}
\Delta a_{\CP}^{\rm dir} = \left(-15.7 \pm 2.9 \right)\times 10^{-4}, \nonumber
\end{equation}
which shows that, as expected, \DACP is primarily sensitive to direct \CP violation. The overall improvement in precision brought by the present analysis to the knowledge of $\Delta a_{\CP}^{\rm dir}$ is apparent when comparing with the value obtained from previous measurements, $\Delta a_{\CP}^{\rm dir} = \left(-13.4 \pm 7.0 \right)\times 10^{-4}$~\cite{HFLAV16}.

In summary, this Letter reports the first observation of a nonzero \CP asymmetry in charm decays, using large samples of \dkk and \dpipi decays collected with the LHCb detector. The result is consistent with, although in magnitude at the upper end of, SM expectations, which lie in the range $10^{-4}$--$10^{-3}$~\cite{Golden:1989qx,Buccella:1994nf,Bianco:2003vb,Grossman:2006jg,Artuso:2008vf,Brod:2011re,Cheng:2012wr,Cheng:2012xb,Li:2012cfa,Franco:2012ck,Pirtskhalava:2011va,Feldmann:2012js,Brod:2012ud,Hiller:2012xm,Grossman:2012ry,Bhattacharya:2012ah,Muller:2015rna,Khodjamirian:2017zdu,Buccella:2019kpn}. In particular, the result challenges predictions based on first-principle QCD dynamics~\cite{Grossman:2006jg,Khodjamirian:2017zdu}. It complies with predictions based on flavor-\grpsuthree symmetry, if one assumes a dynamical enhancement of the penguin amplitude~\cite{Golden:1989qx,Pirtskhalava:2011va,Feldmann:2012js,Brod:2012ud,Hiller:2012xm,Grossman:2012ry,Muller:2015rna}. In the next decade, further measurements with charmed particles, along with possible theoretical improvements, will help clarify the physics picture, and establish whether this result is consistent with the SM or indicates the presence of new dynamics in the up-quark sector.
\section*{Acknowledgements}
%
%
\noindent We express our gratitude to our colleagues in the CERN
accelerator departments for the excellent performance of the LHC. We
thank the technical and administrative staff at the LHCb
institutes.
We acknowledge support from CERN and from the national agencies:
CAPES, CNPq, FAPERJ and FINEP (Brazil); 
MOST and NSFC (China); 
CNRS/IN2P3 (France); 
BMBF, DFG and MPG (Germany); 
INFN (Italy); 
NWO (Netherlands); 
MNiSW and NCN (Poland); 
MEN/IFA (Romania); 
MSHE (Russia); 
MinECo (Spain); 
SNSF and SER (Switzerland); 
NASU (Ukraine); 
STFC (United Kingdom); 
NSF (USA).
We acknowledge the computing resources that are provided by CERN, IN2P3
(France), KIT and DESY (Germany), INFN (Italy), SURF (Netherlands),
PIC (Spain), GridPP (United Kingdom), RRCKI and Yandex
LLC (Russia), CSCS (Switzerland), IFIN-HH (Romania), CBPF (Brazil),
PL-GRID (Poland) and OSC (USA).
We are indebted to the communities behind the multiple open-source
software packages on which we depend.
Individual groups or members have received support from
AvH Foundation (Germany);
EPLANET, Marie Sk\l{}odowska-Curie Actions and ERC (European Union);
ANR, Labex P2IO and OCEVU, and R\'{e}gion Auvergne-Rh\^{o}ne-Alpes (France);
Key Research Program of Frontier Sciences of CAS, CAS PIFI, and the Thousand Talents Program (China);
RFBR, RSF and Yandex LLC (Russia);
GVA, XuntaGal and GENCAT (Spain);
the Royal Society
and the Leverhulme Trust (United Kingdom);
Laboratory Directed Research and Development program of LANL (USA).


\addcontentsline{toc}{section}{References}
\bibliographystyle{LHCb}
\bibliography{main,standard,LHCb-PAPER,LHCb-CONF,LHCb-DP,LHCb-TDR}

\ifx\mcitethebibliography\mciteundefinedmacro
\PackageError{LHCb.bst}{mciteplus.sty has not been loaded}
{This bibstyle requires the use of the mciteplus package.}\fi
\providecommand{\href}[2]{#2}
\begin{mcitethebibliography}{10}
\mciteSetBstSublistMode{n}
\mciteSetBstMaxWidthForm{subitem}{\alph{mcitesubitemcount})}
\mciteSetBstSublistLabelBeginEnd{\mcitemaxwidthsubitemform\space}
{\relax}{\relax}

\bibitem{Sakharov:1967dj}
A.~D. Sakharov, \ifthenelse{\boolean{articletitles}}{\emph{{Violation of CP
  invariance, C asymmetry, and baryon asymmetry of the universe}},
  }{}\href{https://doi.org/10.1070/PU1991v034n05ABEH002497}{Pisma Zh.\ Eksp.\
  Teor.\ Fiz.\  \textbf{5} (1967) 32}\relax
\mciteBstWouldAddEndPuncttrue
\mciteSetBstMidEndSepPunct{\mcitedefaultmidpunct}
{\mcitedefaultendpunct}{\mcitedefaultseppunct}\relax
\EndOfBibitem
\bibitem{Cabibbo:1963yz}
N.~Cabibbo, \ifthenelse{\boolean{articletitles}}{\emph{{Unitary symmetry and
  leptonic decays}},
  }{}\href{https://doi.org/10.1103/PhysRevLett.10.531}{Phys.\ Rev.\ Lett.\
  \textbf{10} (1963) 531}\relax
\mciteBstWouldAddEndPuncttrue
\mciteSetBstMidEndSepPunct{\mcitedefaultmidpunct}
{\mcitedefaultendpunct}{\mcitedefaultseppunct}\relax
\EndOfBibitem
\bibitem{Kobayashi:1973fv}
M.~Kobayashi and T.~Maskawa,
  \ifthenelse{\boolean{articletitles}}{\emph{{\CP-violation in the
  renormalizable theory of weak interaction}},
  }{}\href{https://doi.org/10.1143/PTP.49.652}{Prog.\ Theor.\ Phys.\
  \textbf{49} (1973) 652}\relax
\mciteBstWouldAddEndPuncttrue
\mciteSetBstMidEndSepPunct{\mcitedefaultmidpunct}
{\mcitedefaultendpunct}{\mcitedefaultseppunct}\relax
\EndOfBibitem
\bibitem{Christenson:1964fg}
J.~H. Christenson, J.~W. Cronin, V.~L. Fitch, and R.~Turlay,
  \ifthenelse{\boolean{articletitles}}{\emph{{Evidence for the $2\pi$ decay of
  the $K_2^0$ meson}},
  }{}\href{https://doi.org/10.1103/PhysRevLett.13.138}{Phys.\ Rev.\ Lett.\
  \textbf{13} (1964) 138}\relax
\mciteBstWouldAddEndPuncttrue
\mciteSetBstMidEndSepPunct{\mcitedefaultmidpunct}
{\mcitedefaultendpunct}{\mcitedefaultseppunct}\relax
\EndOfBibitem
\bibitem{AlaviHarati:1999xp}
KTeV collaboration, A.~Alavi-Harati {\em et~al.},
  \ifthenelse{\boolean{articletitles}}{\emph{{Observation of direct CP
  violation in $K_{\rm S,L} \to \pi \pi$ decays}},
  }{}\href{https://doi.org/10.1103/PhysRevLett.83.22}{Phys.\ Rev.\ Lett.\
  \textbf{83} (1999) 22},
  \href{http://arxiv.org/abs/hep-ex/9905060}{{\normalfont\ttfamily
  arXiv:hep-ex/9905060}}\relax
\mciteBstWouldAddEndPuncttrue
\mciteSetBstMidEndSepPunct{\mcitedefaultmidpunct}
{\mcitedefaultendpunct}{\mcitedefaultseppunct}\relax
\EndOfBibitem
\bibitem{Lai:2001ki}
NA48 collaboration, A.~Lai {\em et~al.},
  \ifthenelse{\boolean{articletitles}}{\emph{{A precise measurement of the
  direct CP violation parameter $Re(\eps^\prime / \eps)$}},
  }{}\href{https://doi.org/10.1007/s100520100822}{Eur.\ Phys.\ J.\
  \textbf{C22} (2001) 231},
  \href{http://arxiv.org/abs/hep-ex/0110019}{{\normalfont\ttfamily
  arXiv:hep-ex/0110019}}\relax
\mciteBstWouldAddEndPuncttrue
\mciteSetBstMidEndSepPunct{\mcitedefaultmidpunct}
{\mcitedefaultendpunct}{\mcitedefaultseppunct}\relax
\EndOfBibitem
\bibitem{Aubert:2001nu}
BaBar collaboration, B.~Aubert {\em et~al.},
  \ifthenelse{\boolean{articletitles}}{\emph{{Observation of \CP violation in
  the $B^0$ meson system}},
  }{}\href{https://doi.org/10.1103/PhysRevLett.87.091801}{Phys.\ Rev.\ Lett.\
  \textbf{87} (2001) 091801},
  \href{http://arxiv.org/abs/hep-ex/0107013}{{\normalfont\ttfamily
  arXiv:hep-ex/0107013}}\relax
\mciteBstWouldAddEndPuncttrue
\mciteSetBstMidEndSepPunct{\mcitedefaultmidpunct}
{\mcitedefaultendpunct}{\mcitedefaultseppunct}\relax
\EndOfBibitem
\bibitem{Abe:2001xe}
Belle collaboration, K.~Abe {\em et~al.},
  \ifthenelse{\boolean{articletitles}}{\emph{{Observation of large \CP
  violation in the neutral $B$ meson system}},
  }{}\href{https://doi.org/10.1103/PhysRevLett.87.091802}{Phys.\ Rev.\ Lett.\
  \textbf{87} (2001) 091802},
  \href{http://arxiv.org/abs/hep-ex/0107061}{{\normalfont\ttfamily
  arXiv:hep-ex/0107061}}\relax
\mciteBstWouldAddEndPuncttrue
\mciteSetBstMidEndSepPunct{\mcitedefaultmidpunct}
{\mcitedefaultendpunct}{\mcitedefaultseppunct}\relax
\EndOfBibitem
\bibitem{Aubert:2004qm}
BaBar collaboration, B.~Aubert {\em et~al.},
  \ifthenelse{\boolean{articletitles}}{\emph{{Direct \CP violating asymmetry in
  \mbox{$B^0 \to K^+\pi^-$} decays}},
  }{}\href{https://doi.org/10.1103/PhysRevLett.93.131801}{Phys.\ Rev.\ Lett.\
  \textbf{93} (2004) 131801},
  \href{http://arxiv.org/abs/hep-ex/0407057}{{\normalfont\ttfamily
  arXiv:hep-ex/0407057}}\relax
\mciteBstWouldAddEndPuncttrue
\mciteSetBstMidEndSepPunct{\mcitedefaultmidpunct}
{\mcitedefaultendpunct}{\mcitedefaultseppunct}\relax
\EndOfBibitem
\bibitem{Chao:2004mn}
Belle collaboration, Y.~Chao {\em et~al.},
  \ifthenelse{\boolean{articletitles}}{\emph{{Evidence for direct CP violation
  in $B^0\to \Kp \pim$ decays}},
  }{}\href{https://doi.org/10.1103/PhysRevLett.93.191802}{Phys.\ Rev.\ Lett.\
  \textbf{93} (2004) 191802},
  \href{http://arxiv.org/abs/hep-ex/0408100}{{\normalfont\ttfamily
  arXiv:hep-ex/0408100}}\relax
\mciteBstWouldAddEndPuncttrue
\mciteSetBstMidEndSepPunct{\mcitedefaultmidpunct}
{\mcitedefaultendpunct}{\mcitedefaultseppunct}\relax
\EndOfBibitem
\bibitem{LHCb-PAPER-2013-018}
LHCb collaboration, R.~Aaij {\em et~al.},
  \ifthenelse{\boolean{articletitles}}{\emph{{First observation of \CP
  violation in the decays of $\Bs$ mesons}},
  }{}\href{https://doi.org/10.1103/PhysRevLett.110.221601}{Phys.\ Rev.\ Lett.\
  \textbf{110} (2013) 221601},
  \href{http://arxiv.org/abs/1304.6173}{{\normalfont\ttfamily
  arXiv:1304.6173}}\relax
\mciteBstWouldAddEndPuncttrue
\mciteSetBstMidEndSepPunct{\mcitedefaultmidpunct}
{\mcitedefaultendpunct}{\mcitedefaultseppunct}\relax
\EndOfBibitem
\bibitem{LHCb-PAPER-2012-001}
LHCb collaboration, R.~Aaij {\em et~al.},
  \ifthenelse{\boolean{articletitles}}{\emph{{Observation of \CP violation in
  $\Bpm\to \D\Kpm$ decays}},
  }{}\href{https://doi.org/10.1016/j.physletb.2012.04.060}{Phys.\ Lett.\
  \textbf{B712} (2012) 203}, Erratum
  \href{https://doi.org/10.1016/j.physletb.2012.05.060}{ibid.\   \textbf{B713}
  (2012) 351}, \href{http://arxiv.org/abs/1203.3662}{{\normalfont\ttfamily
  arXiv:1203.3662}}\relax
\mciteBstWouldAddEndPuncttrue
\mciteSetBstMidEndSepPunct{\mcitedefaultmidpunct}
{\mcitedefaultendpunct}{\mcitedefaultseppunct}\relax
\EndOfBibitem
\bibitem{Cohen:1993nk}
A.~G. Cohen, D.~B. Kaplan, and A.~E. Nelson,
  \ifthenelse{\boolean{articletitles}}{\emph{{Progress in electroweak
  baryogenesis}},
  }{}\href{https://doi.org/10.1146/annurev.ns.43.120193.000331}{Ann.\ Rev.\
  Nucl.\ Part.\ Sci.\  \textbf{43} (1993) 27},
  \href{http://arxiv.org/abs/hep-ph/9302210}{{\normalfont\ttfamily
  arXiv:hep-ph/9302210}}\relax
\mciteBstWouldAddEndPuncttrue
\mciteSetBstMidEndSepPunct{\mcitedefaultmidpunct}
{\mcitedefaultendpunct}{\mcitedefaultseppunct}\relax
\EndOfBibitem
\bibitem{Riotto:1999yt}
A.~Riotto and M.~Trodden, \ifthenelse{\boolean{articletitles}}{\emph{{Recent
  progress in baryogenesis}},
  }{}\href{https://doi.org/10.1146/annurev.nucl.49.1.35}{Ann.\ Rev.\ Nucl.\
  Part.\ Sci.\  \textbf{49} (1999) 35},
  \href{http://arxiv.org/abs/hep-ph/9901362}{{\normalfont\ttfamily
  arXiv:hep-ph/9901362}}\relax
\mciteBstWouldAddEndPuncttrue
\mciteSetBstMidEndSepPunct{\mcitedefaultmidpunct}
{\mcitedefaultendpunct}{\mcitedefaultseppunct}\relax
\EndOfBibitem
\bibitem{Hou:2008xd}
W.-S. Hou, \ifthenelse{\boolean{articletitles}}{\emph{{Source of CP violation
  for the baryon asymmetry of the universe}},
  }{}\href{https://www.ps-taiwan.org/cjp/issues.php?vol=47&num=2}{Chin.\ J.\
  Phys.\  \textbf{47} (2009) 134},
  \href{http://arxiv.org/abs/0803.1234}{{\normalfont\ttfamily
  arXiv:0803.1234}}\relax
\mciteBstWouldAddEndPuncttrue
\mciteSetBstMidEndSepPunct{\mcitedefaultmidpunct}
{\mcitedefaultendpunct}{\mcitedefaultseppunct}\relax
\EndOfBibitem
\bibitem{Golden:1989qx}
M.~Golden and B.~Grinstein,
  \ifthenelse{\boolean{articletitles}}{\emph{{Enhanced CP violations in
  hadronic charm decays}},
  }{}\href{https://doi.org/10.1016/0370-2693(89)90353-5}{Phys.\ Lett.\
  \textbf{B222} (1989) 501}\relax
\mciteBstWouldAddEndPuncttrue
\mciteSetBstMidEndSepPunct{\mcitedefaultmidpunct}
{\mcitedefaultendpunct}{\mcitedefaultseppunct}\relax
\EndOfBibitem
\bibitem{Buccella:1994nf}
F.~Buccella {\em et~al.},
  \ifthenelse{\boolean{articletitles}}{\emph{{Nonleptonic weak decays of
  charmed mesons}}, }{}\href{https://doi.org/10.1103/PhysRevD.51.3478}{Phys.\
  Rev.\  \textbf{D51} (1995) 3478},
  \href{http://arxiv.org/abs/hep-ph/9411286}{{\normalfont\ttfamily
  arXiv:hep-ph/9411286}}\relax
\mciteBstWouldAddEndPuncttrue
\mciteSetBstMidEndSepPunct{\mcitedefaultmidpunct}
{\mcitedefaultendpunct}{\mcitedefaultseppunct}\relax
\EndOfBibitem
\bibitem{Bianco:2003vb}
S.~Bianco, F.~L. Fabbri, D.~Benson, and I.~Bigi,
  \ifthenelse{\boolean{articletitles}}{\emph{{A Cicerone for the physics of
  charm}}, }{}\href{https://doi.org/10.1393/ncr/i2003-10003-1}{Riv.\ Nuovo
  Cim.\  \textbf{26N7} (2003) 1},
  \href{http://arxiv.org/abs/hep-ex/0309021}{{\normalfont\ttfamily
  arXiv:hep-ex/0309021}}\relax
\mciteBstWouldAddEndPuncttrue
\mciteSetBstMidEndSepPunct{\mcitedefaultmidpunct}
{\mcitedefaultendpunct}{\mcitedefaultseppunct}\relax
\EndOfBibitem
\bibitem{Grossman:2006jg}
Y.~Grossman, A.~L. Kagan, and Y.~Nir,
  \ifthenelse{\boolean{articletitles}}{\emph{{New physics and CP violation in
  singly Cabibbo suppressed D decays}},
  }{}\href{https://doi.org/10.1103/PhysRevD.75.036008}{Phys.\ Rev.\
  \textbf{D75} (2007) 036008},
  \href{http://arxiv.org/abs/hep-ph/0609178}{{\normalfont\ttfamily
  arXiv:hep-ph/0609178}}\relax
\mciteBstWouldAddEndPuncttrue
\mciteSetBstMidEndSepPunct{\mcitedefaultmidpunct}
{\mcitedefaultendpunct}{\mcitedefaultseppunct}\relax
\EndOfBibitem
\bibitem{Artuso:2008vf}
M.~Artuso, B.~Meadows, and A.~A. Petrov,
  \ifthenelse{\boolean{articletitles}}{\emph{{Charm meson decays}},
  }{}\href{https://doi.org/10.1146/annurev.nucl.58.110707.171131}{Ann.\ Rev.\
  Nucl.\ Part.\ Sci.\  \textbf{58} (2008) 249},
  \href{http://arxiv.org/abs/0802.2934}{{\normalfont\ttfamily
  arXiv:0802.2934}}\relax
\mciteBstWouldAddEndPuncttrue
\mciteSetBstMidEndSepPunct{\mcitedefaultmidpunct}
{\mcitedefaultendpunct}{\mcitedefaultseppunct}\relax
\EndOfBibitem
\bibitem{Brod:2011re}
J.~Brod, A.~L. Kagan, and J.~Zupan,
  \ifthenelse{\boolean{articletitles}}{\emph{{Size of direct CP violation in
  singly Cabibbo-suppressed D decays}},
  }{}\href{https://doi.org/10.1103/PhysRevD.86.014023}{Phys.\ Rev.\
  \textbf{D86} (2012) 014023},
  \href{http://arxiv.org/abs/1111.5000}{{\normalfont\ttfamily
  arXiv:1111.5000}}\relax
\mciteBstWouldAddEndPuncttrue
\mciteSetBstMidEndSepPunct{\mcitedefaultmidpunct}
{\mcitedefaultendpunct}{\mcitedefaultseppunct}\relax
\EndOfBibitem
\bibitem{Cheng:2012wr}
H.-Y. Cheng and C.-W. Chiang,
  \ifthenelse{\boolean{articletitles}}{\emph{{Direct \CP violation in two-body
  hadronic charmed meson decays}},
  }{}\href{https://doi.org/10.1103/PhysRevD.85.034036}{Phys.\ Rev.\
  \textbf{D85} (2012) 034036}, Erratum
  \href{https://doi.org/10.1103/PhysRevD.85.079903}{ibid.\   \textbf{D85}
  (2012) 079903}, \href{http://arxiv.org/abs/1201.0785}{{\normalfont\ttfamily
  arXiv:1201.0785}}\relax
\mciteBstWouldAddEndPuncttrue
\mciteSetBstMidEndSepPunct{\mcitedefaultmidpunct}
{\mcitedefaultendpunct}{\mcitedefaultseppunct}\relax
\EndOfBibitem
\bibitem{Cheng:2012xb}
H.-Y. Cheng and C.-W. Chiang, \ifthenelse{\boolean{articletitles}}{\emph{{SU(3)
  symmetry breaking and \CP violation in $D \to PP$ decays}},
  }{}\href{https://doi.org/10.1103/PhysRevD.86.014014}{Phys.\ Rev.\
  \textbf{D86} (2012) 014014},
  \href{http://arxiv.org/abs/1205.0580}{{\normalfont\ttfamily
  arXiv:1205.0580}}\relax
\mciteBstWouldAddEndPuncttrue
\mciteSetBstMidEndSepPunct{\mcitedefaultmidpunct}
{\mcitedefaultendpunct}{\mcitedefaultseppunct}\relax
\EndOfBibitem
\bibitem{Li:2012cfa}
H.-n. Li, C.-D. Lu, and F.-S. Yu,
  \ifthenelse{\boolean{articletitles}}{\emph{{Branching ratios and direct CP
  asymmetries in $D\to PP$ decays}},
  }{}\href{https://doi.org/10.1103/PhysRevD.86.036012}{Phys.\ Rev.\
  \textbf{D86} (2012) 036012},
  \href{http://arxiv.org/abs/1203.3120}{{\normalfont\ttfamily
  arXiv:1203.3120}}\relax
\mciteBstWouldAddEndPuncttrue
\mciteSetBstMidEndSepPunct{\mcitedefaultmidpunct}
{\mcitedefaultendpunct}{\mcitedefaultseppunct}\relax
\EndOfBibitem
\bibitem{Franco:2012ck}
E.~Franco, S.~Mishima, and L.~Silvestrini,
  \ifthenelse{\boolean{articletitles}}{\emph{{The Standard Model confronts \CP
  violation in $D^0 \to \pi^+\pi^-$ and $D^0 \to K^+K^-$}},
  }{}\href{https://doi.org/10.1007/JHEP05(2012)140}{JHEP \textbf{05} (2012)
  140}, \href{http://arxiv.org/abs/1203.3131}{{\normalfont\ttfamily
  arXiv:1203.3131}}\relax
\mciteBstWouldAddEndPuncttrue
\mciteSetBstMidEndSepPunct{\mcitedefaultmidpunct}
{\mcitedefaultendpunct}{\mcitedefaultseppunct}\relax
\EndOfBibitem
\bibitem{Pirtskhalava:2011va}
D.~Pirtskhalava and P.~Uttayarat,
  \ifthenelse{\boolean{articletitles}}{\emph{{CP Violation and flavor SU(3)
  breaking in D-meson decays}},
  }{}\href{https://doi.org/10.1016/j.physletb.2012.04.039}{Phys.\ Lett.\
  \textbf{B712} (2012) 81},
  \href{http://arxiv.org/abs/1112.5451}{{\normalfont\ttfamily
  arXiv:1112.5451}}\relax
\mciteBstWouldAddEndPuncttrue
\mciteSetBstMidEndSepPunct{\mcitedefaultmidpunct}
{\mcitedefaultendpunct}{\mcitedefaultseppunct}\relax
\EndOfBibitem
\bibitem{Feldmann:2012js}
T.~Feldmann, S.~Nandi, and A.~Soni,
  \ifthenelse{\boolean{articletitles}}{\emph{{Repercussions of flavour symmetry
  breaking on CP violation in D-meson decays}},
  }{}\href{https://doi.org/10.1007/JHEP06(2012)007}{JHEP \textbf{06} (2012)
  007}, \href{http://arxiv.org/abs/1202.3795}{{\normalfont\ttfamily
  arXiv:1202.3795}}\relax
\mciteBstWouldAddEndPuncttrue
\mciteSetBstMidEndSepPunct{\mcitedefaultmidpunct}
{\mcitedefaultendpunct}{\mcitedefaultseppunct}\relax
\EndOfBibitem
\bibitem{Brod:2012ud}
J.~Brod, Y.~Grossman, A.~L. Kagan, and J.~Zupan,
  \ifthenelse{\boolean{articletitles}}{\emph{{A consistent picture for large
  penguins in $D \to \pip\pim,\,\Kp\Km$}},
  }{}\href{https://doi.org/10.1007/JHEP10(2012)161}{JHEP \textbf{10} (2012)
  161}, \href{http://arxiv.org/abs/1203.6659}{{\normalfont\ttfamily
  arXiv:1203.6659}}\relax
\mciteBstWouldAddEndPuncttrue
\mciteSetBstMidEndSepPunct{\mcitedefaultmidpunct}
{\mcitedefaultendpunct}{\mcitedefaultseppunct}\relax
\EndOfBibitem
\bibitem{Hiller:2012xm}
G.~Hiller, M.~Jung, and S.~Schacht,
  \ifthenelse{\boolean{articletitles}}{\emph{{SU(3)-flavor anatomy of
  nonleptonic charm decays}},
  }{}\href{https://doi.org/10.1103/PhysRevD.87.014024}{Phys.\ Rev.\
  \textbf{D87} (2013) 014024},
  \href{http://arxiv.org/abs/1211.3734}{{\normalfont\ttfamily
  arXiv:1211.3734}}\relax
\mciteBstWouldAddEndPuncttrue
\mciteSetBstMidEndSepPunct{\mcitedefaultmidpunct}
{\mcitedefaultendpunct}{\mcitedefaultseppunct}\relax
\EndOfBibitem
\bibitem{Grossman:2012ry}
Y.~Grossman and D.~J. Robinson,
  \ifthenelse{\boolean{articletitles}}{\emph{{SU(3) sum rules for charm
  decay}}, }{}\href{https://doi.org/10.1007/JHEP04(2013)067}{JHEP \textbf{04}
  (2013) 067}, \href{http://arxiv.org/abs/1211.3361}{{\normalfont\ttfamily
  arXiv:1211.3361}}\relax
\mciteBstWouldAddEndPuncttrue
\mciteSetBstMidEndSepPunct{\mcitedefaultmidpunct}
{\mcitedefaultendpunct}{\mcitedefaultseppunct}\relax
\EndOfBibitem
\bibitem{Bhattacharya:2012ah}
B.~Bhattacharya, M.~Gronau, and J.~L. Rosner,
  \ifthenelse{\boolean{articletitles}}{\emph{{CP asymmetries in
  singly-Cabibbo-suppressed $D$ decays to two pseudoscalar mesons}},
  }{}\href{https://doi.org/10.1103/PhysRevD.85.079901}{Phys.\ Rev.\
  \textbf{D85} (2012) 054014},
  \href{http://arxiv.org/abs/1201.2351}{{\normalfont\ttfamily
  arXiv:1201.2351}}\relax
\mciteBstWouldAddEndPuncttrue
\mciteSetBstMidEndSepPunct{\mcitedefaultmidpunct}
{\mcitedefaultendpunct}{\mcitedefaultseppunct}\relax
\EndOfBibitem
\bibitem{Muller:2015rna}
S.~M{\"u}ller, U.~Nierste, and S.~Schacht,
  \ifthenelse{\boolean{articletitles}}{\emph{{Sum rules of charm \CP
  asymmetries beyond the SU(3)$_F$ limit}},
  }{}\href{https://doi.org/10.1103/PhysRevLett.115.251802}{Phys.\ Rev.\ Lett.\
  \textbf{115} (2015) 251802},
  \href{http://arxiv.org/abs/1506.04121}{{\normalfont\ttfamily
  arXiv:1506.04121}}\relax
\mciteBstWouldAddEndPuncttrue
\mciteSetBstMidEndSepPunct{\mcitedefaultmidpunct}
{\mcitedefaultendpunct}{\mcitedefaultseppunct}\relax
\EndOfBibitem
\bibitem{Khodjamirian:2017zdu}
A.~Khodjamirian and A.~A. Petrov,
  \ifthenelse{\boolean{articletitles}}{\emph{{Direct CP asymmetry in $D\to
  \pi^-\pi^+$ and {\mbox{$D\to K^-K^+$}} in QCD-based approach}},
  }{}\href{https://doi.org/10.1016/j.physletb.2017.09.070}{Phys.\ Lett.\
  \textbf{B774} (2017) 235},
  \href{http://arxiv.org/abs/1706.07780}{{\normalfont\ttfamily
  arXiv:1706.07780}}\relax
\mciteBstWouldAddEndPuncttrue
\mciteSetBstMidEndSepPunct{\mcitedefaultmidpunct}
{\mcitedefaultendpunct}{\mcitedefaultseppunct}\relax
\EndOfBibitem
\bibitem{Buccella:2019kpn}
F.~Buccella, A.~Paul, and P.~Santorelli,
  \ifthenelse{\boolean{articletitles}}{\emph{{On $SU(3)_{F}$ breaking through
  final state interactions and CP asymmetries in $D\to P P$ decays}},
  }{}\href{http://arxiv.org/abs/1902.05564}{{\normalfont\ttfamily
  arXiv:1902.05564}}\relax
\mciteBstWouldAddEndPuncttrue
\mciteSetBstMidEndSepPunct{\mcitedefaultmidpunct}
{\mcitedefaultendpunct}{\mcitedefaultseppunct}\relax
\EndOfBibitem
\bibitem{Atwood:2012ac}
D.~Atwood and A.~Soni, \ifthenelse{\boolean{articletitles}}{\emph{{Searching
  for the origin of \CP violation in Cabibbo-suppressed \D-meson decays}},
  }{}\href{https://doi.org/10.1093/ptep/ptt065}{PTEP \textbf{2013} (2013)
  093B05}, \href{http://arxiv.org/abs/1211.1026}{{\normalfont\ttfamily
  arXiv:1211.1026}}\relax
\mciteBstWouldAddEndPuncttrue
\mciteSetBstMidEndSepPunct{\mcitedefaultmidpunct}
{\mcitedefaultendpunct}{\mcitedefaultseppunct}\relax
\EndOfBibitem
\bibitem{bib:babarpaper2008}
BaBar collaboration, B.~Aubert {\em et~al.},
  \ifthenelse{\boolean{articletitles}}{\emph{{Search for CP violation in the
  decays {\mbox{$D^0 \to K^{-}K^{+}$}} and $D^0 \to \pi^{-} \pi^{+}$}},
  }{}\href{https://doi.org/10.1103/PhysRevLett.100.061803}{Phys.\ Rev.\ Lett.\
  \textbf{100} (2008) 061803},
  \href{http://arxiv.org/abs/0709.2715}{{\normalfont\ttfamily
  arXiv:0709.2715}}\relax
\mciteBstWouldAddEndPuncttrue
\mciteSetBstMidEndSepPunct{\mcitedefaultmidpunct}
{\mcitedefaultendpunct}{\mcitedefaultseppunct}\relax
\EndOfBibitem
\bibitem{bib:bellepaper2008}
Belle collaboration, M.~Stari\v{c} {\em et~al.},
  \ifthenelse{\boolean{articletitles}}{\emph{{Search for a CP asymmetry in
  Cabibbo-suppressed $D^0$ decays}},
  }{}\href{https://doi.org/10.1016/j.physletb.2008.10.052}{Phys.\ Lett.\
  \textbf{B670} (2008) 190},
  \href{http://arxiv.org/abs/0807.0148}{{\normalfont\ttfamily
  arXiv:0807.0148}}\relax
\mciteBstWouldAddEndPuncttrue
\mciteSetBstMidEndSepPunct{\mcitedefaultmidpunct}
{\mcitedefaultendpunct}{\mcitedefaultseppunct}\relax
\EndOfBibitem
\bibitem{bib:cdfpaper}
CDF collaboration, T.~Aaltonen {\em et~al.},
  \ifthenelse{\boolean{articletitles}}{\emph{{Measurement of CP-violating
  asymmetries in $D^0\to\pi^+\pi^-$ and $D^0\to K^+K^-$ decays at CDF}},
  }{}\href{https://doi.org/10.1103/PhysRevD.85.012009}{Phys.\ Rev.\
  \textbf{D85} (2012) 012009},
  \href{http://arxiv.org/abs/1111.5023}{{\normalfont\ttfamily
  arXiv:1111.5023}}\relax
\mciteBstWouldAddEndPuncttrue
\mciteSetBstMidEndSepPunct{\mcitedefaultmidpunct}
{\mcitedefaultendpunct}{\mcitedefaultseppunct}\relax
\EndOfBibitem
\bibitem{CDF:2012qw}
CDF collaboration, T.~Aaltonen {\em et~al.},
  \ifthenelse{\boolean{articletitles}}{\emph{{Measurement of the difference of
  CP-violating asymmetries in $D^0 \to K^+K^-$ and $D^0 \to \pi^+\pi^-$ decays
  at CDF}}, }{}\href{https://doi.org/10.1103/PhysRevLett.109.111801}{Phys.\
  Rev.\ Lett.\  \textbf{109} (2012) 111801},
  \href{http://arxiv.org/abs/1207.2158}{{\normalfont\ttfamily
  arXiv:1207.2158}}\relax
\mciteBstWouldAddEndPuncttrue
\mciteSetBstMidEndSepPunct{\mcitedefaultmidpunct}
{\mcitedefaultendpunct}{\mcitedefaultseppunct}\relax
\EndOfBibitem
\bibitem{LHCB-PAPER-2011-023}
LHCb collaboration, R.~Aaij {\em et~al.},
  \ifthenelse{\boolean{articletitles}}{\emph{{Evidence for \CP violation in
  time-integrated $\Dz\to h^-h^+$ decay rates}},
  }{}\href{https://doi.org/10.1103/PhysRevLett.108.111602}{Phys.\ Rev.\ Lett.\
  \textbf{108} (2012) 111602},
  \href{http://arxiv.org/abs/1112.0938}{{\normalfont\ttfamily
  arXiv:1112.0938}}\relax
\mciteBstWouldAddEndPuncttrue
\mciteSetBstMidEndSepPunct{\mcitedefaultmidpunct}
{\mcitedefaultendpunct}{\mcitedefaultseppunct}\relax
\EndOfBibitem
\bibitem{LHCB-PAPER-2013-003}
LHCb collaboration, R.~Aaij {\em et~al.},
  \ifthenelse{\boolean{articletitles}}{\emph{{Search for direct \CP violation
  in $\Dz\to h^- h^+$ modes using semileptonic $\B$ decays}},
  }{}\href{https://doi.org/10.1016/j.physletb.2013.04.061}{Phys.\ Lett.\
  \textbf{B723} (2013) 33},
  \href{http://arxiv.org/abs/1303.2614}{{\normalfont\ttfamily
  arXiv:1303.2614}}\relax
\mciteBstWouldAddEndPuncttrue
\mciteSetBstMidEndSepPunct{\mcitedefaultmidpunct}
{\mcitedefaultendpunct}{\mcitedefaultseppunct}\relax
\EndOfBibitem
\bibitem{LHCB-PAPER-2014-013}
LHCb collaboration, R.~Aaij {\em et~al.},
  \ifthenelse{\boolean{articletitles}}{\emph{{Measurement of \CP asymmetry in
  $\Dz\to \Km\Kp$ and $\Dz\to \pim\pip$ decays}},
  }{}\href{https://doi.org/10.1007/JHEP07(2014)041}{JHEP \textbf{07} (2014)
  041}, \href{http://arxiv.org/abs/1405.2797}{{\normalfont\ttfamily
  arXiv:1405.2797}}\relax
\mciteBstWouldAddEndPuncttrue
\mciteSetBstMidEndSepPunct{\mcitedefaultmidpunct}
{\mcitedefaultendpunct}{\mcitedefaultseppunct}\relax
\EndOfBibitem
\bibitem{LHCB-PAPER-2015-055}
LHCb collaboration, R.~Aaij {\em et~al.},
  \ifthenelse{\boolean{articletitles}}{\emph{{Measurement of the difference of
  time-integrated \CP asymmetries in $\Dz\to \Km\Kp$ and $\Dz\to \pim\pip$
  decays}}, }{}\href{https://doi.org/10.1103/PhysRevLett.116.191601}{Phys.\
  Rev.\ Lett.\  \textbf{116} (2016) 191601},
  \href{http://arxiv.org/abs/1602.03160}{{\normalfont\ttfamily
  arXiv:1602.03160}}\relax
\mciteBstWouldAddEndPuncttrue
\mciteSetBstMidEndSepPunct{\mcitedefaultmidpunct}
{\mcitedefaultendpunct}{\mcitedefaultseppunct}\relax
\EndOfBibitem
\bibitem{LHCB-PAPER-2016-035}
LHCb collaboration, R.~Aaij {\em et~al.},
  \ifthenelse{\boolean{articletitles}}{\emph{{Measurement of \CP asymmetry in
  $\Dz\to \Kp\Km$ decays}},
  }{}\href{https://doi.org/10.1016/j.physletb.2017.01.061}{Phys.\ Lett.\
  \textbf{B767} (2017) 177},
  \href{http://arxiv.org/abs/1610.09476}{{\normalfont\ttfamily
  arXiv:1610.09476}}\relax
\mciteBstWouldAddEndPuncttrue
\mciteSetBstMidEndSepPunct{\mcitedefaultmidpunct}
{\mcitedefaultendpunct}{\mcitedefaultseppunct}\relax
\EndOfBibitem
\bibitem{Gersabeck:2011xj}
M.~Gersabeck {\em et~al.}, \ifthenelse{\boolean{articletitles}}{\emph{{On the
  interplay of direct and indirect CP violation in the charm sector}},
  }{}\href{https://doi.org/10.1088/0954-3899/39/4/045005}{J.\ Phys.\
  \textbf{G39} (2012) 045005},
  \href{http://arxiv.org/abs/1111.6515}{{\normalfont\ttfamily
  arXiv:1111.6515}}\relax
\mciteBstWouldAddEndPuncttrue
\mciteSetBstMidEndSepPunct{\mcitedefaultmidpunct}
{\mcitedefaultendpunct}{\mcitedefaultseppunct}\relax
\EndOfBibitem
\bibitem{LHCb-PAPER-2014-069}
LHCb collaboration, R.~Aaij {\em et~al.},
  \ifthenelse{\boolean{articletitles}}{\emph{{Measurement of indirect \CP
  asymmetries in $\Dz\to \Km\Kp$ and $\Dz\to \pim\pip$ decays using
  semileptonic $B$ decays}},
  }{}\href{https://doi.org/10.1007/JHEP04(2015)043}{JHEP \textbf{04} (2015)
  043}, \href{http://arxiv.org/abs/1501.06777}{{\normalfont\ttfamily
  arXiv:1501.06777}}\relax
\mciteBstWouldAddEndPuncttrue
\mciteSetBstMidEndSepPunct{\mcitedefaultmidpunct}
{\mcitedefaultendpunct}{\mcitedefaultseppunct}\relax
\EndOfBibitem
\bibitem{LHCb-PAPER-2016-063}
LHCb collaboration, R.~Aaij {\em et~al.},
  \ifthenelse{\boolean{articletitles}}{\emph{{Measurement of the \CP violation
  parameter $A_\Gamma$ in $\Dz\to \Kp\Km$ and $\Dz\to \pip\pim$ decays}},
  }{}\href{https://doi.org/10.1103/PhysRevLett.118.261803}{Phys.\ Rev.\ Lett.\
  \textbf{118} (2017) 261803},
  \href{http://arxiv.org/abs/1702.06490}{{\normalfont\ttfamily
  arXiv:1702.06490}}\relax
\mciteBstWouldAddEndPuncttrue
\mciteSetBstMidEndSepPunct{\mcitedefaultmidpunct}
{\mcitedefaultendpunct}{\mcitedefaultseppunct}\relax
\EndOfBibitem
\bibitem{bib:kagansokoloff}
A.~L. Kagan and M.~D. Sokoloff,
  \ifthenelse{\boolean{articletitles}}{\emph{{Indirect \CP violation and
  implications for \Dz--\Dzb and $B_s$--$\overline{B}_s$ mixing}},
  }{}\href{https://doi.org/10.1103/PhysRevD.80.076008}{Phys.\ Rev.\
  \textbf{D80} (2009) 076008},
  \href{http://arxiv.org/abs/0907.3917}{{\normalfont\ttfamily
  arXiv:0907.3917}}\relax
\mciteBstWouldAddEndPuncttrue
\mciteSetBstMidEndSepPunct{\mcitedefaultmidpunct}
{\mcitedefaultendpunct}{\mcitedefaultseppunct}\relax
\EndOfBibitem
\bibitem{Du:2006jc}
D.-S. Du, \ifthenelse{\boolean{articletitles}}{\emph{{CP violation for neutral
  charmed meson decays to CP eigenstates}},
  }{}\href{https://doi.org/10.1140/epjc/s10052-007-0242-6}{Eur.\ Phys.\ J.\
  \textbf{C50} (2007) 579},
  \href{http://arxiv.org/abs/hep-ph/0608313}{{\normalfont\ttfamily
  arXiv:hep-ph/0608313}}\relax
\mciteBstWouldAddEndPuncttrue
\mciteSetBstMidEndSepPunct{\mcitedefaultmidpunct}
{\mcitedefaultendpunct}{\mcitedefaultseppunct}\relax
\EndOfBibitem
\bibitem{LHCb-PAPER-2016-062}
LHCb collaboration, R.~Aaij {\em et~al.},
  \ifthenelse{\boolean{articletitles}}{\emph{{Measurement of \Bd, \Bs, \Bp and
  \Lb production asymmetries in $7$ and $8$\tev\ proton-proton collisions}},
  }{}\href{https://doi.org/10.1016/j.physletb.2017.09.023}{Phys.\ Lett.\
  \textbf{B774} (2017) 139},
  \href{http://arxiv.org/abs/1703.08464}{{\normalfont\ttfamily
  arXiv:1703.08464}}\relax
\mciteBstWouldAddEndPuncttrue
\mciteSetBstMidEndSepPunct{\mcitedefaultmidpunct}
{\mcitedefaultendpunct}{\mcitedefaultseppunct}\relax
\EndOfBibitem
\bibitem{LHCb-PAPER-2013-033}
LHCb collaboration, R.~Aaij {\em et~al.},
  \ifthenelse{\boolean{articletitles}}{\emph{{Measurement of the
  flavour-specific \CP-violating asymmetry $a_{\mathrm{sl}}^s$ in $\Bs$
  decays}}, }{}\href{https://doi.org/10.1016/j.physletb.2013.12.030}{Phys.\
  Lett.\  \textbf{B728} (2014) 607},
  \href{http://arxiv.org/abs/1308.1048}{{\normalfont\ttfamily
  arXiv:1308.1048}}\relax
\mciteBstWouldAddEndPuncttrue
\mciteSetBstMidEndSepPunct{\mcitedefaultmidpunct}
{\mcitedefaultendpunct}{\mcitedefaultseppunct}\relax
\EndOfBibitem
\bibitem{LHCb-PAPER-2012-026}
LHCb collaboration, R.~Aaij {\em et~al.},
  \ifthenelse{\boolean{articletitles}}{\emph{{Measurement of the $\Dpm$
  production asymmetry in $7$\tev $\proton\proton$ collisions}},
  }{}\href{https://doi.org/10.1016/j.physletb.2012.11.038}{Phys.\ Lett.\
  \textbf{B718} (2013) 902},
  \href{http://arxiv.org/abs/1210.4112}{{\normalfont\ttfamily
  arXiv:1210.4112}}\relax
\mciteBstWouldAddEndPuncttrue
\mciteSetBstMidEndSepPunct{\mcitedefaultmidpunct}
{\mcitedefaultendpunct}{\mcitedefaultseppunct}\relax
\EndOfBibitem
\bibitem{LHCb-PAPER-2012-009}
LHCb collaboration, R.~Aaij {\em et~al.},
  \ifthenelse{\boolean{articletitles}}{\emph{{Measurement of the $\Dsp$--$\Dsm$
  production asymmetry in $7$\tev $\proton\proton$ collisions}},
  }{}\href{https://doi.org/10.1016/j.physletb.2012.06.001}{Phys.\ Lett.\
  \textbf{B713} (2012) 186},
  \href{http://arxiv.org/abs/1205.0897}{{\normalfont\ttfamily
  arXiv:1205.0897}}\relax
\mciteBstWouldAddEndPuncttrue
\mciteSetBstMidEndSepPunct{\mcitedefaultmidpunct}
{\mcitedefaultendpunct}{\mcitedefaultseppunct}\relax
\EndOfBibitem
\bibitem{Alves:2008zz}
LHCb collaboration, A.~A. Alves~Jr.\ {\em et~al.},
  \ifthenelse{\boolean{articletitles}}{\emph{{The \lhcb detector at the LHC}},
  }{}\href{https://doi.org/10.1088/1748-0221/3/08/S08005}{JINST \textbf{3}
  (2008) S08005}\relax
\mciteBstWouldAddEndPuncttrue
\mciteSetBstMidEndSepPunct{\mcitedefaultmidpunct}
{\mcitedefaultendpunct}{\mcitedefaultseppunct}\relax
\EndOfBibitem
\bibitem{LHCb-DP-2014-002}
LHCb collaboration, R.~Aaij {\em et~al.},
  \ifthenelse{\boolean{articletitles}}{\emph{{LHCb detector performance}},
  }{}\href{https://doi.org/10.1142/S0217751X15300227}{Int.\ J.\ Mod.\ Phys.\
  \textbf{A30} (2015) 1530022},
  \href{http://arxiv.org/abs/1412.6352}{{\normalfont\ttfamily
  arXiv:1412.6352}}\relax
\mciteBstWouldAddEndPuncttrue
\mciteSetBstMidEndSepPunct{\mcitedefaultmidpunct}
{\mcitedefaultendpunct}{\mcitedefaultseppunct}\relax
\EndOfBibitem
\bibitem{BBDT}
V.~V. Gligorov and M.~Williams,
  \ifthenelse{\boolean{articletitles}}{\emph{{Efficient, reliable and fast
  high-level triggering using a bonsai boosted decision tree}},
  }{}\href{https://doi.org/10.1088/1748-0221/8/02/P02013}{JINST \textbf{8}
  (2013) P02013}, \href{http://arxiv.org/abs/1210.6861}{{\normalfont\ttfamily
  arXiv:1210.6861}}\relax
\mciteBstWouldAddEndPuncttrue
\mciteSetBstMidEndSepPunct{\mcitedefaultmidpunct}
{\mcitedefaultendpunct}{\mcitedefaultseppunct}\relax
\EndOfBibitem
\bibitem{Likhomanenko:2015aba}
T.~Likhomanenko {\em et~al.}, \ifthenelse{\boolean{articletitles}}{\emph{{LHCb
  topological trigger reoptimization}},
  }{}\href{https://doi.org/10.1088/1742-6596/664/8/082025}{J.\ Phys.\ Conf.\
  Ser.\  \textbf{664} (2015) 082025},
  \href{http://arxiv.org/abs/1510.00572}{{\normalfont\ttfamily
  arXiv:1510.00572}}\relax
\mciteBstWouldAddEndPuncttrue
\mciteSetBstMidEndSepPunct{\mcitedefaultmidpunct}
{\mcitedefaultendpunct}{\mcitedefaultseppunct}\relax
\EndOfBibitem
\bibitem{LHCb-PROC-2015-011}
G.~Dujany and B.~Storaci, \ifthenelse{\boolean{articletitles}}{\emph{{Real-time
  alignment and calibration of the LHCb Detector in Run II}},
  }{}\href{https://doi.org/10.1088/1742-6596/664/8/082010}{J.\ Phys.\ Conf.\
  Ser.\  \textbf{664} (2015) 082010}\relax
\mciteBstWouldAddEndPuncttrue
\mciteSetBstMidEndSepPunct{\mcitedefaultmidpunct}
{\mcitedefaultendpunct}{\mcitedefaultseppunct}\relax
\EndOfBibitem
\bibitem{Kodama:1991ij}
E653 collaboration, K.~Kodama {\em et~al.},
  \ifthenelse{\boolean{articletitles}}{\emph{{Measurement of the relative
  branching fraction $\Gamma (D^0 \to K \mu \nu) / \Gamma (D^0 \to \mu X)$}},
  }{}\href{https://doi.org/10.1103/PhysRevLett.66.1819}{Phys.\ Rev.\ Lett.\
  \textbf{66} (1991) 1819}\relax
\mciteBstWouldAddEndPuncttrue
\mciteSetBstMidEndSepPunct{\mcitedefaultmidpunct}
{\mcitedefaultendpunct}{\mcitedefaultseppunct}\relax
\EndOfBibitem
\bibitem{PRL:supplemental}
\ifthenelse{\boolean{articletitles}}{{See
  \hyperref[sec:supplemental]{supplemental material} for additional numerical
  values and plots}}{}\relax
\mciteBstWouldAddEndPuncttrue
\mciteSetBstMidEndSepPunct{\mcitedefaultmidpunct}
{\mcitedefaultendpunct}{\mcitedefaultseppunct}\relax
\EndOfBibitem
\bibitem{Hulsbergen2005566}
W.~D. Hulsbergen, \ifthenelse{\boolean{articletitles}}{\emph{{Decay chain
  fitting with a Kalman filter}},
  }{}\href{https://doi.org/10.1016/j.nima.2005.06.078}{Nucl.\ Instrum.\ Meth.\
  \textbf{A552} (2005) 566},
  \href{http://arxiv.org/abs/physics/0503191}{{\normalfont\ttfamily
  arXiv:physics/0503191}}\relax
\mciteBstWouldAddEndPuncttrue
\mciteSetBstMidEndSepPunct{\mcitedefaultmidpunct}
{\mcitedefaultendpunct}{\mcitedefaultseppunct}\relax
\EndOfBibitem
\bibitem{Johnson:1949zj}
N.~L. Johnson, \ifthenelse{\boolean{articletitles}}{\emph{{Systems of frequency
  curves generated by methods of translation}},
  }{}\href{https://doi.org/10.1093/biomet/36.1-2.149}{Biometrika \textbf{36}
  (1949) 149}\relax
\mciteBstWouldAddEndPuncttrue
\mciteSetBstMidEndSepPunct{\mcitedefaultmidpunct}
{\mcitedefaultendpunct}{\mcitedefaultseppunct}\relax
\EndOfBibitem
\bibitem{LHCb-PAPER-2017-046}
LHCb collaboration, R.~Aaij {\em et~al.},
  \ifthenelse{\boolean{articletitles}}{\emph{{Updated determination of
  $\Dz$-$\Dbar^0$ mixing and \CP violation parameters with $\Dz\to K^+\pi^-$
  decays}}, }{}\href{https://doi.org/10.1103/PhysRevD.97.031101}{Phys.\ Rev.\
  \textbf{D97} (2018) 031101},
  \href{http://arxiv.org/abs/1712.03220}{{\normalfont\ttfamily
  arXiv:1712.03220}}\relax
\mciteBstWouldAddEndPuncttrue
\mciteSetBstMidEndSepPunct{\mcitedefaultmidpunct}
{\mcitedefaultendpunct}{\mcitedefaultseppunct}\relax
\EndOfBibitem
\bibitem{Cowan:2016tnm}
G.~A. Cowan, D.~C. Craik, and M.~D. Needham,
  \ifthenelse{\boolean{articletitles}}{\emph{{RapidSim: An application for the
  fast simulation of heavy-quark hadron decays}},
  }{}\href{https://doi.org/10.1016/j.cpc.2017.01.029}{Comput.\ Phys.\ Commun.\
  \textbf{214} (2017) 239},
  \href{http://arxiv.org/abs/1612.07489}{{\normalfont\ttfamily
  arXiv:1612.07489}}\relax
\mciteBstWouldAddEndPuncttrue
\mciteSetBstMidEndSepPunct{\mcitedefaultmidpunct}
{\mcitedefaultendpunct}{\mcitedefaultseppunct}\relax
\EndOfBibitem
\bibitem{HFLAV16}
Heavy Flavor Averaging Group, Y.~Amhis {\em et~al.},
  \ifthenelse{\boolean{articletitles}}{\emph{{Averages of $b$-hadron,
  $c$-hadron, and $\tau$-lepton properties as of summer 2016}},
  }{}\href{https://doi.org/10.1140/epjc/s10052-017-5058-4}{Eur.\ Phys.\ J.\
  \textbf{C77} (2017) 895},
  \href{http://arxiv.org/abs/1612.07233}{{\normalfont\ttfamily
  arXiv:1612.07233}}, {updated results and plots available at
  \href{https://hflav.web.cern.ch}{{\texttt{https://hflav.web.cern.ch}}}}\relax
\mciteBstWouldAddEndPuncttrue
\mciteSetBstMidEndSepPunct{\mcitedefaultmidpunct}
{\mcitedefaultendpunct}{\mcitedefaultseppunct}\relax
\EndOfBibitem
\bibitem{PDG2018}
Particle Data Group, M.~Tanabashi {\em et~al.},
  \ifthenelse{\boolean{articletitles}}{\emph{{\href{http://pdg.lbl.gov/}{Review
  of particle physics}}},
  }{}\href{https://doi.org/10.1103/PhysRevD.98.030001}{Phys.\ Rev.\
  \textbf{D98} (2018) 030001}\relax
\mciteBstWouldAddEndPuncttrue
\mciteSetBstMidEndSepPunct{\mcitedefaultmidpunct}
{\mcitedefaultendpunct}{\mcitedefaultseppunct}\relax
\EndOfBibitem
\end{mcitethebibliography}

 
\clearpage
\centerline{\large\bf LHCb collaboration}
\begin{flushleft}
\small
R.~Aaij$^{28}$,
C.~Abell{\'a}n~Beteta$^{46}$,
B.~Adeva$^{43}$,
M.~Adinolfi$^{50}$,
C.A.~Aidala$^{78}$,
Z.~Ajaltouni$^{6}$,
S.~Akar$^{61}$,
P.~Albicocco$^{19}$,
J.~Albrecht$^{11}$,
F.~Alessio$^{44}$
% ...

{\footnotesize \it
% ...
$ ^{6}$Universit{\'e} Clermont Auvergne, CNRS/IN2P3, LPC, Clermont-Ferrand, France\\
% ...
$ ^{11}$Fakult{\"a}t Physik, Technische Universit{\"a}t Dortmund, Dortmund, Germany\\
% ...
$ ^{19}$INFN Laboratori Nazionali di Frascati, Frascati, Italy\\
% ...
$ ^{28}$Nikhef National Institute for Subatomic Physics, Amsterdam, Netherlands\\
% ...
$ ^{43}$Instituto Galego de F{\'\i}sica de Altas Enerx{\'\i}as (IGFAE), Universidade de Santiago de Compostela, Santiago de Compostela, Spain\\
$ ^{44}$European Organization for Nuclear Research (CERN), Geneva, Switzerland\\
% ...
$ ^{46}$Physik-Institut, Universit{\"a}t Z{\"u}rich, Z{\"u}rich, Switzerland\\
% ...
$ ^{50}$H.H. Wills Physics Laboratory, University of Bristol, Bristol, United Kingdom\\
% ...
$ ^{61}$University of Cincinnati, Cincinnati, OH, United States\\
% ...
$ ^{63}$Syracuse University, Syracuse, NY, United States\\
% ...
$ ^{78}$University of Michigan, Ann Arbor, United States, associated to $^{63}$\\
% ...
}
\end{flushleft}
\clearpage
\appendix
\section*{Supplemental material}
\label{sec:supplemental}
\subsection*{Arithmetic average of mean decay times}
The arithmetic average of the reconstructed mean decay times for \dkk and \dpipi decays, $\overbar{\langle t \rangle}$, can be useful when interpreting the measurement of \DACP. The values corresponding to the present measurements are $\overbar{\langle t \rangle}^{\pi{\mbox{-}}\rm{tagged}}/\tau(\Dz)  =  1.74 \pm 0.10$ and $\overbar{\langle t \rangle}^{\mu{\mbox{-}}\rm{tagged}}/\tau(\Dz)  =  1.21 \pm 0.01$, 
whereas that corresponding to the combination with previous LHCb measurements is $\overbar{\langle t \rangle}/\tau(\Dz)  =  1.71 \pm 0.10$. The uncertainties include statistical and systematic contributions, and the world average of the \Dz lifetime is used.

\subsection*{Additional plots}
\begin{figure}[htb]
\centering
\includegraphics[width=0.48\textwidth]{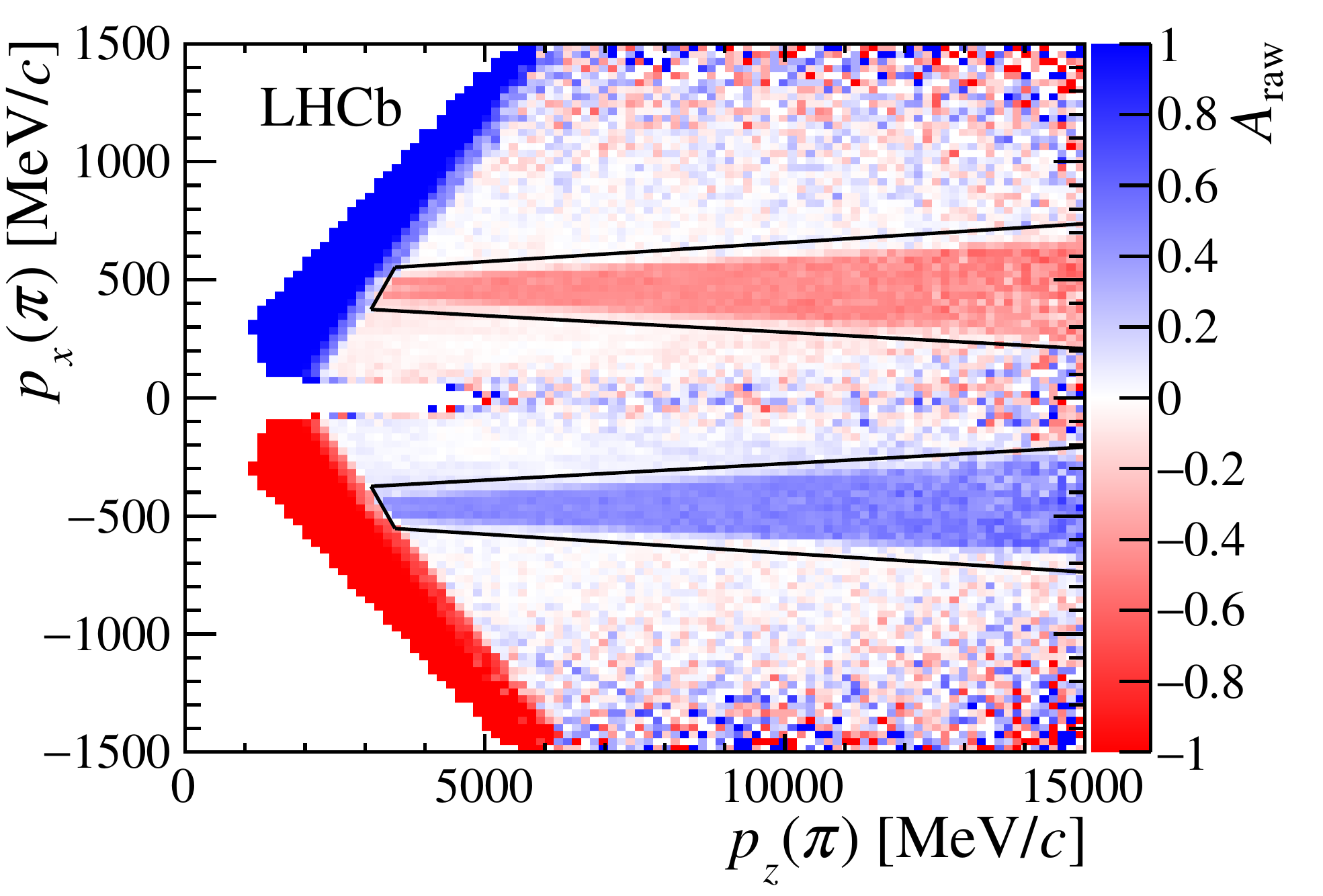}
\includegraphics[width=0.48\textwidth]{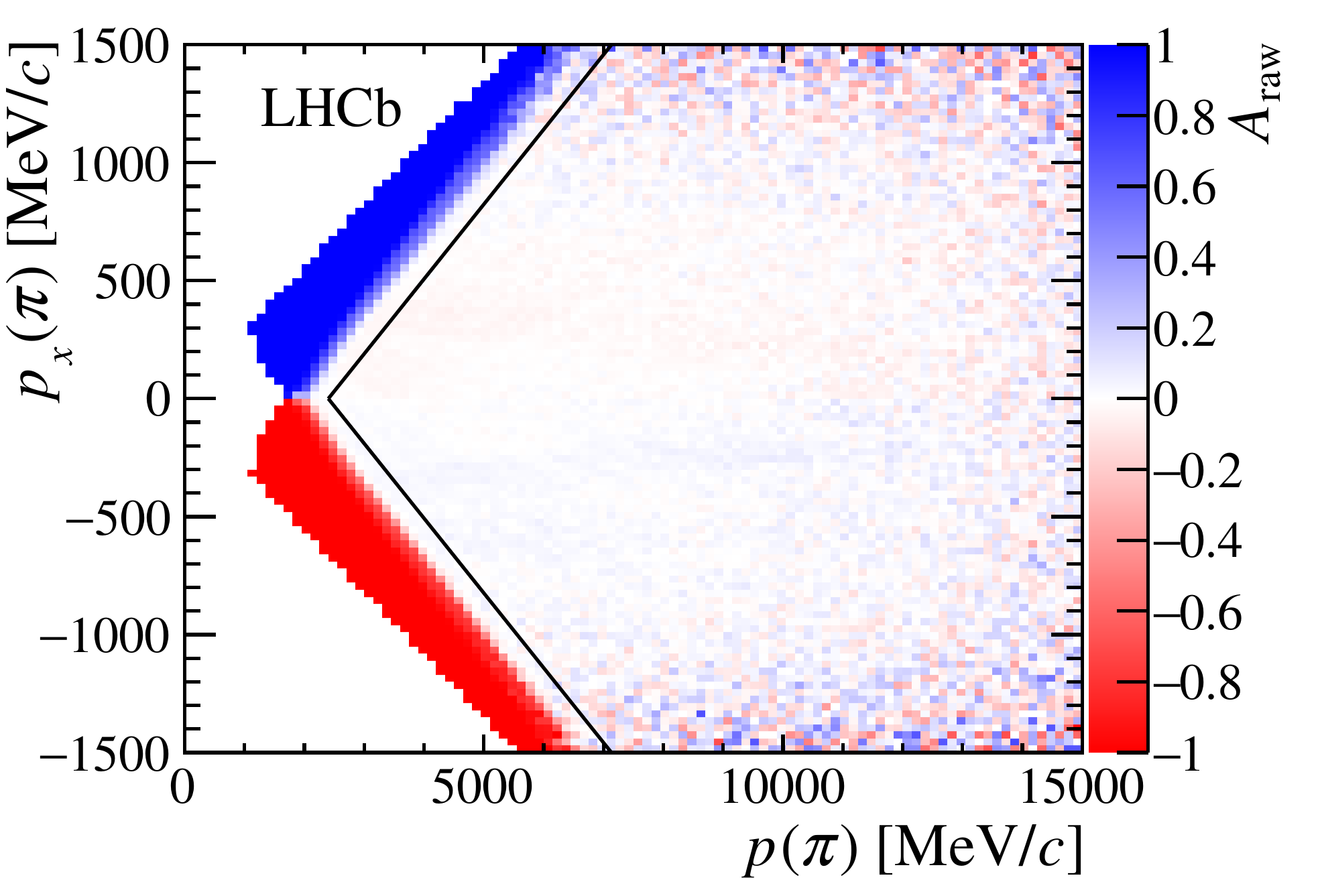}
\includegraphics[width=0.48\textwidth]{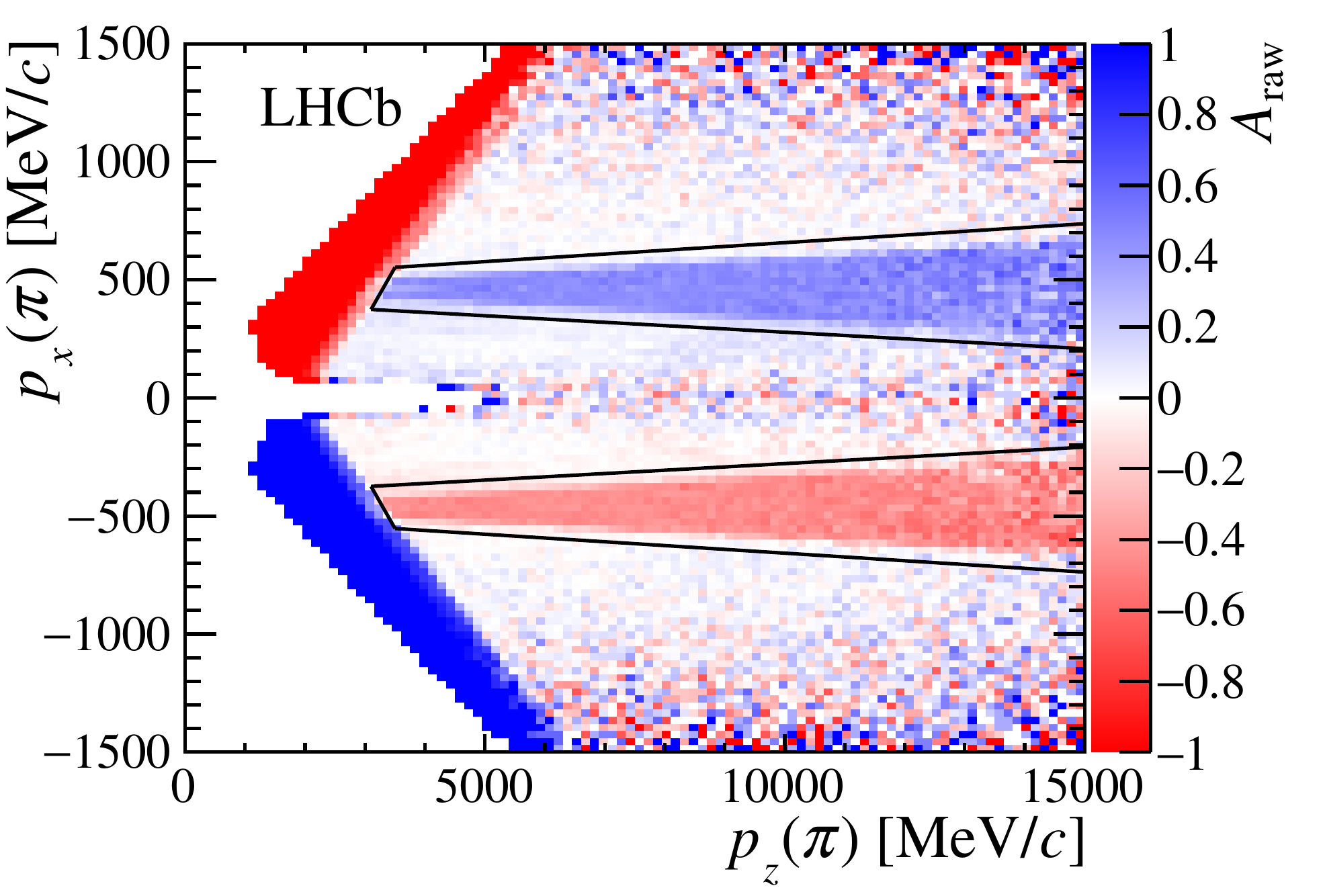}
\includegraphics[width=0.48\textwidth]{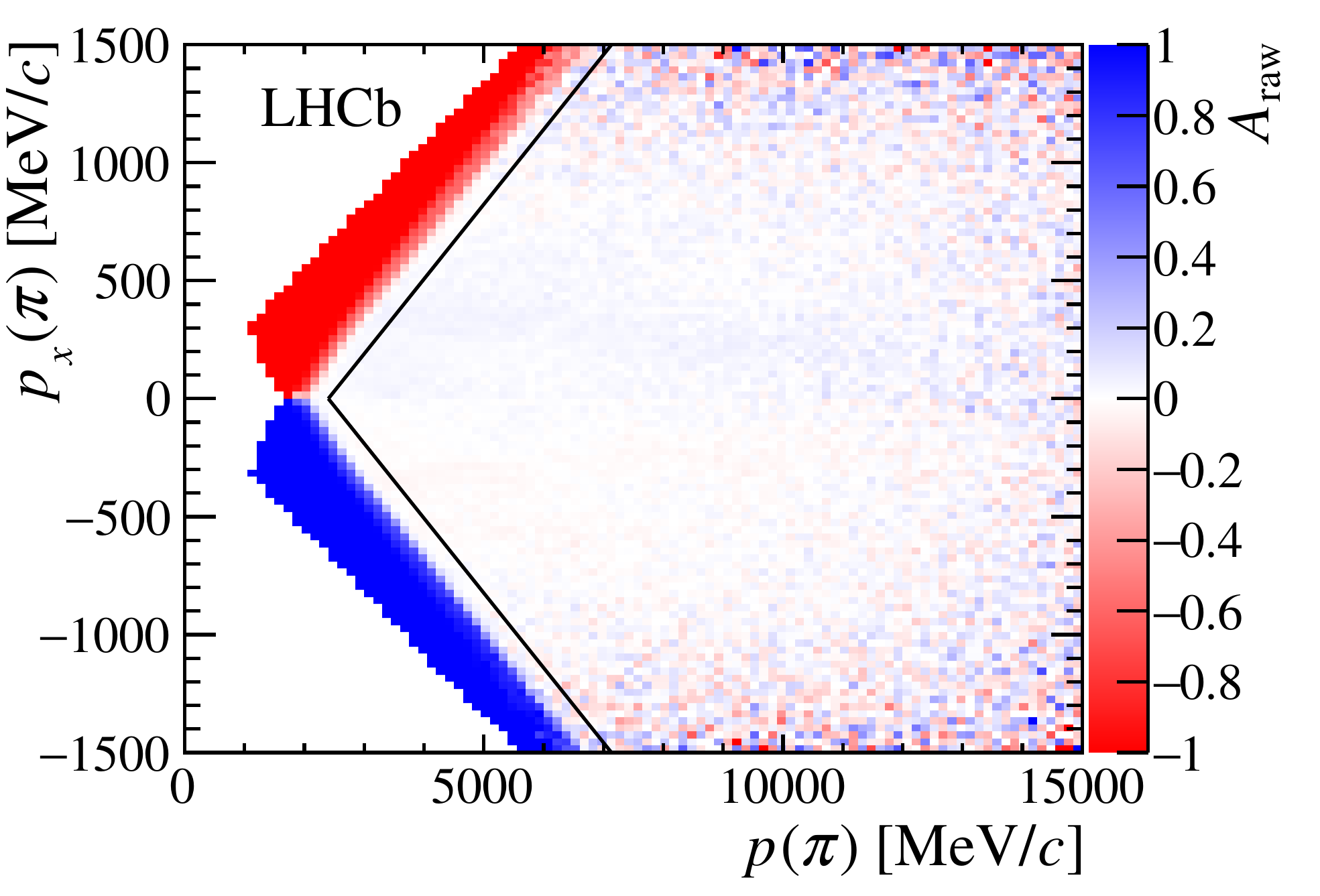}
\caption{Raw asymmetries of the tagging pion for the $\pi$-tagged \dkk sample, with the polarity of the magnet (top)~up and (bottom)~down. The plots on the left include only candidates with $|p_y / p_z| < 0.02$, \emph{i.e.}, close to the horizontal plane, and the fiducial requirements used to exclude the kinematic region surrounding the beam pipe, characterized by large values of the raw asymmetry, are indicated as black lines (in addition to the forementioned requirement $|p_y / p_z| < 0.02$). The plots on the right include all candidates except those excluded by the beam-pipe fiducial requirements, and the black lines indicate the fiducial requirements used to exclude regions at the boundary of the detector acceptance, which are also characterized by large values of the raw asymmetry. Distributions for the \dpipi sample are very similar.}
\label{fig:FiducialKK}
\end{figure}

\begin{figure}[htb]
\centering
\includegraphics[width=0.48\textwidth]{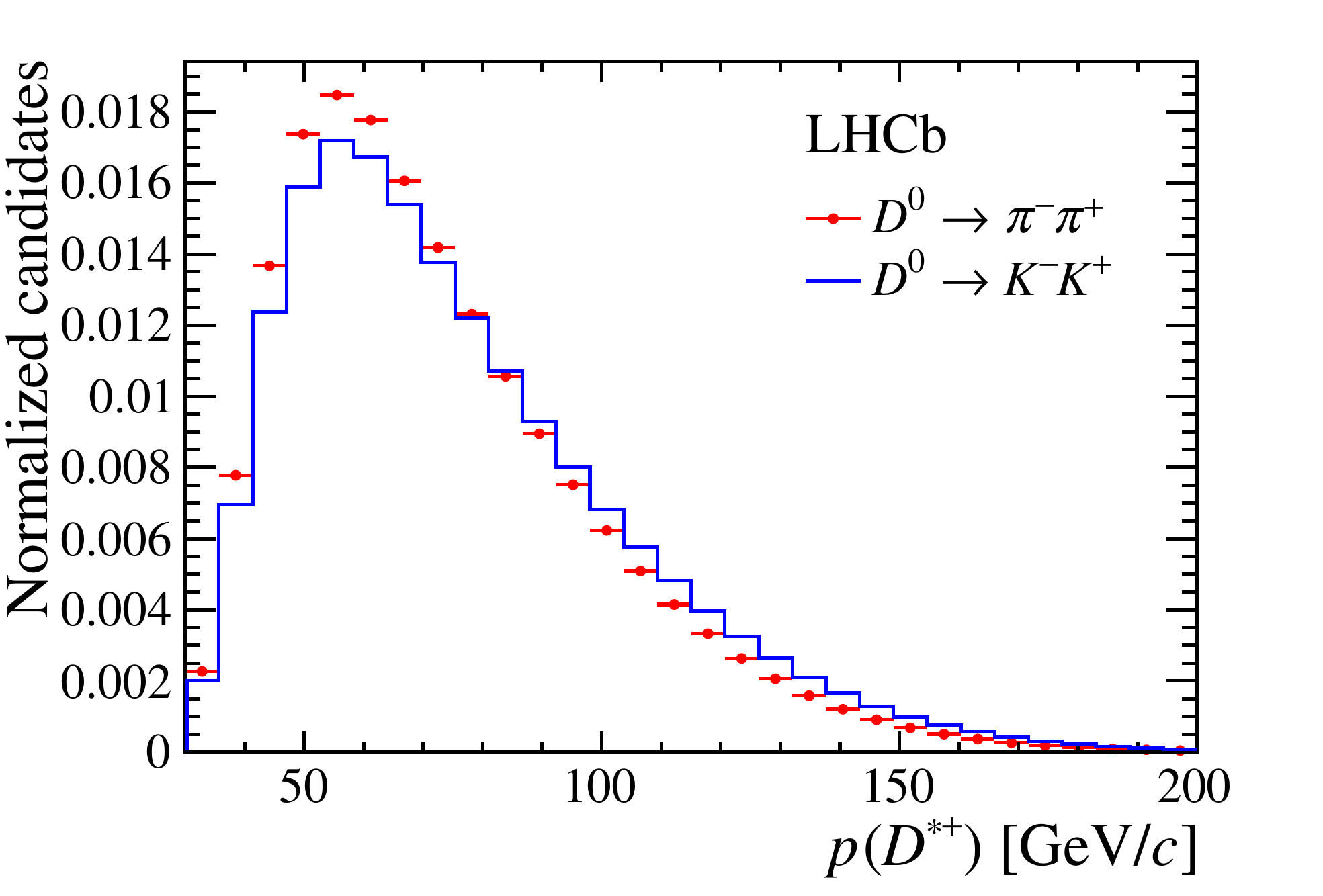}
\includegraphics[width=0.48\textwidth]{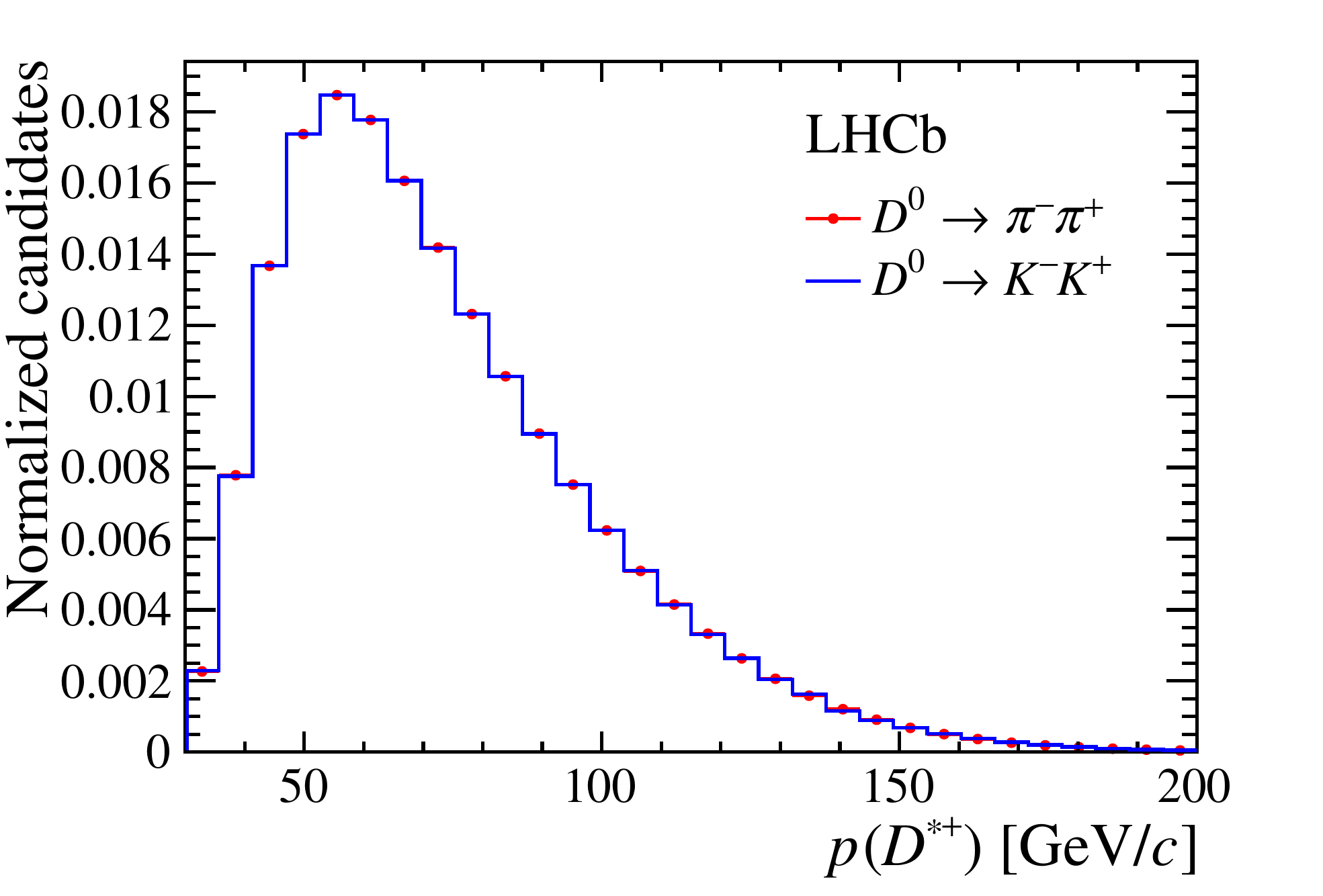}
\includegraphics[width=0.48\textwidth]{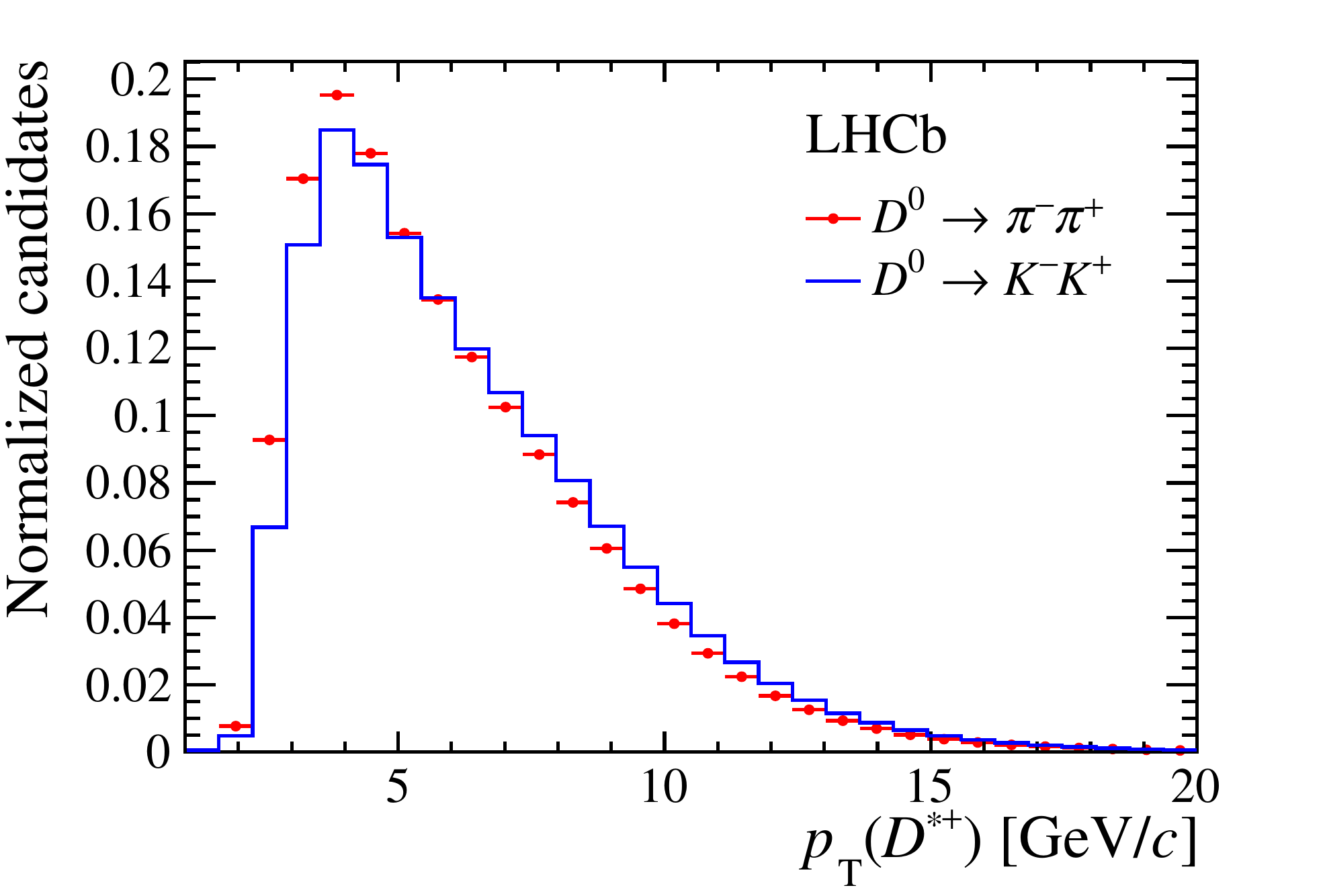}
\includegraphics[width=0.48\textwidth]{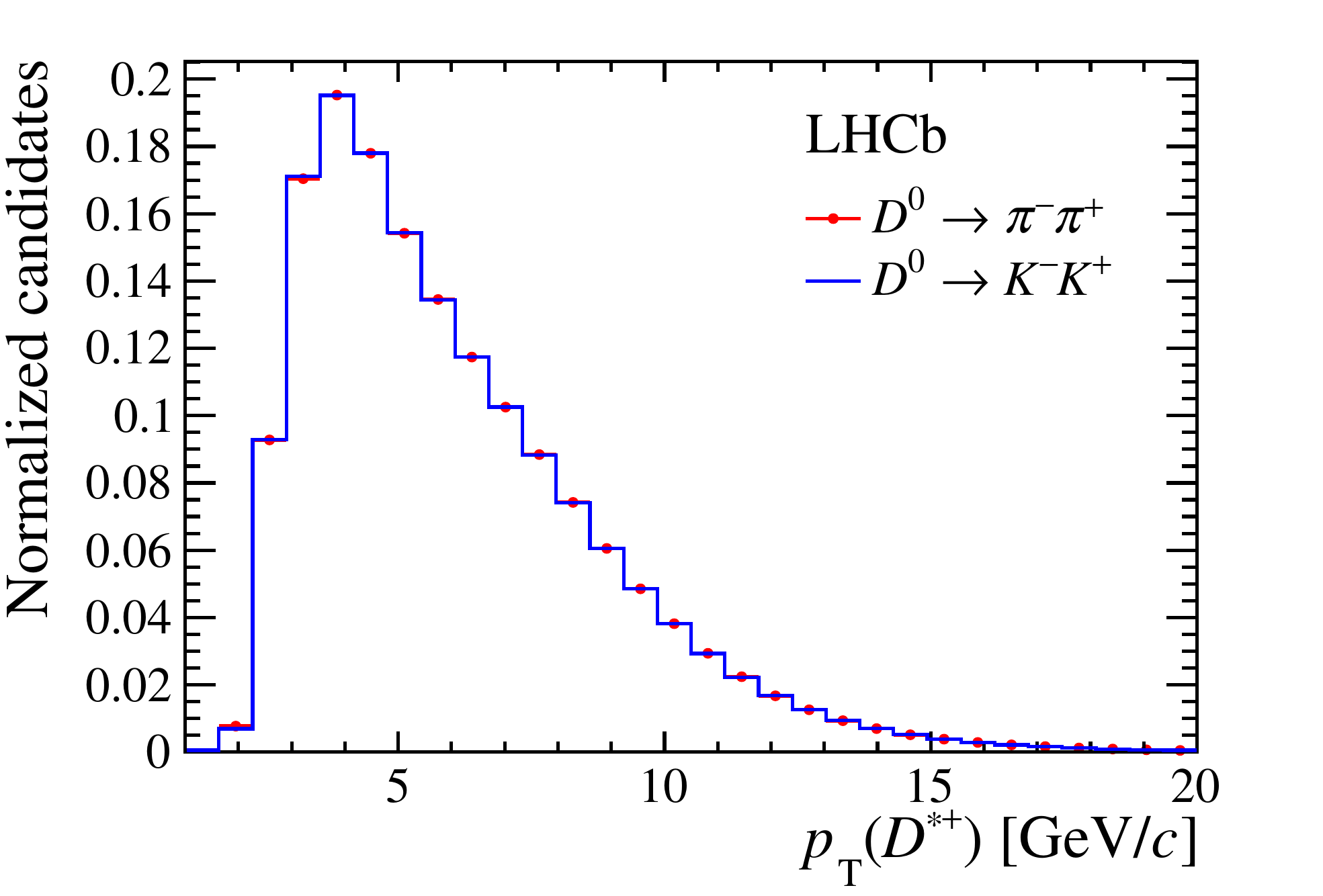}
\includegraphics[width=0.48\textwidth]{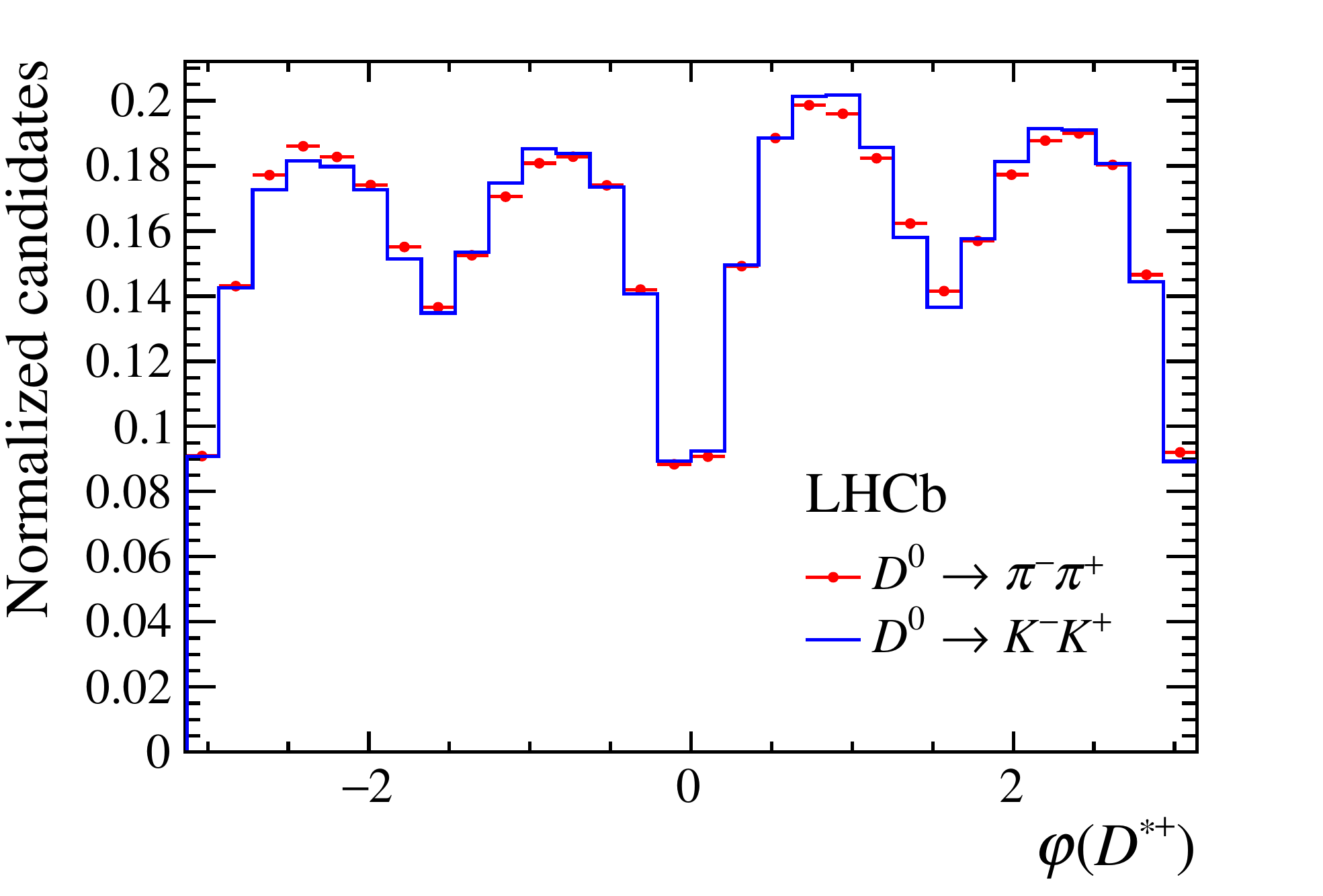}
\includegraphics[width=0.48\textwidth]{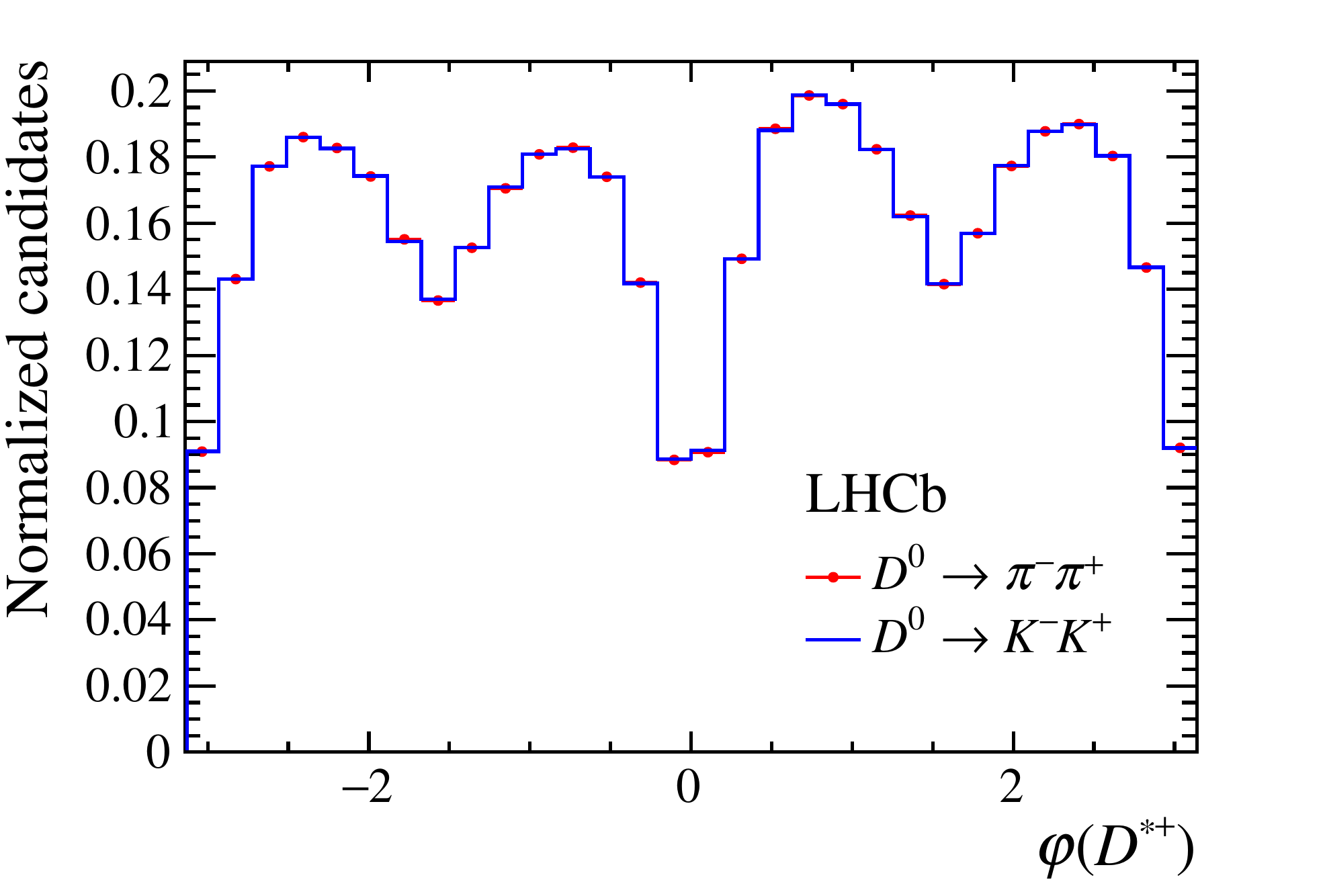}
\includegraphics[width=0.48\textwidth]{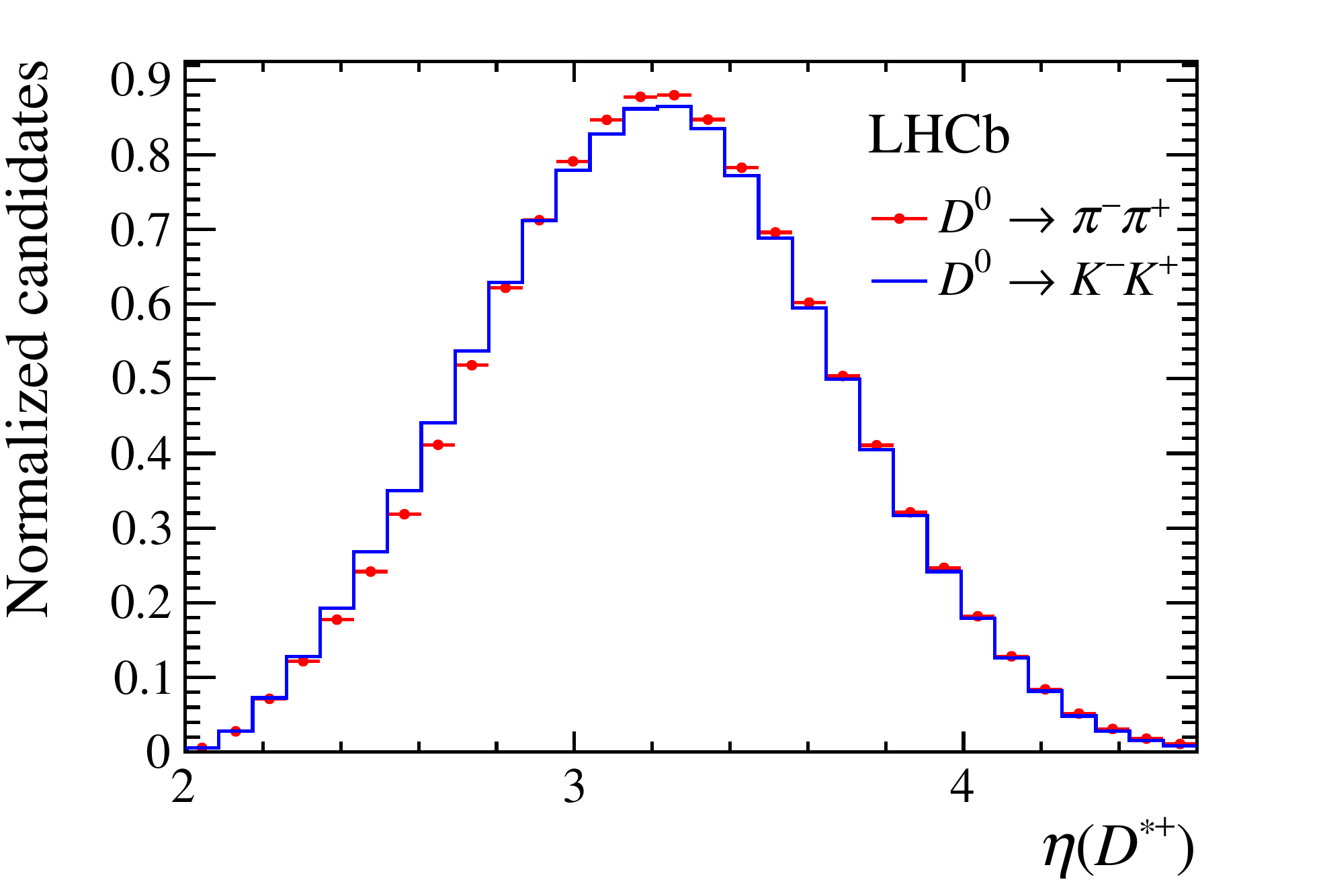}
\includegraphics[width=0.48\textwidth]{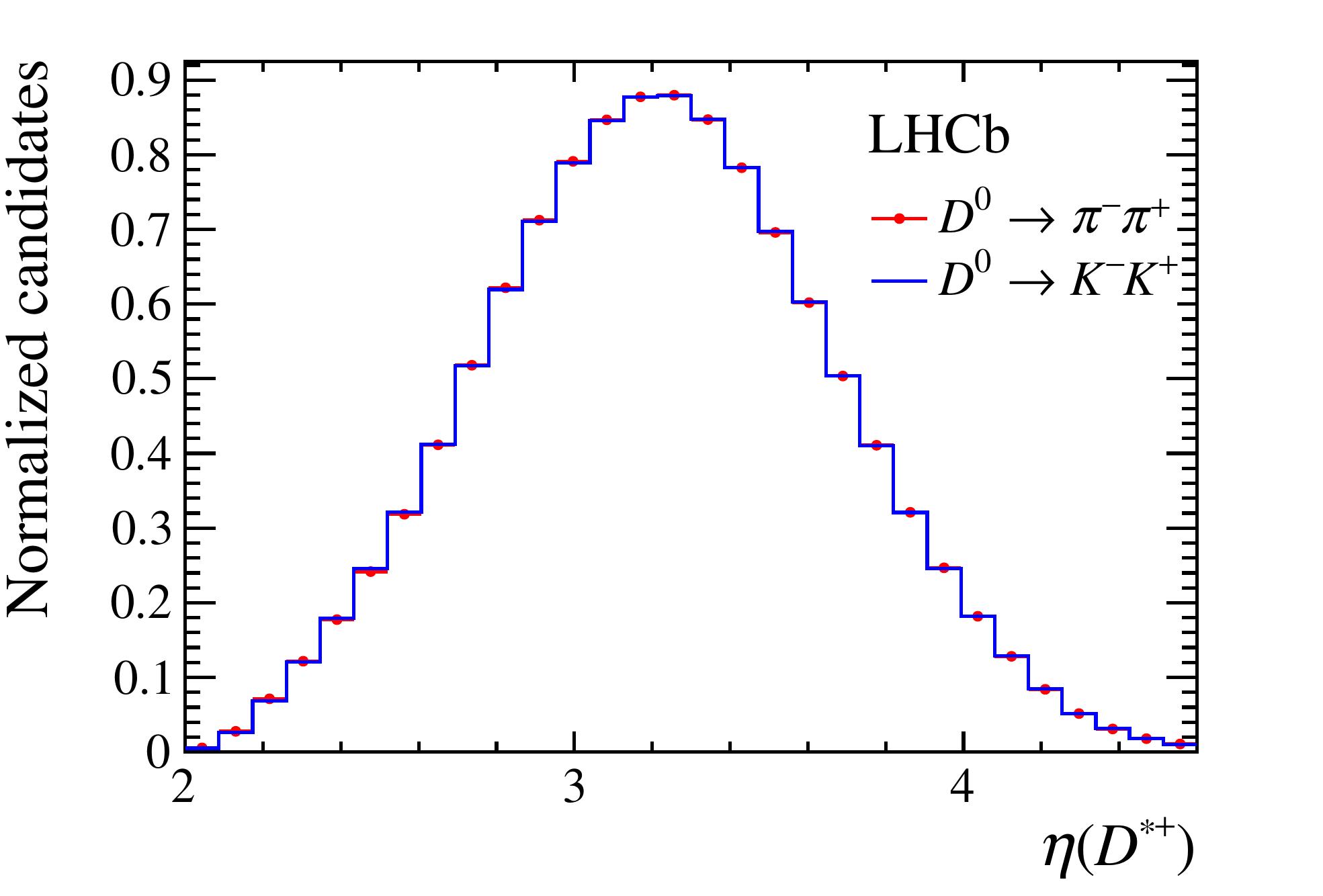}
\caption{Background-subtracted distributions of momentum~($p$), transverse momentum~($\pt$), azimuthal angle~($\varphi$) and pseudorapidity~($\eta$) of \Dstarp mesons for the prompt sample: (left column)~before and (right column)~after the weighting procedure for \dkk and \dpipi decays, as indicated in the legends. The distributions are normalized to unit area.}
\label{fig:PR_dstar_weight}
\end{figure}

\begin{figure}[htb]
\centering
\includegraphics[width=0.48\textwidth]{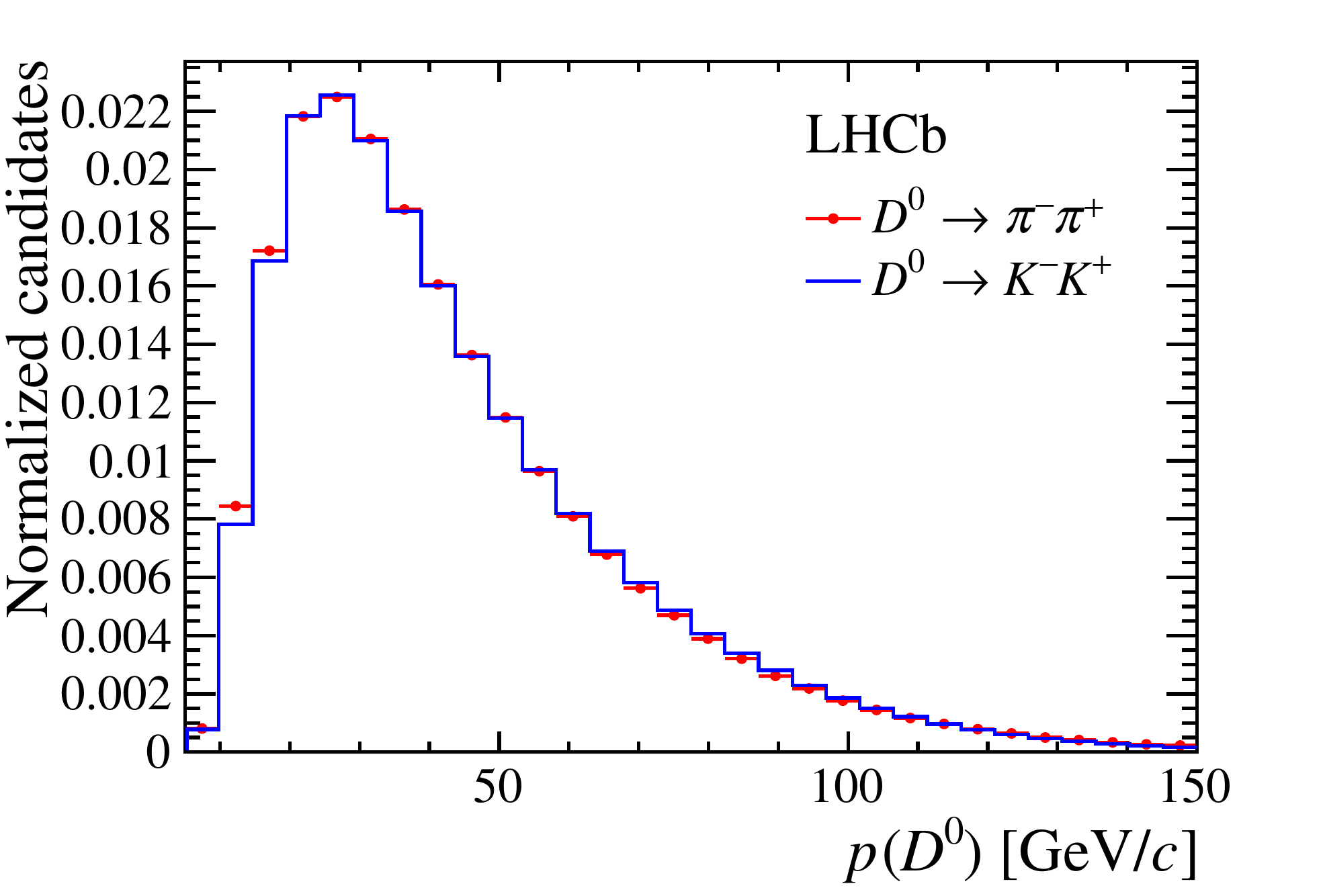}
\includegraphics[width=0.48\textwidth]{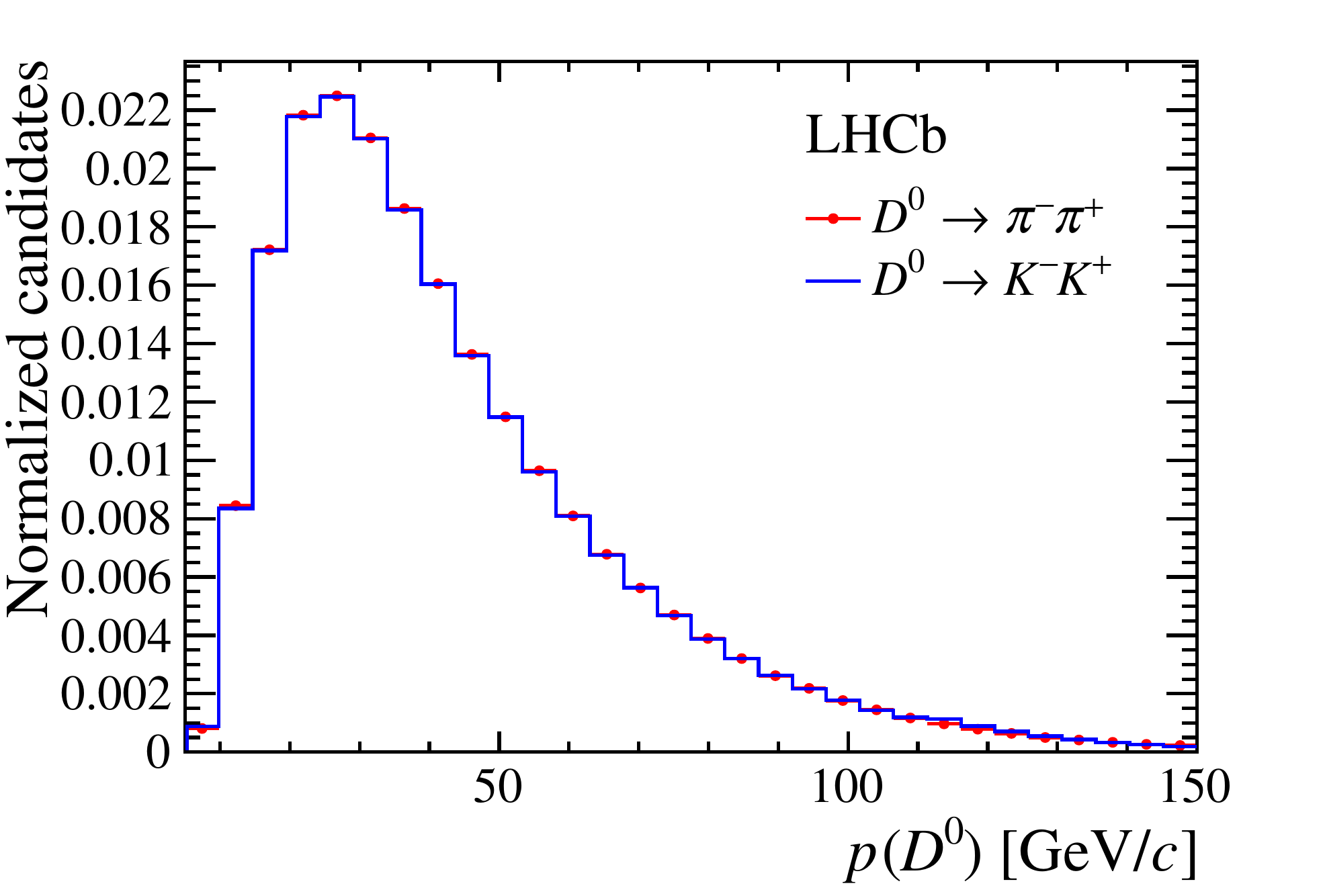}
\includegraphics[width=0.48\textwidth]{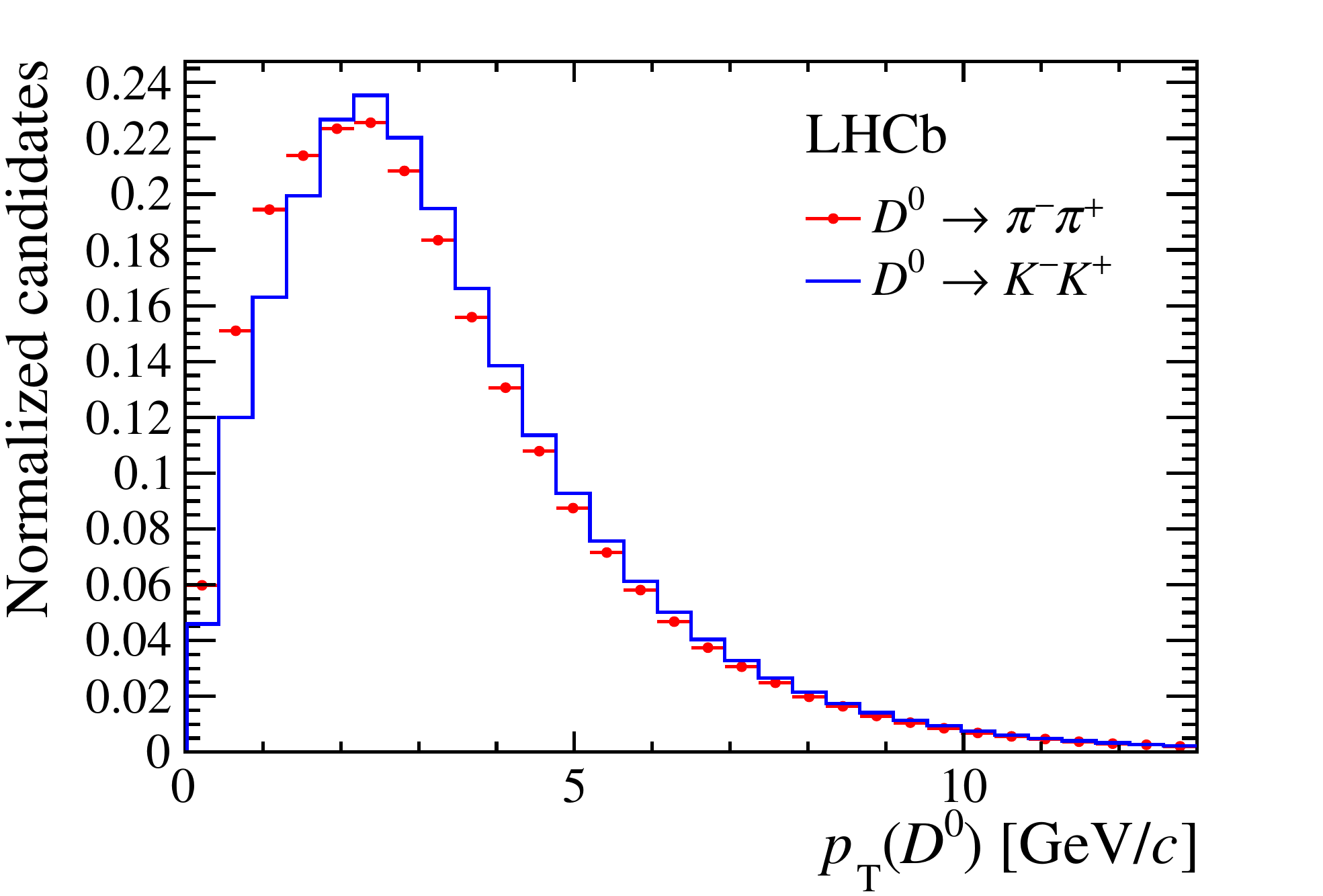}
\includegraphics[width=0.48\textwidth]{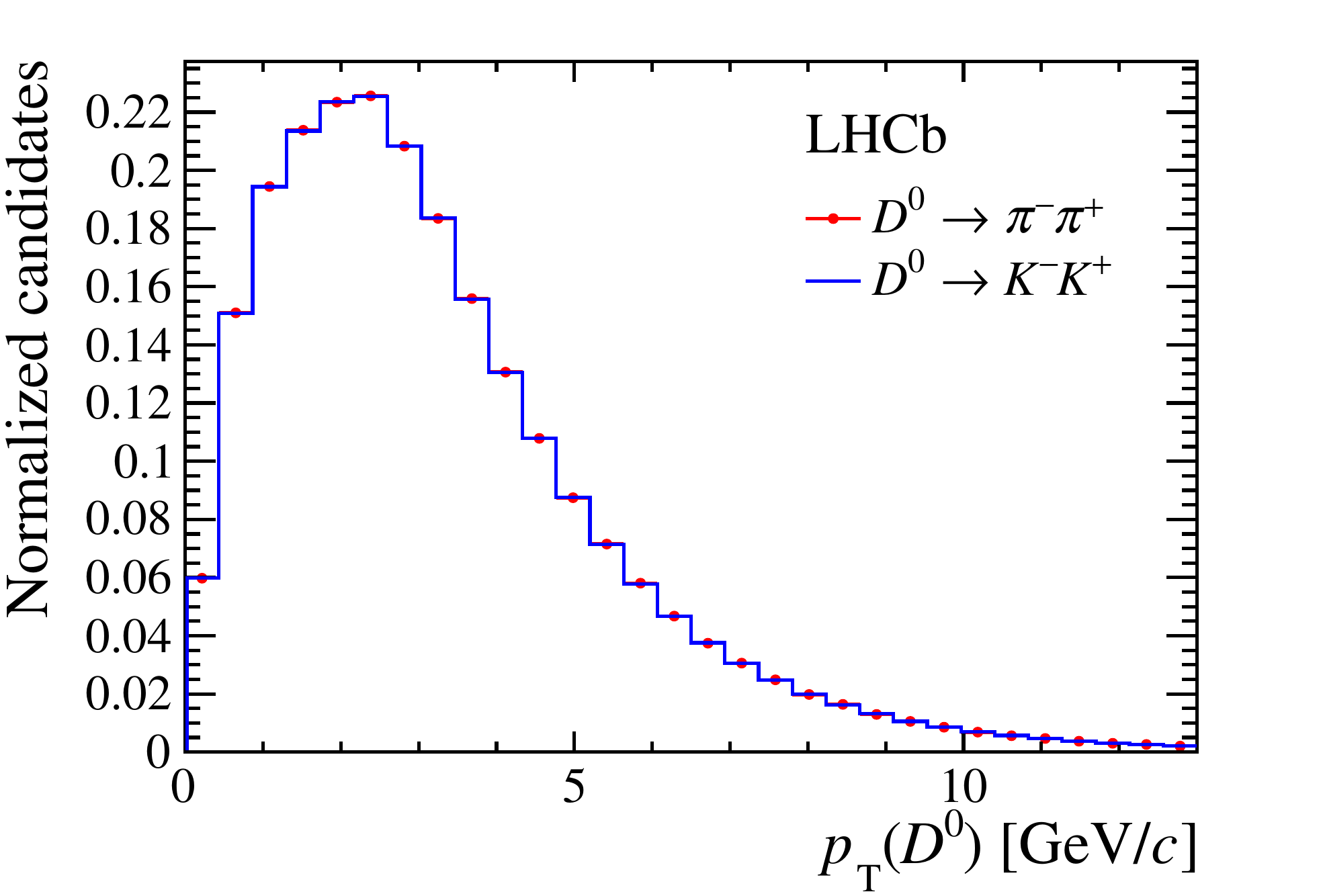}
\includegraphics[width=0.48\textwidth]{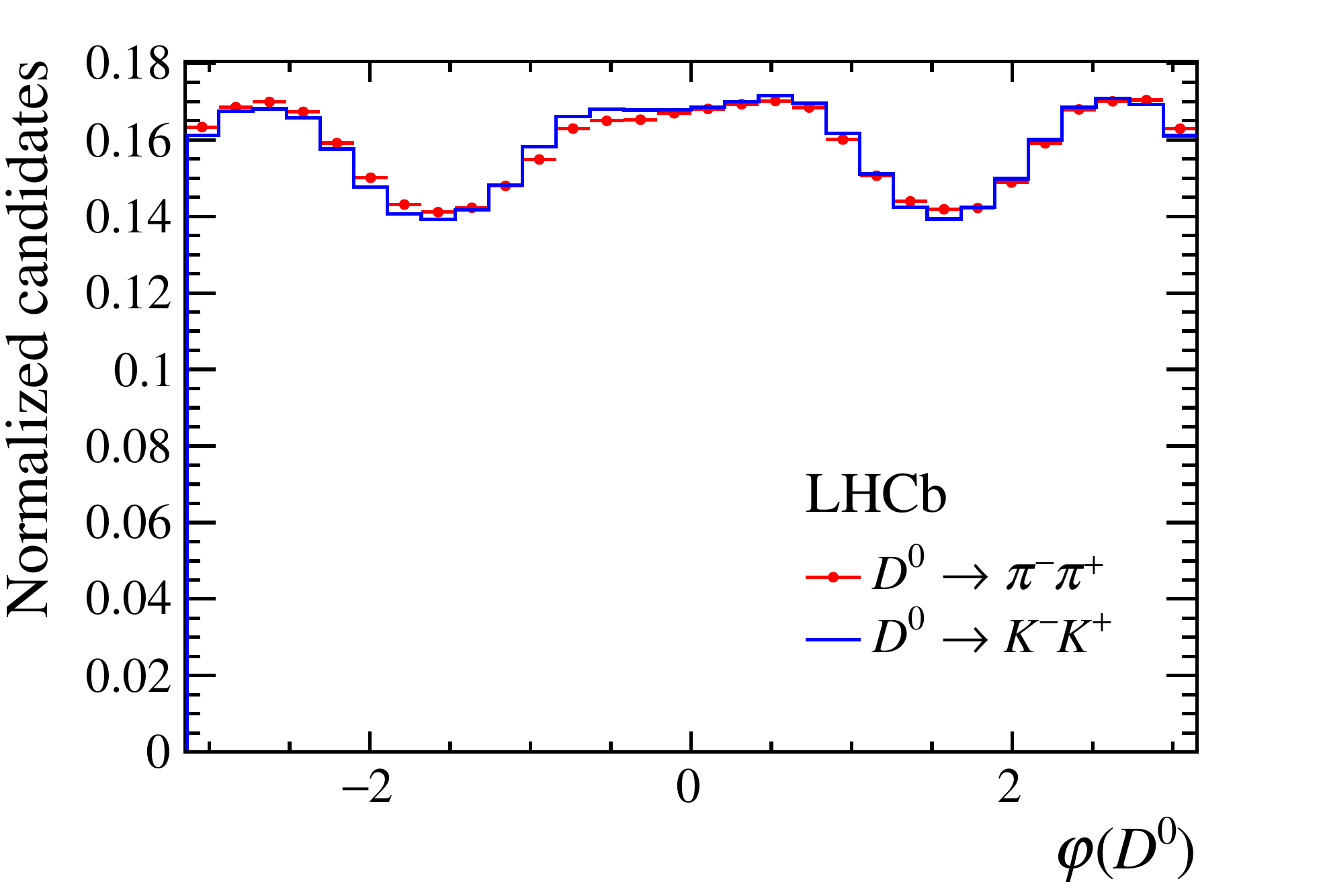}
\includegraphics[width=0.48\textwidth]{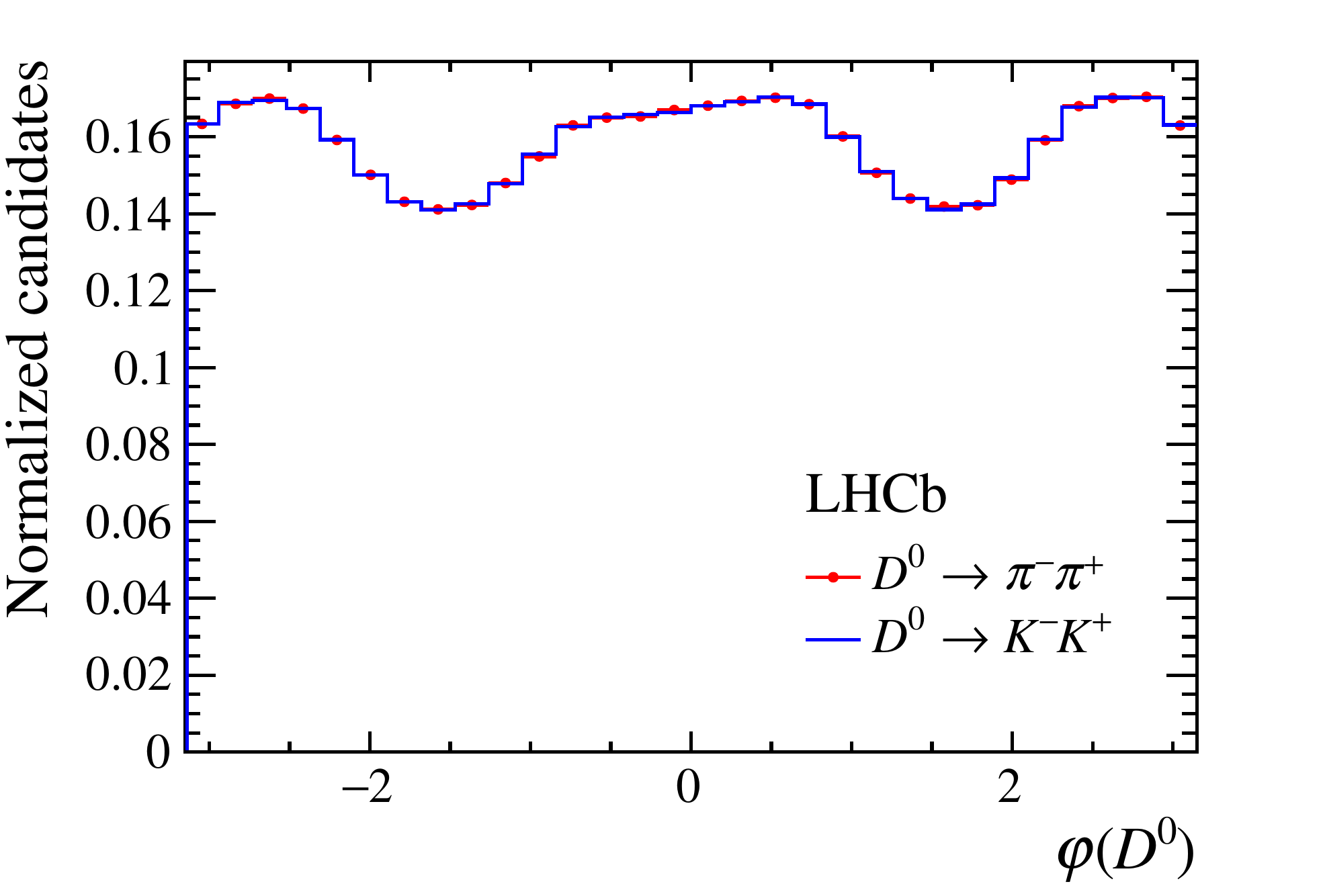}
\includegraphics[width=0.48\textwidth]{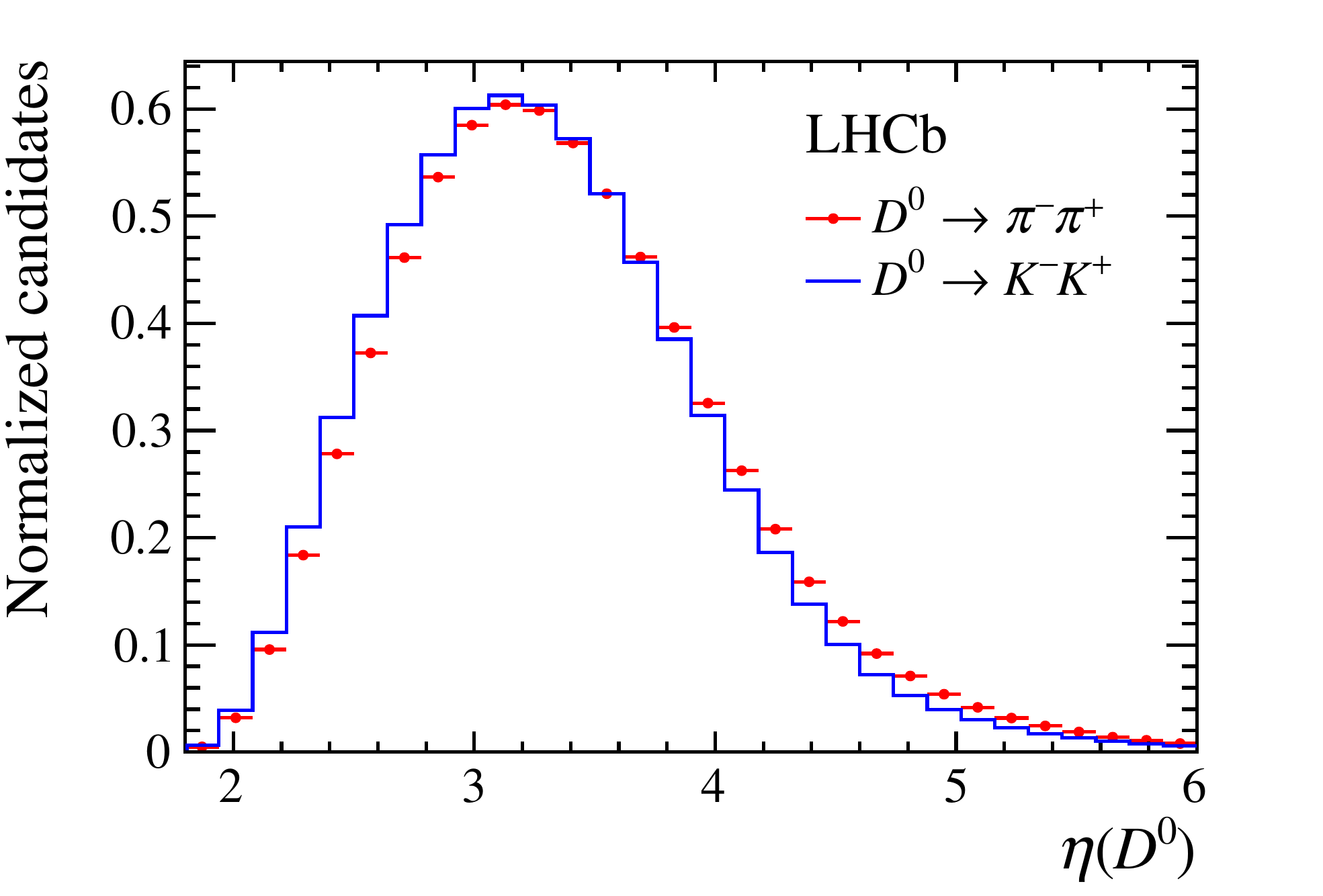}
\includegraphics[width=0.48\textwidth]{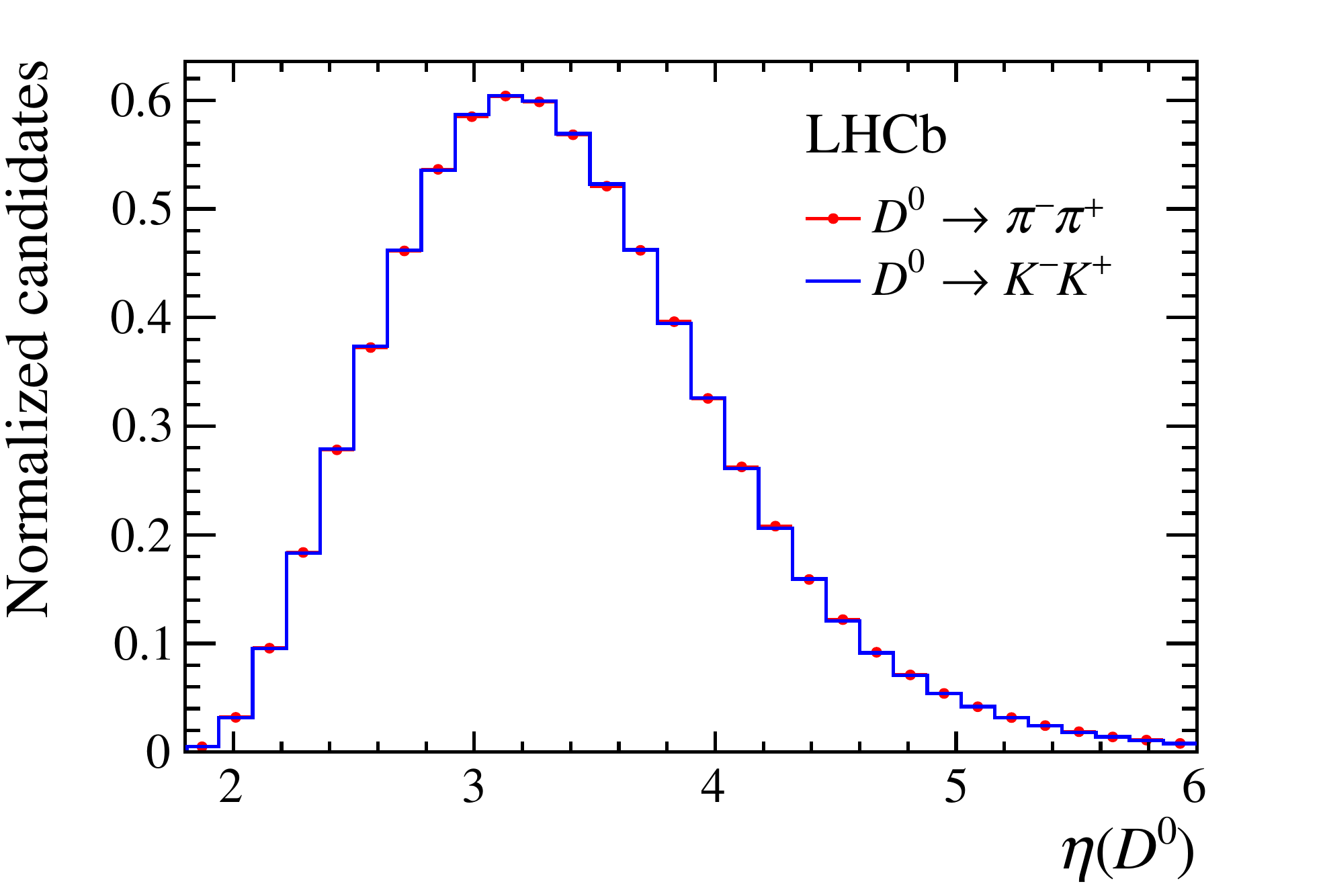}
\caption{Background-subtracted distributions of momentum~($p$), transverse momentum~($\pt$), azimuthal angle~($\varphi$) and pseudorapidity~($\eta$) of \Dz mesons for the semileptonic sample: (left column)~before and (right column)~after the weighting procedure for \dkk and \dpipi decays, as indicated in the legends. The distributions are normalized to unit area.}
\label{fig:SL_D0_weight}
\end{figure}

\begin{figure}[htb]
\centering
\includegraphics[width=0.48\textwidth]{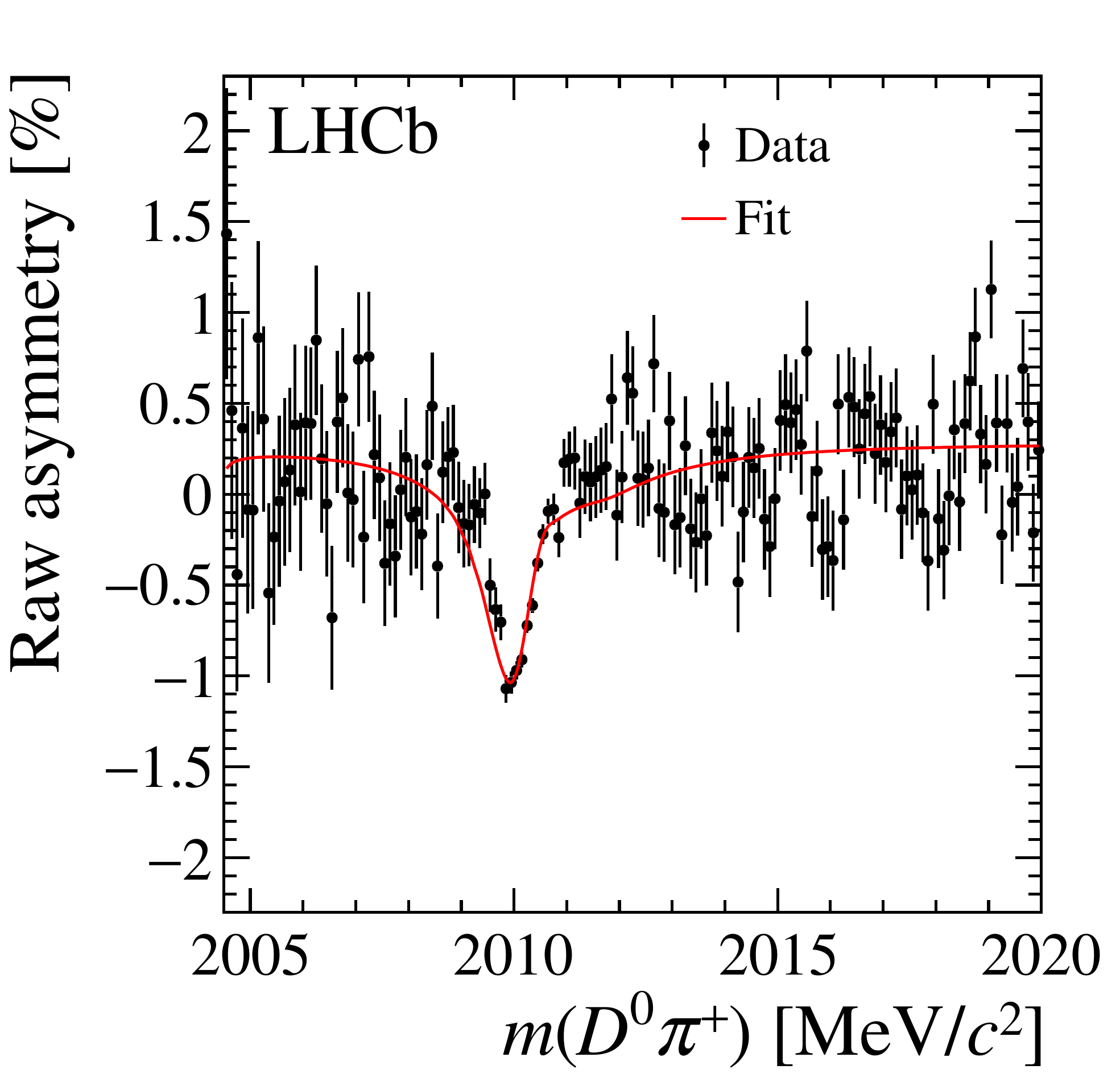}
\includegraphics[width=0.48\textwidth]{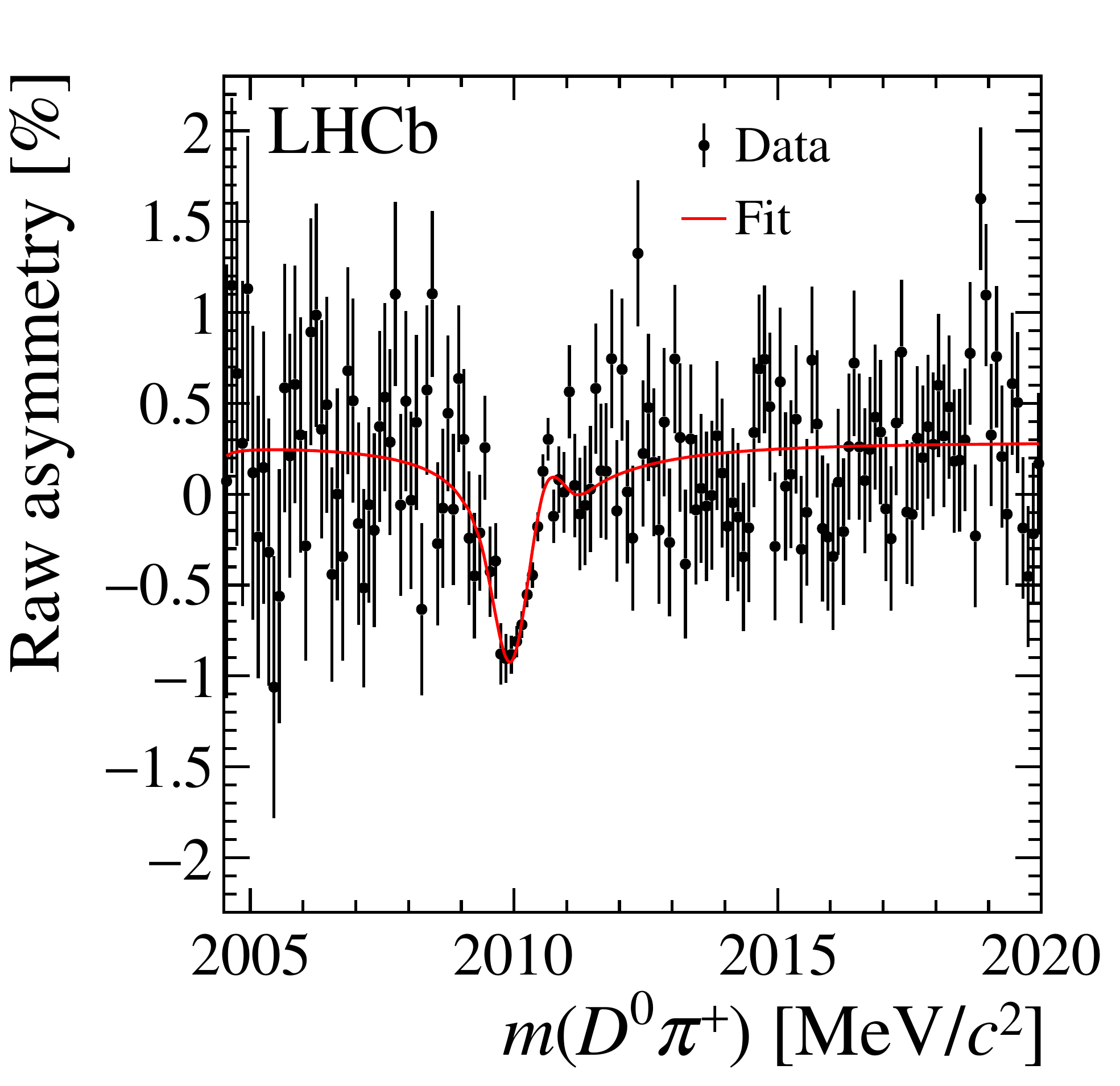}
\caption{Raw asymmetries in the $\pi$-tagged sample for (left)~\dkk and (right)~\dpipi candidates as a function of $m(\Dz\pip)$.}
\label{fig:rawAsymPR}
\end{figure}

\begin{figure}[htb]
\centering
\includegraphics[width=0.48\textwidth]{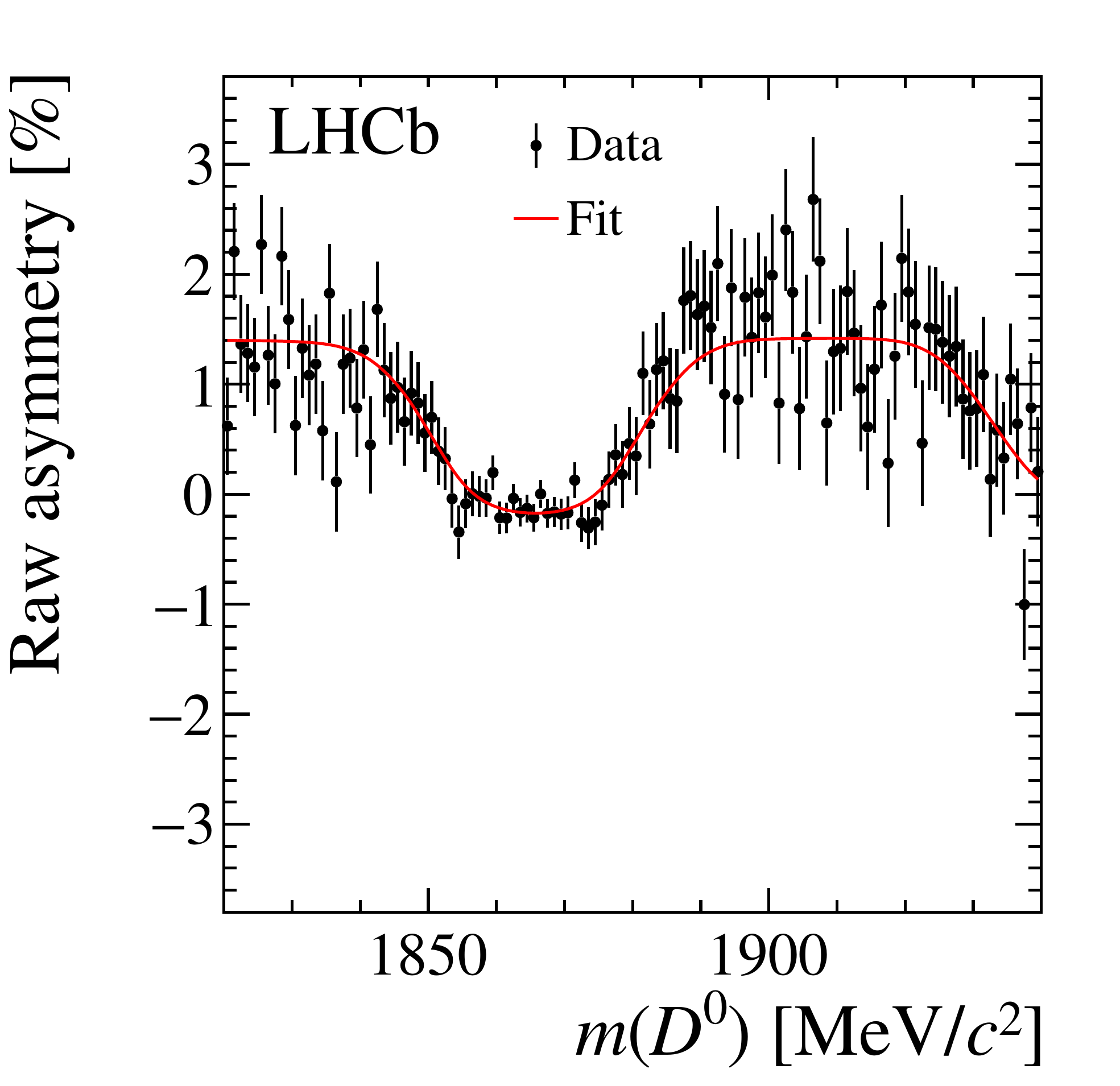}
\includegraphics[width=0.48\textwidth]{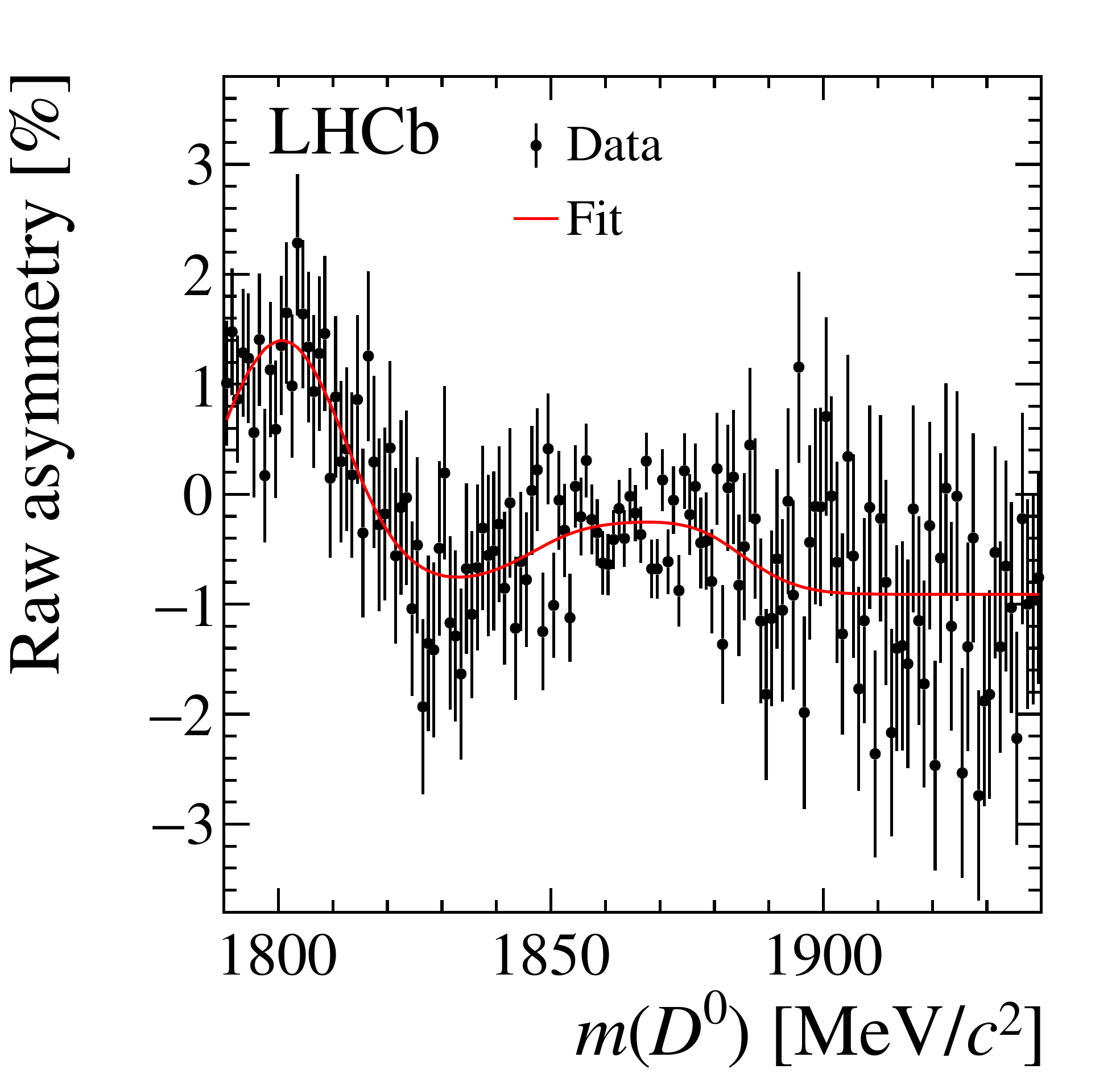}
\caption{Raw asymmetries in the $\mu$-tagged sample for (left)~\dkk and (right)~\dpipi candidates as a function of $m(\Dz)$.}
\label{fig:rawAsymSL}
\end{figure}

\begin{figure}[htb]
\centering
\includegraphics[width=0.48\textwidth]{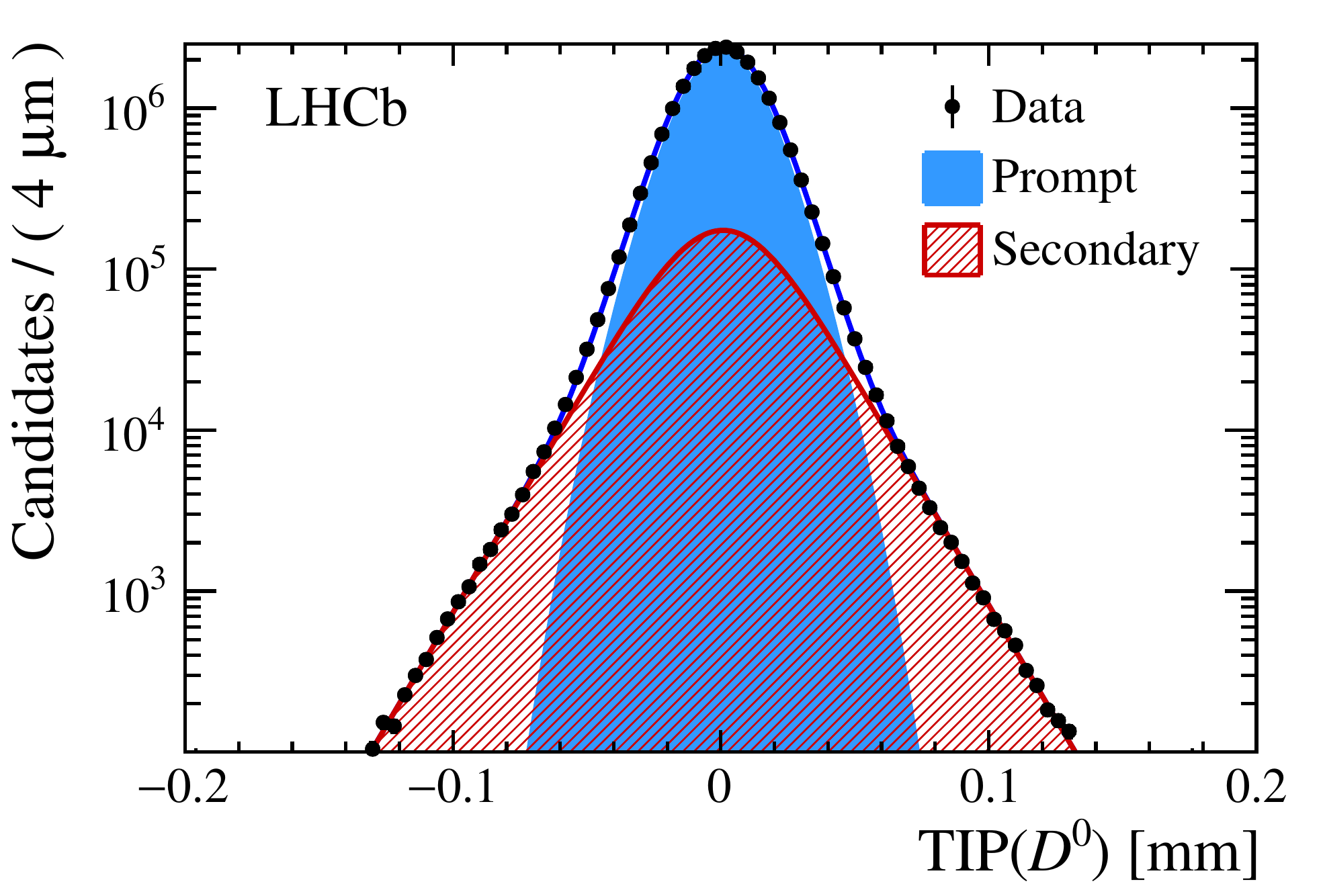}
\includegraphics[width=0.48\textwidth]{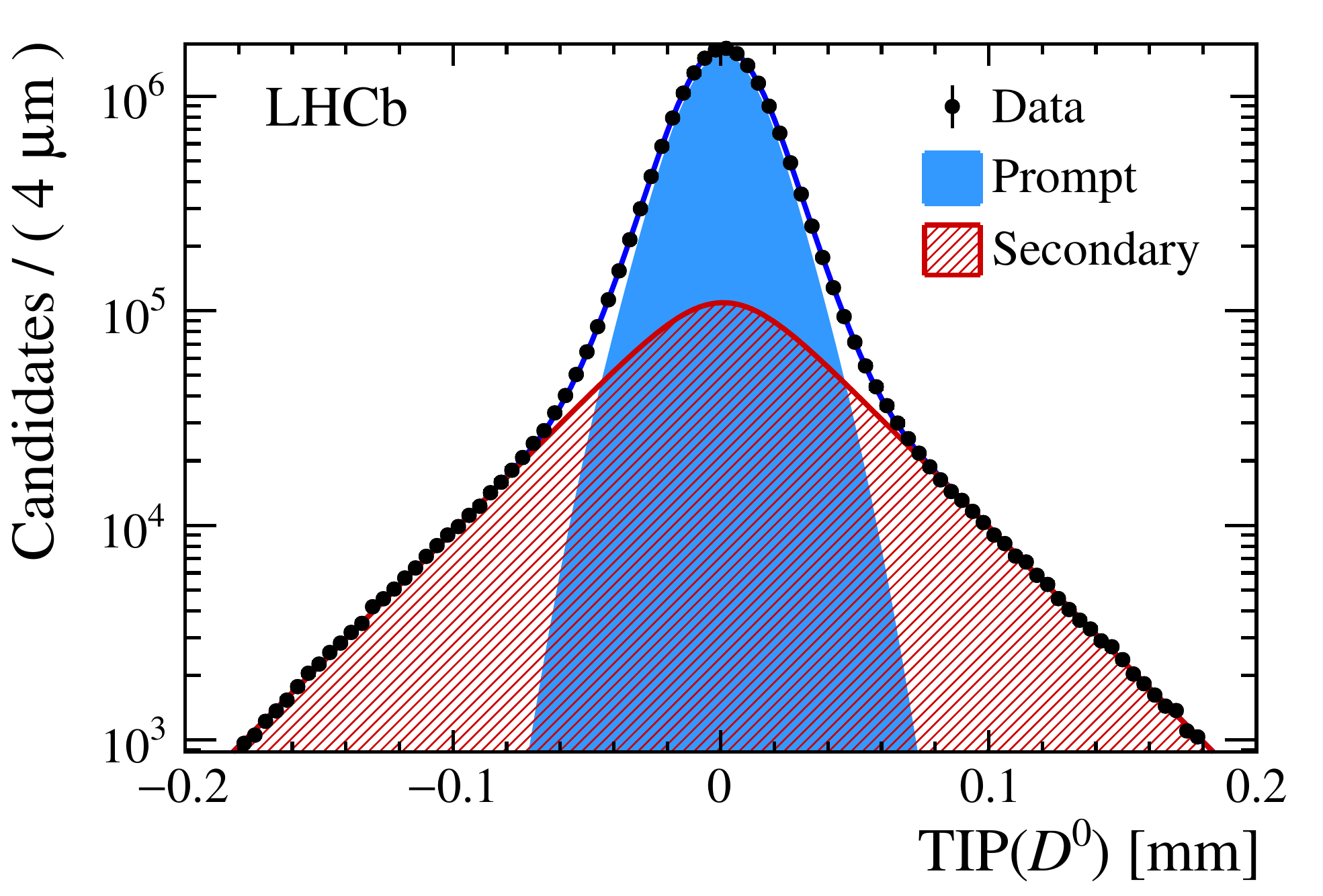}
\includegraphics[width=0.48\textwidth]{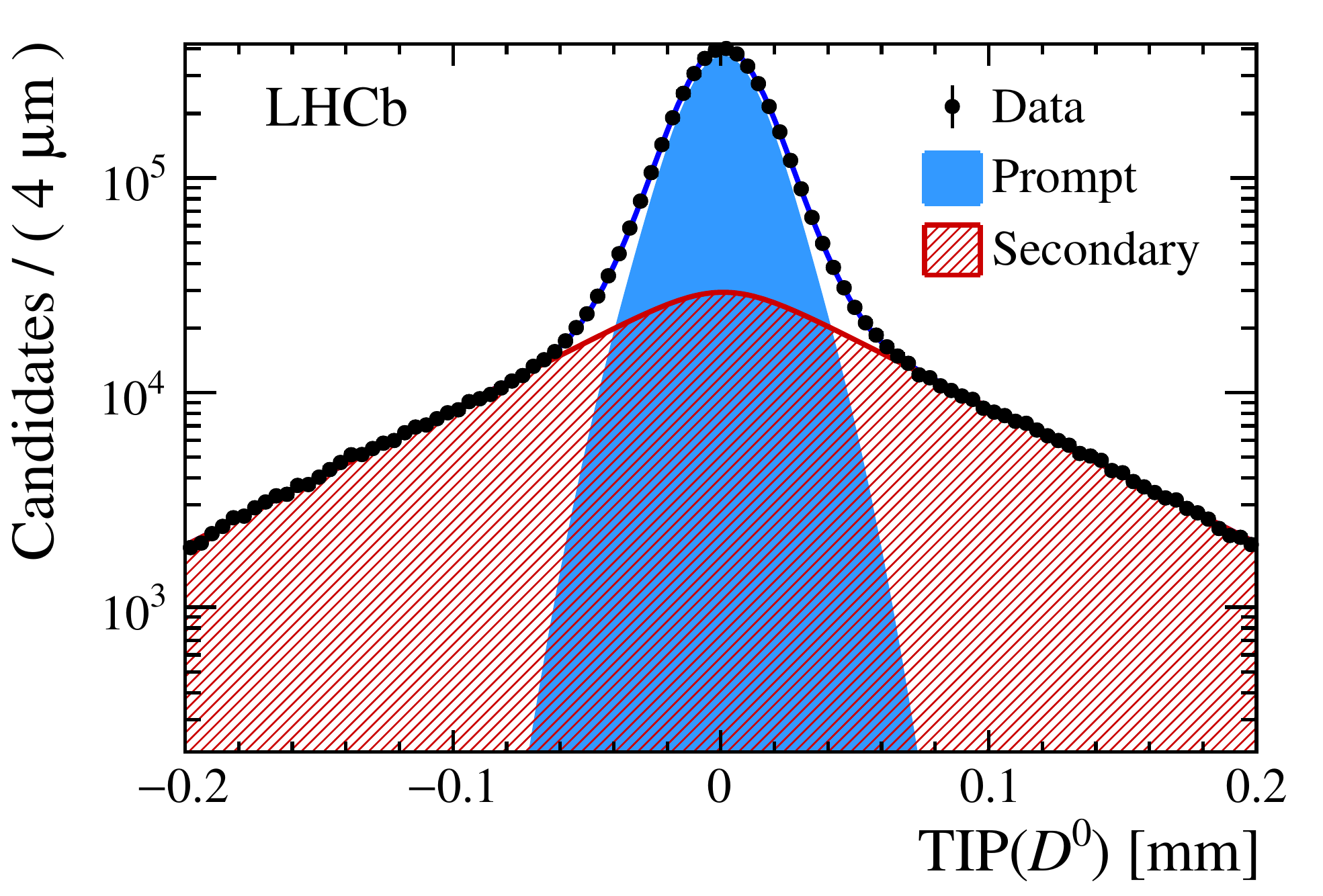}
\includegraphics[width=0.48\textwidth]{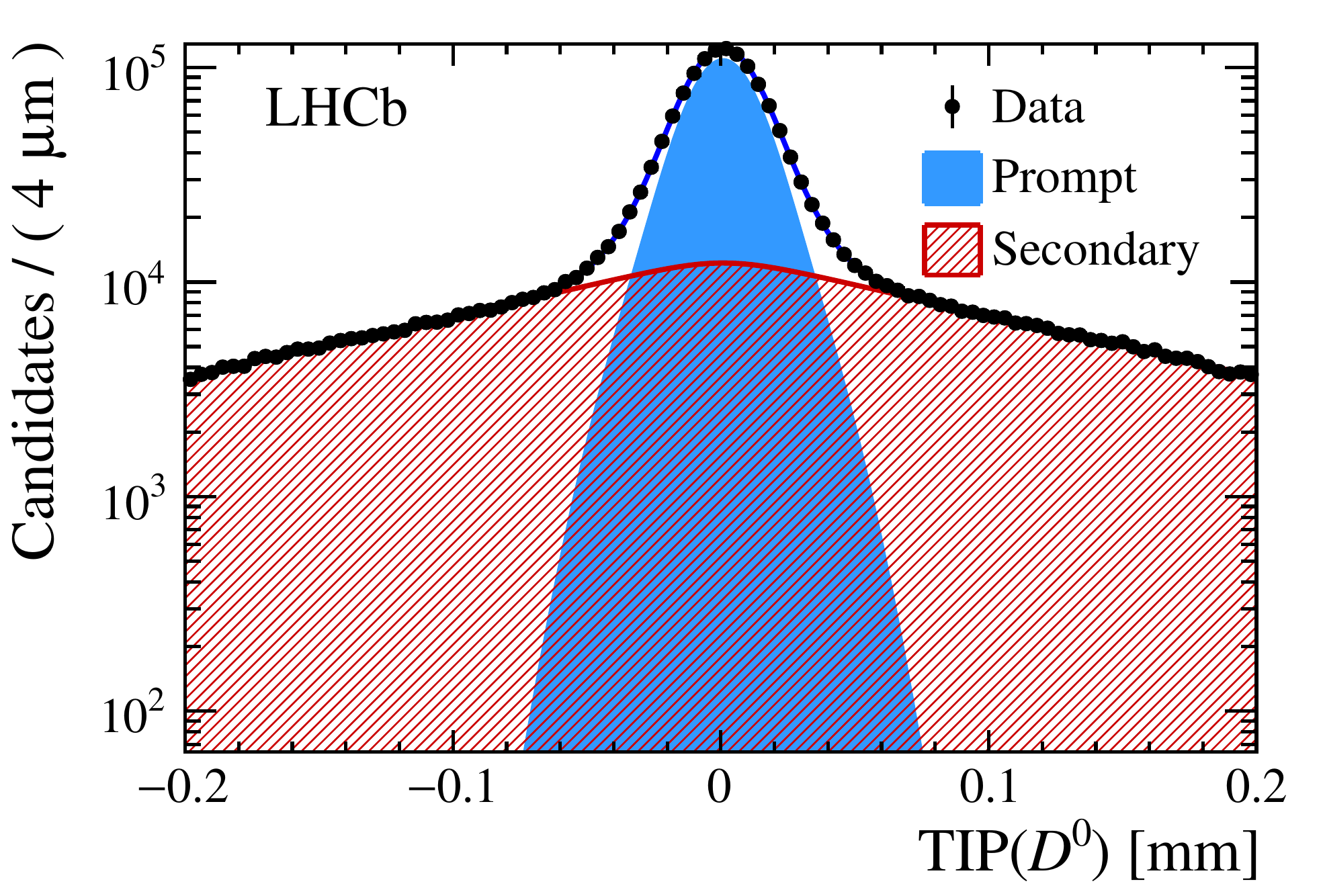}
\caption{Distributions of the signed \Dz impact parameter in the plane transverse to the beam direction, \tip, in bins of reconstructed \Dz decay time, for the $\pi$-tagged \dkk sample: (top left)~$\tilde{t} < 1.5$, (top right)~$1.5 < \tilde{t} < 3.0$, (bottom left)~$3.0 < \tilde{t} < 4.5$ and (bottom right)~$\tilde{t} > 4.5$, where $\tilde{t} \equiv t / \tau(\Dz)$. The fit results are overlaid and the contributions from prompt and secondary decays are shown, as indicated in the legends. Distributions for the \dpipi sample are very similar.}
\label{fig:TIPKK}
\end{figure}

\begin{figure}[htb]
\centering
\includegraphics[width=0.7\textwidth]{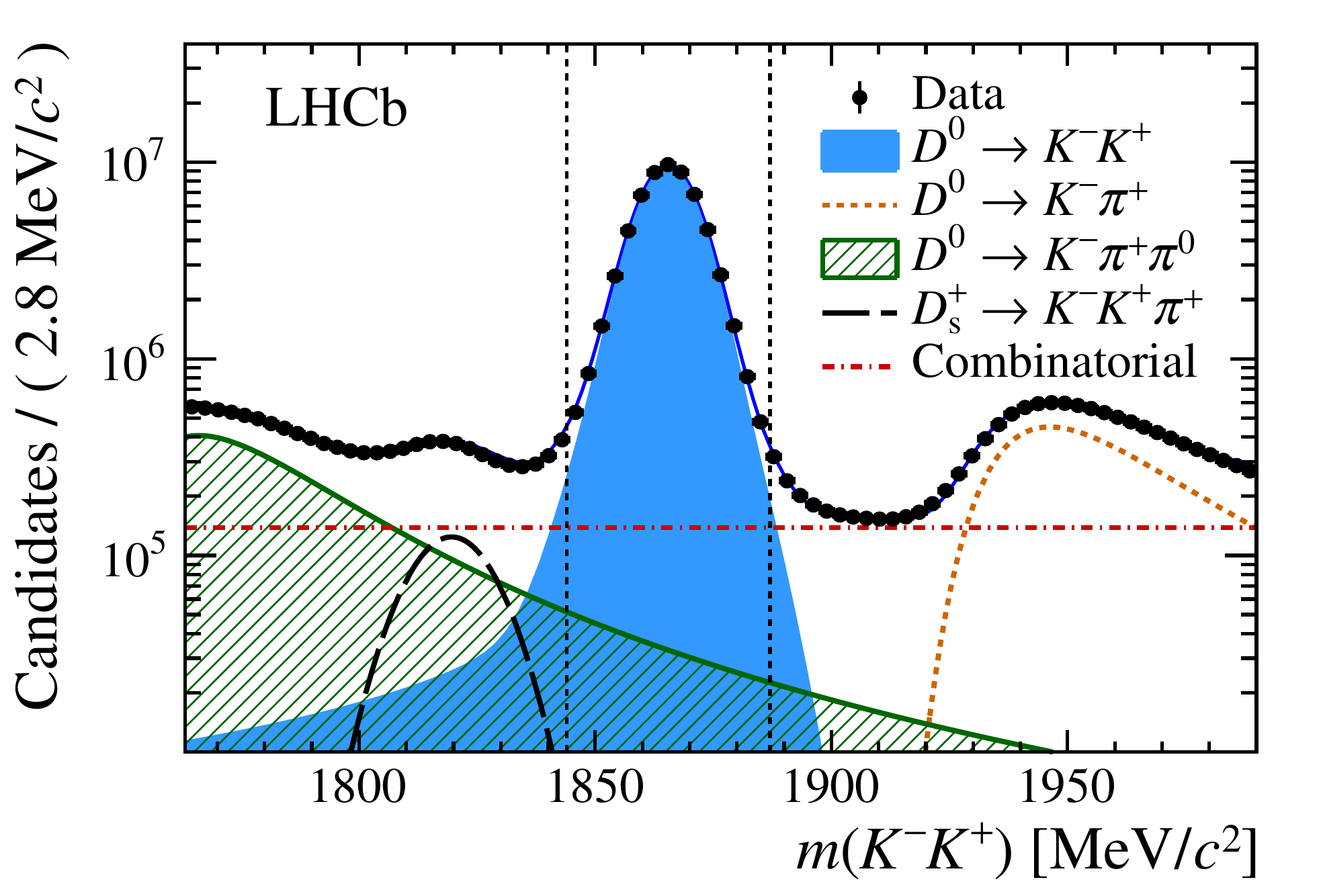}
\includegraphics[width=0.7\textwidth]{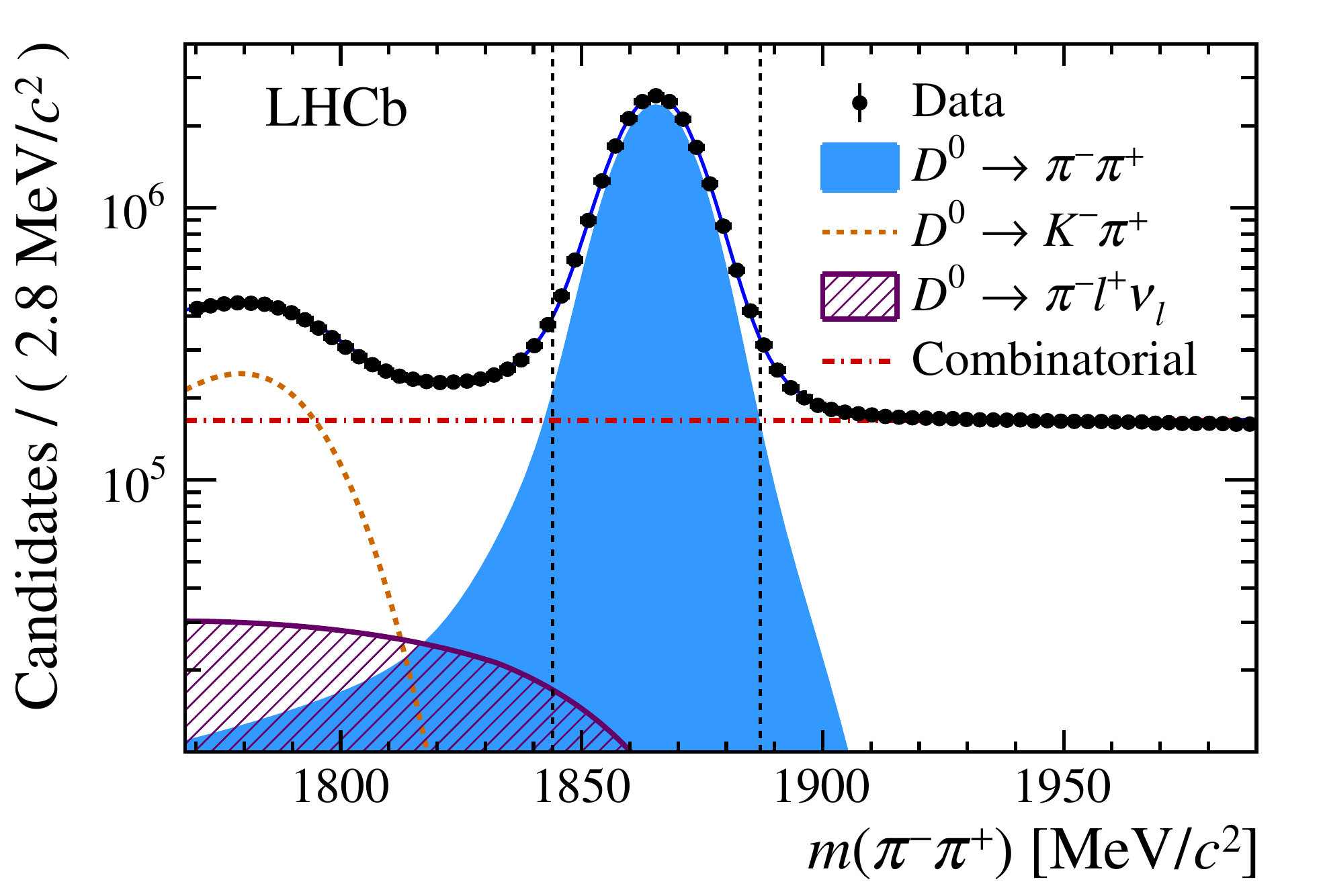}
\caption{Invariant-mass distributions of (top)~\dkk and (bottom)~\dpipi candidates in the prompt sample with fit results overlaid. These fits are used to determine the yields and raw asymmetries of (top)~$\Dz \to K^-\pi^+\pi^0$ and (bottom)~$\Dz \to \pi^-l^+\nu_l$ backgrounds, whose mass shapes extend to the \Dz-signal mass region. The various components included in the fit model are indicated in the legends.}
\label{fig:peakingbkg}
\end{figure}

\begin{figure}[htb]
\centering
\includegraphics[width=0.48\textwidth]{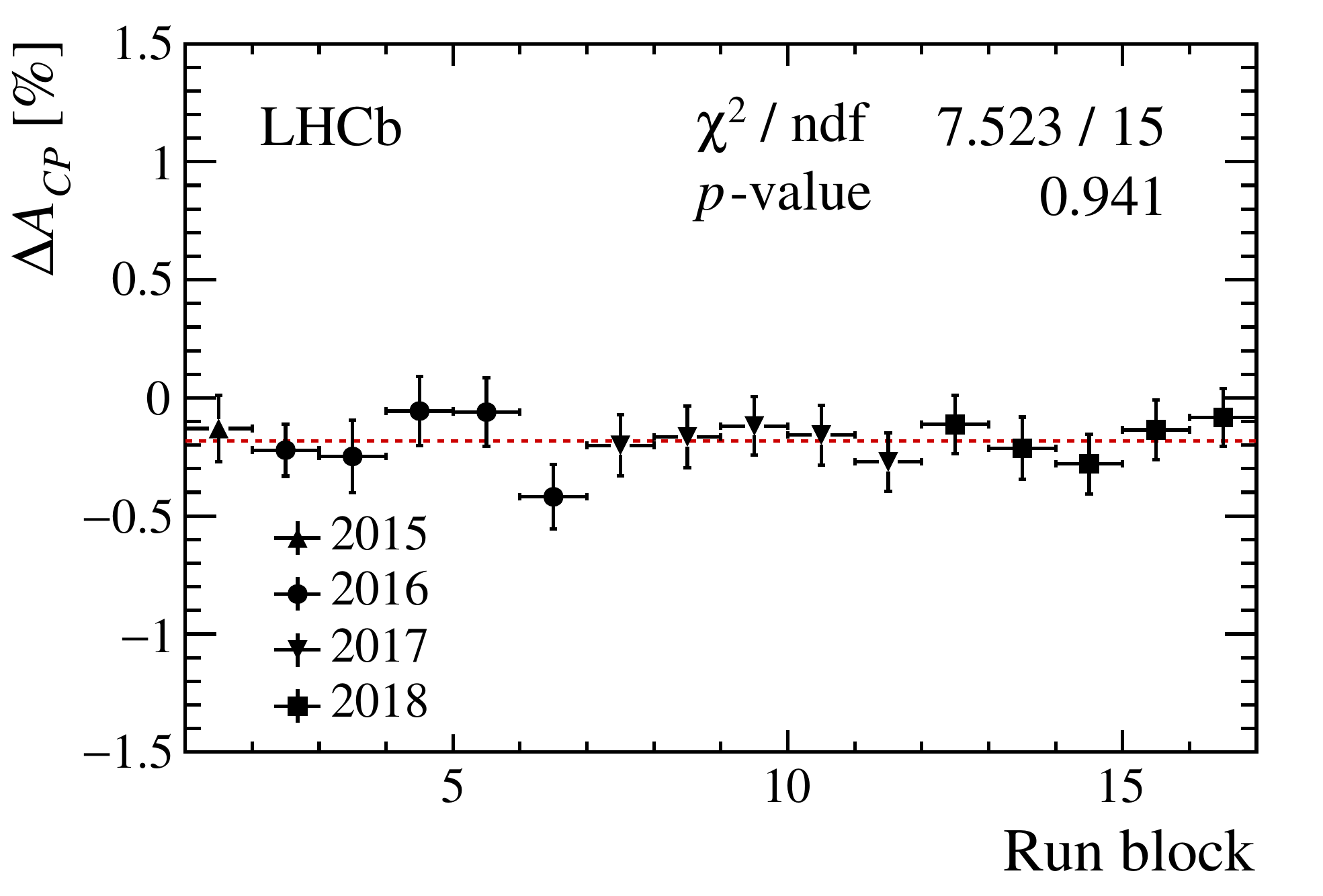}
\includegraphics[width=0.48\textwidth]{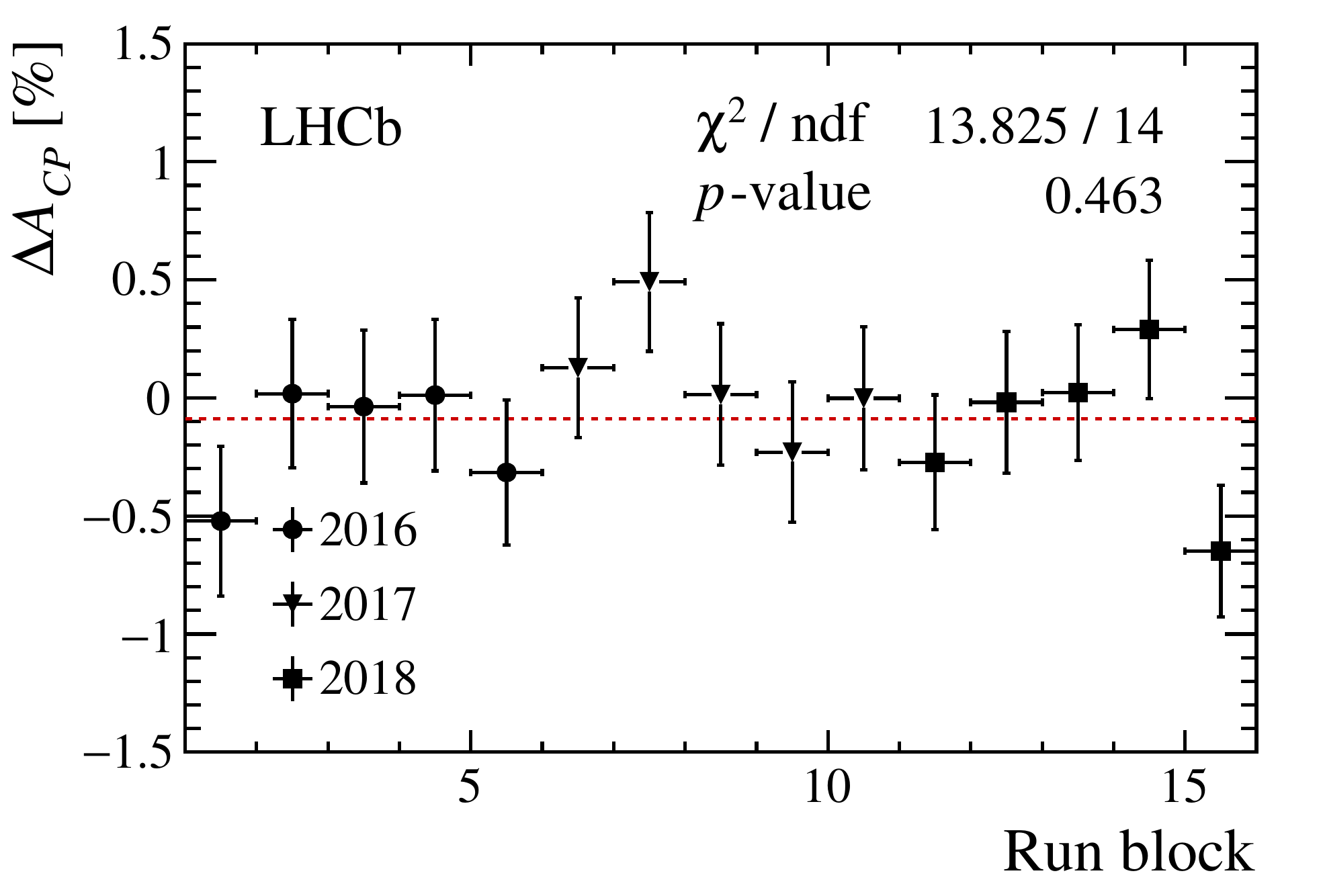}
\caption{Measurements of \DACP in time-ordered data-taking subsamples (referred to as run blocks) for (left)~prompt and (right)~semileptonic samples. The uncertainties are statistical only. The horizontal red-dashed lines show the central values of the nominal results.}
\label{fig:runNumber}
\end{figure}

\begin{figure}[htb]
\centering
\includegraphics[width=0.48\textwidth]{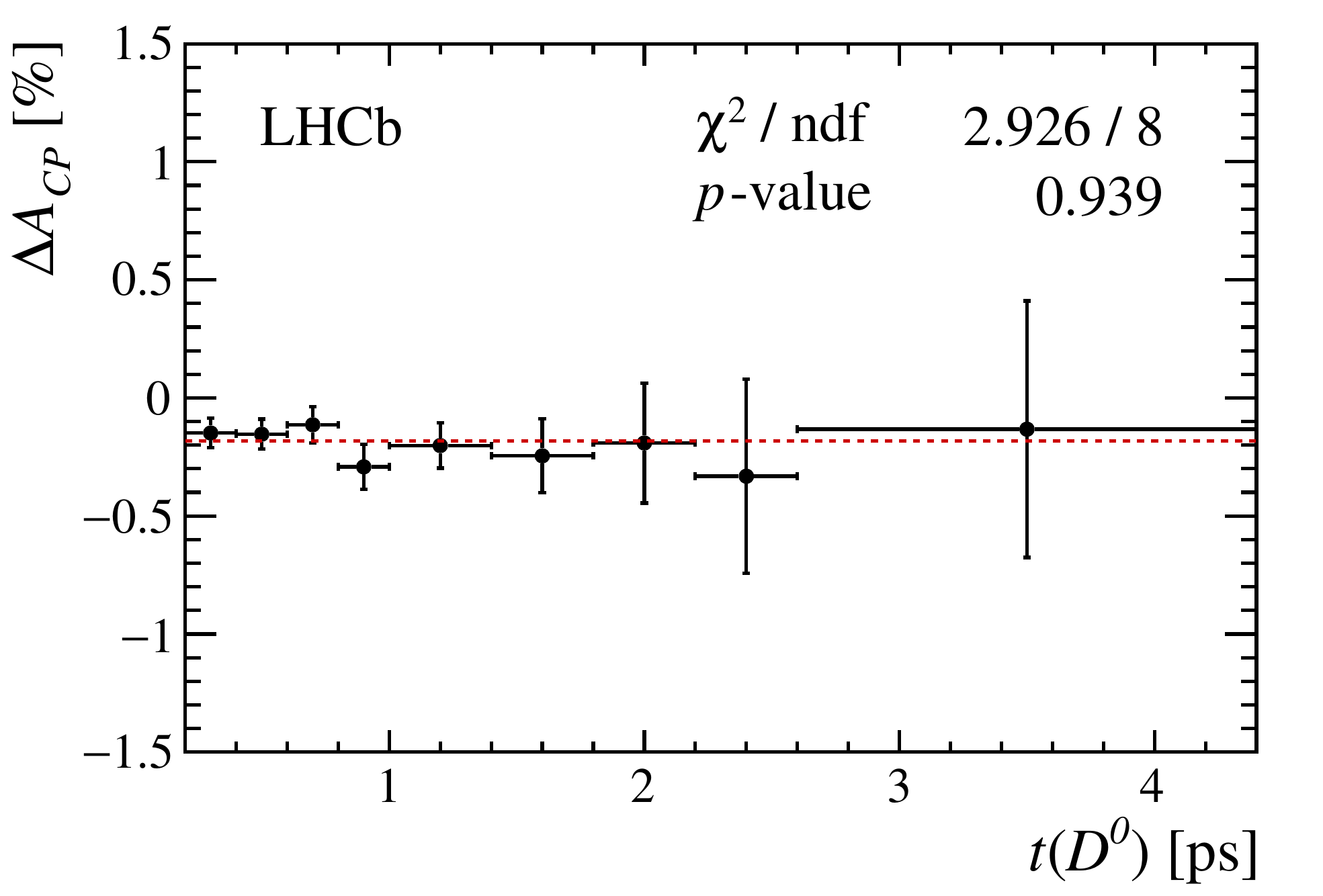}
\includegraphics[width=0.48\textwidth]{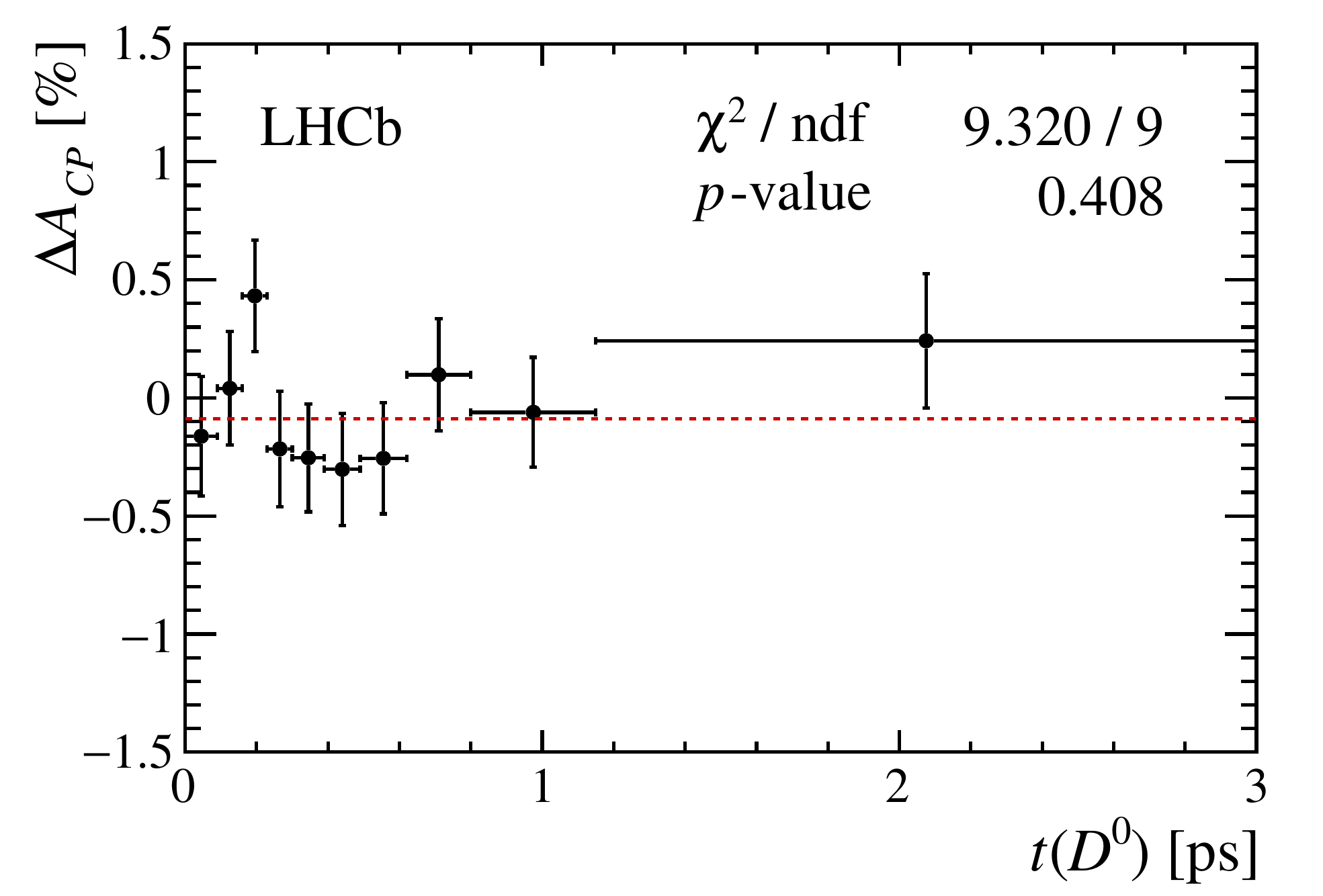}

\caption{Measurements of \DACP in bins of \Dz decay time for (left)~prompt and (right)~semileptonic samples. In each plot, the last bin on the right also includes a few overflow candidates. The uncertainties are statistical only. The horizontal red-dashed lines show the central values of the nominal results.}
\label{fig:D0tau}
\end{figure}

\begin{figure}[htb]
\centering
\includegraphics[width=0.48\textwidth]{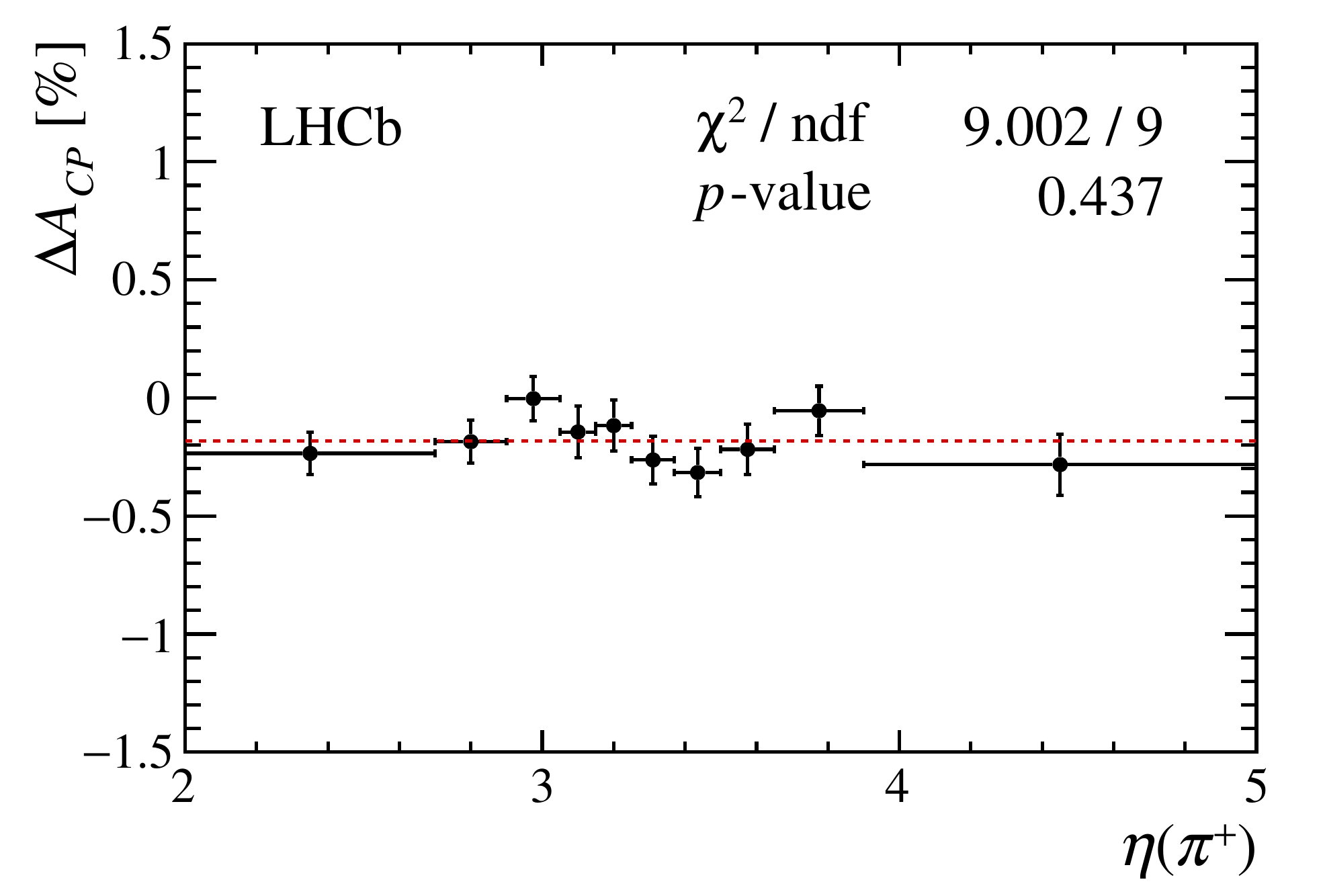}
\includegraphics[width=0.48\textwidth]{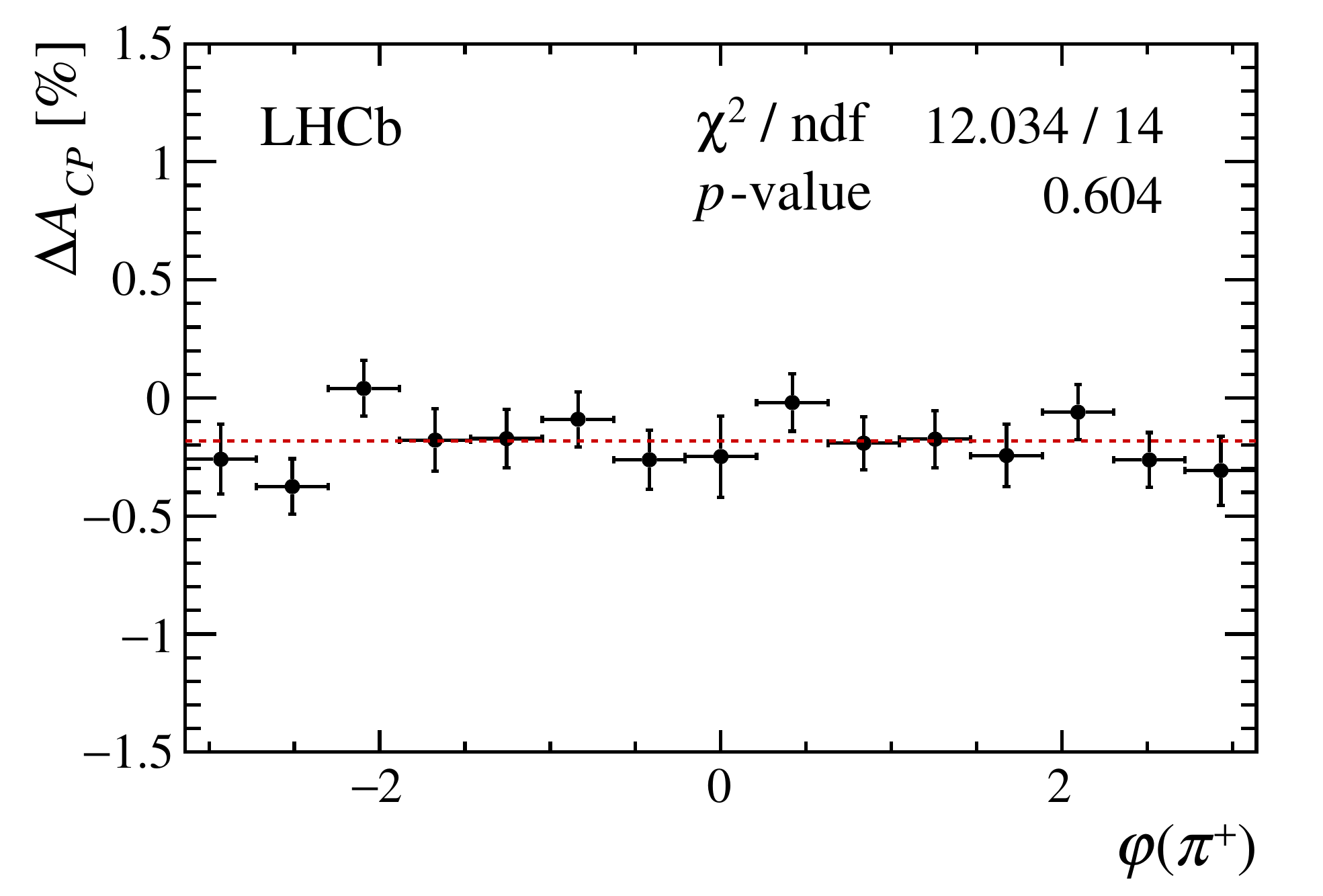}
\includegraphics[width=0.48\textwidth]{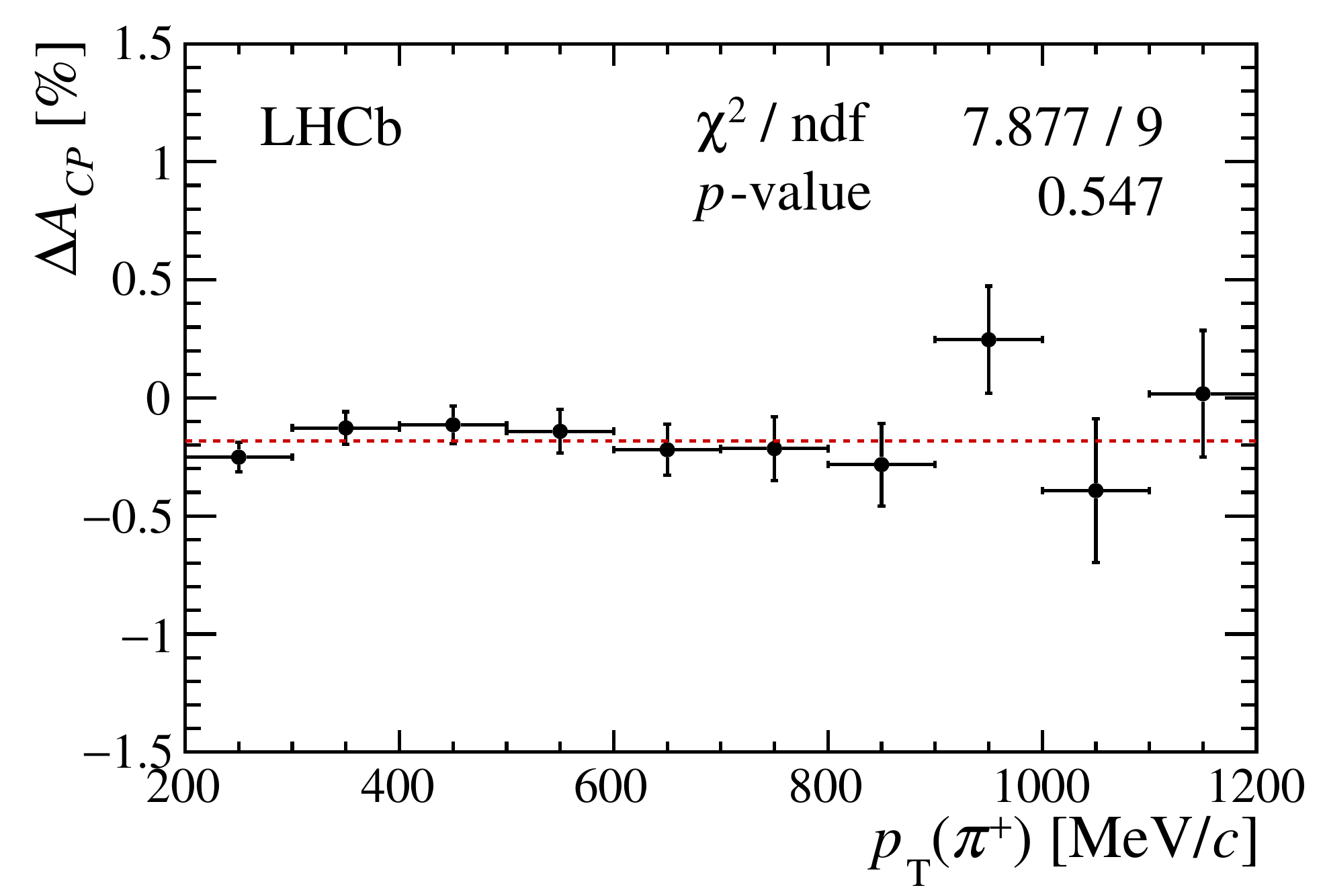}
\includegraphics[width=0.48\textwidth]{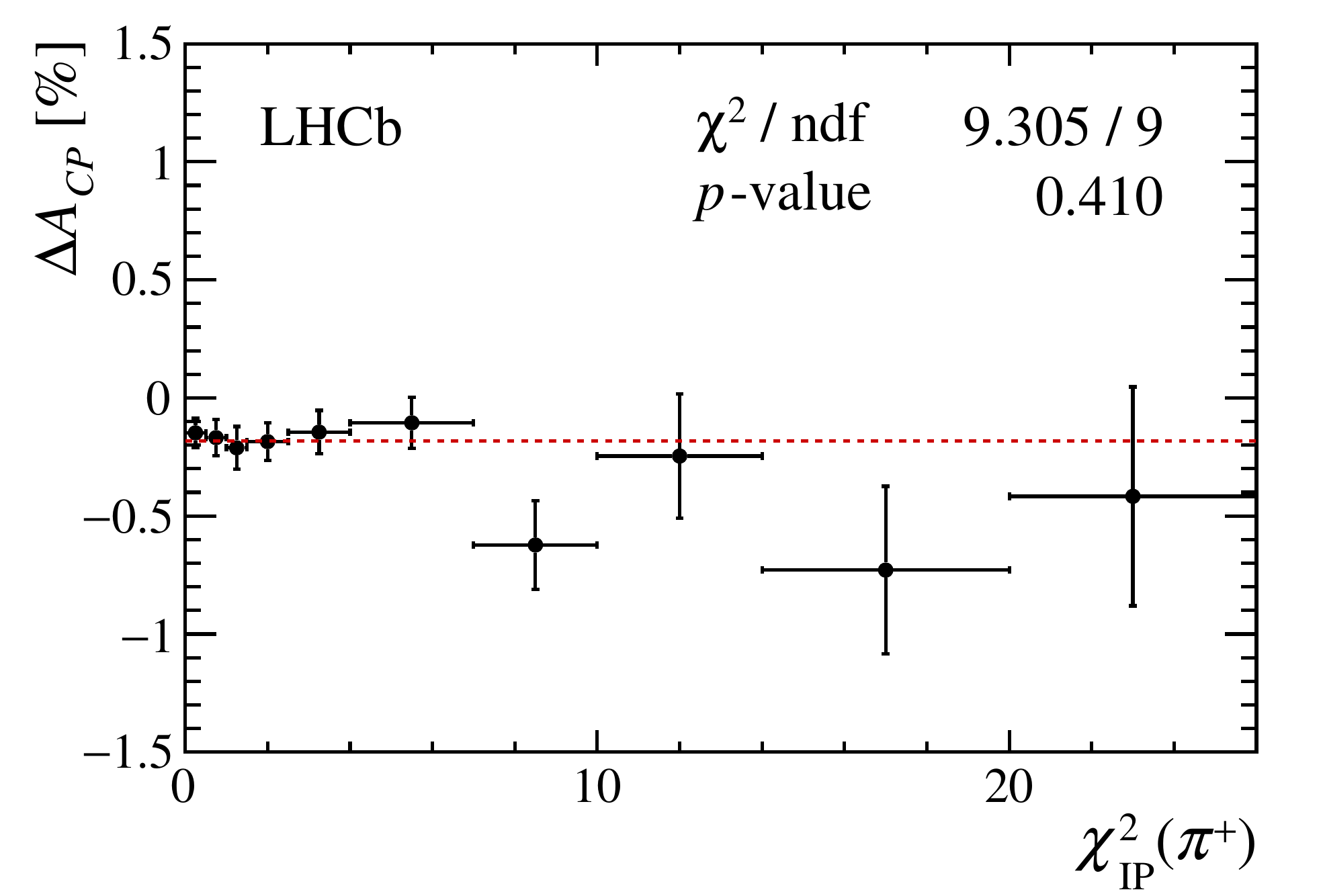}
\caption{Measurements of \DACP in bins of (top left)~pseudorapidity, (top right)~azimuthal angle, (bottom left)~transverse momentum and (bottom right)~\chisqip of tagging pions for the prompt sample. In each plot but that of the azimuthal angle, the last bin on the right also includes a few overflow candidates. The uncertainties are statistical only. The horizontal red-dashed lines show the central value of the nominal result.}
\label{fig:taggingpion}
\end{figure}

\begin{figure}[htb]
\centering
\includegraphics[width=0.48\textwidth]{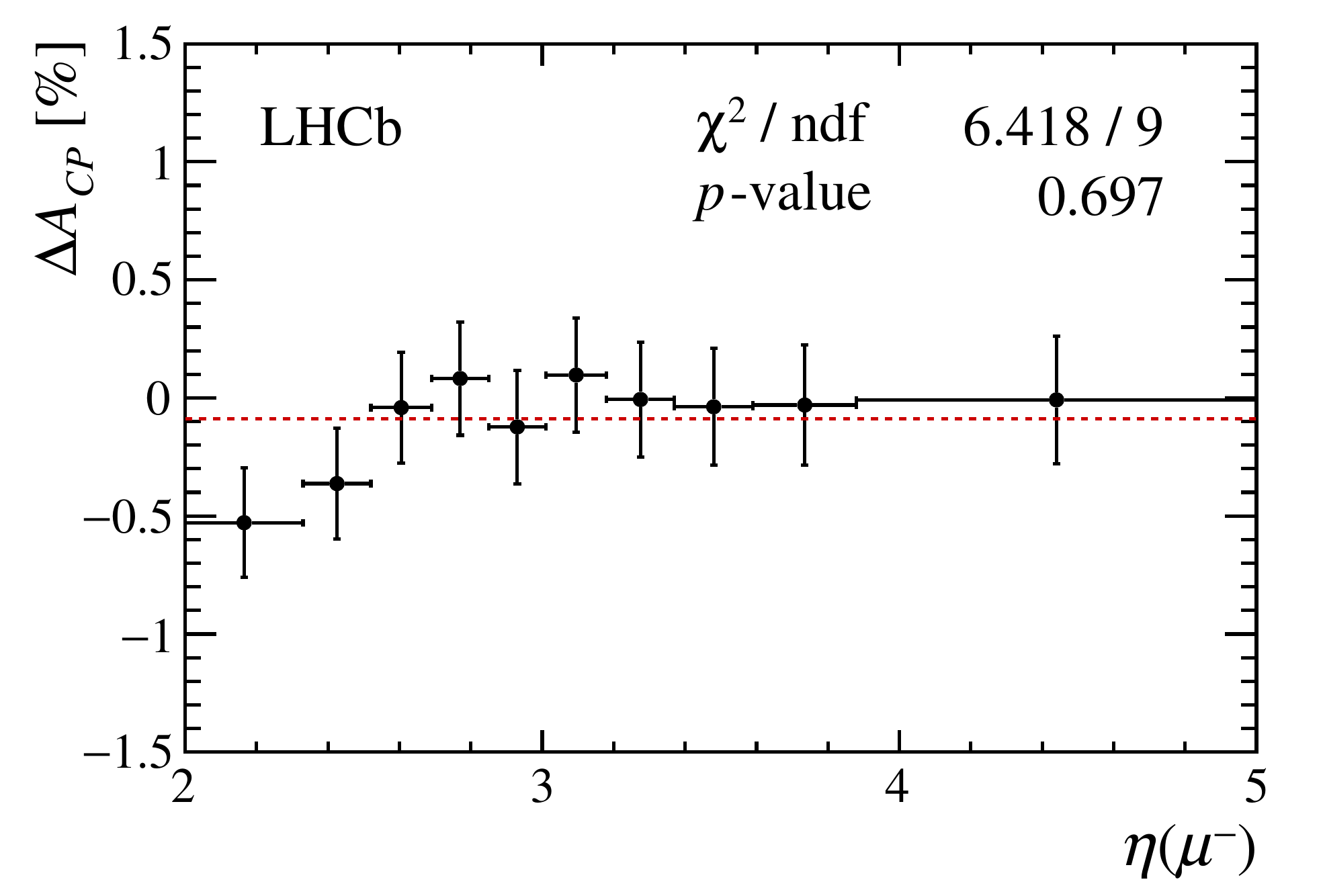}
\includegraphics[width=0.48\textwidth]{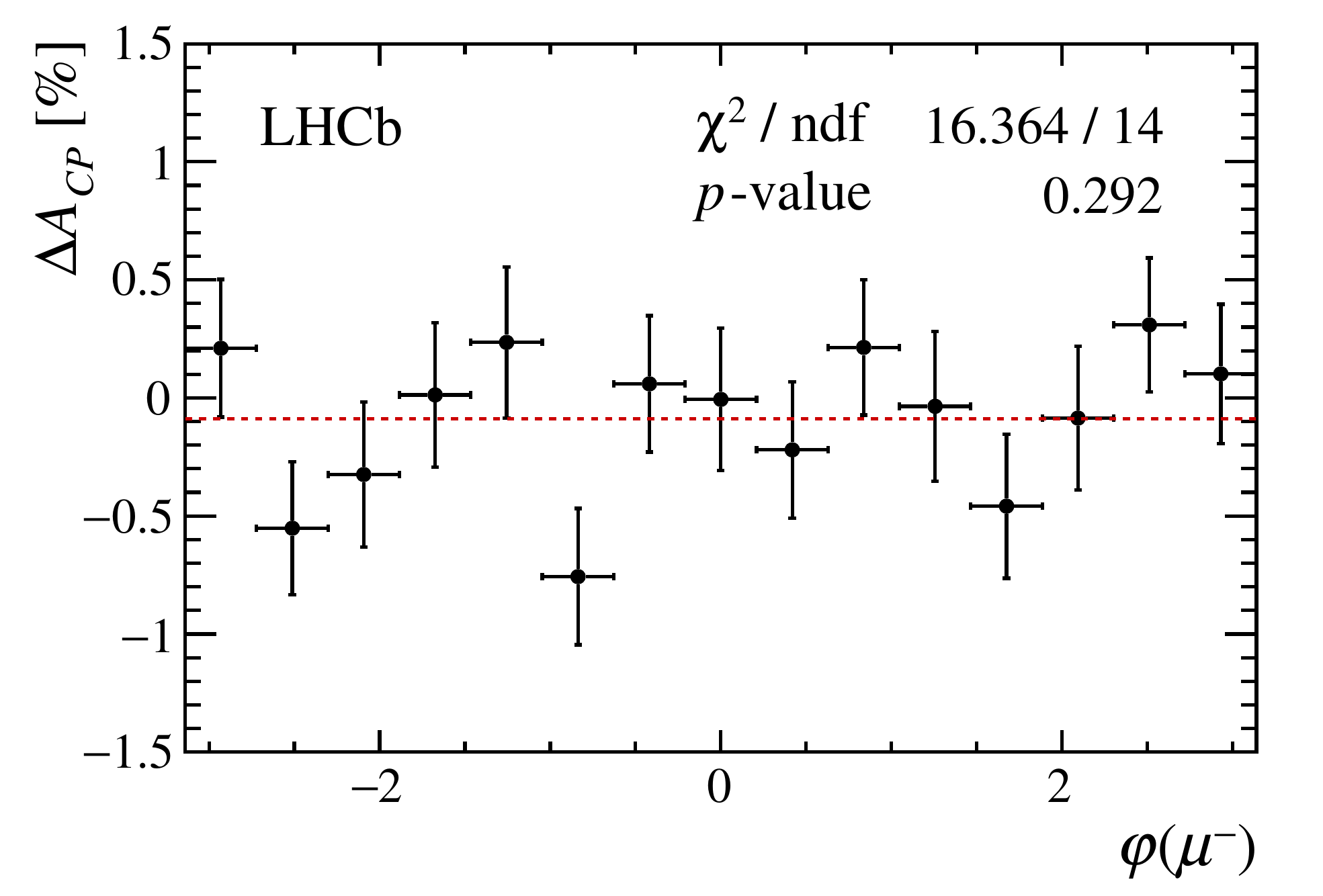}
\includegraphics[width=0.48\textwidth]{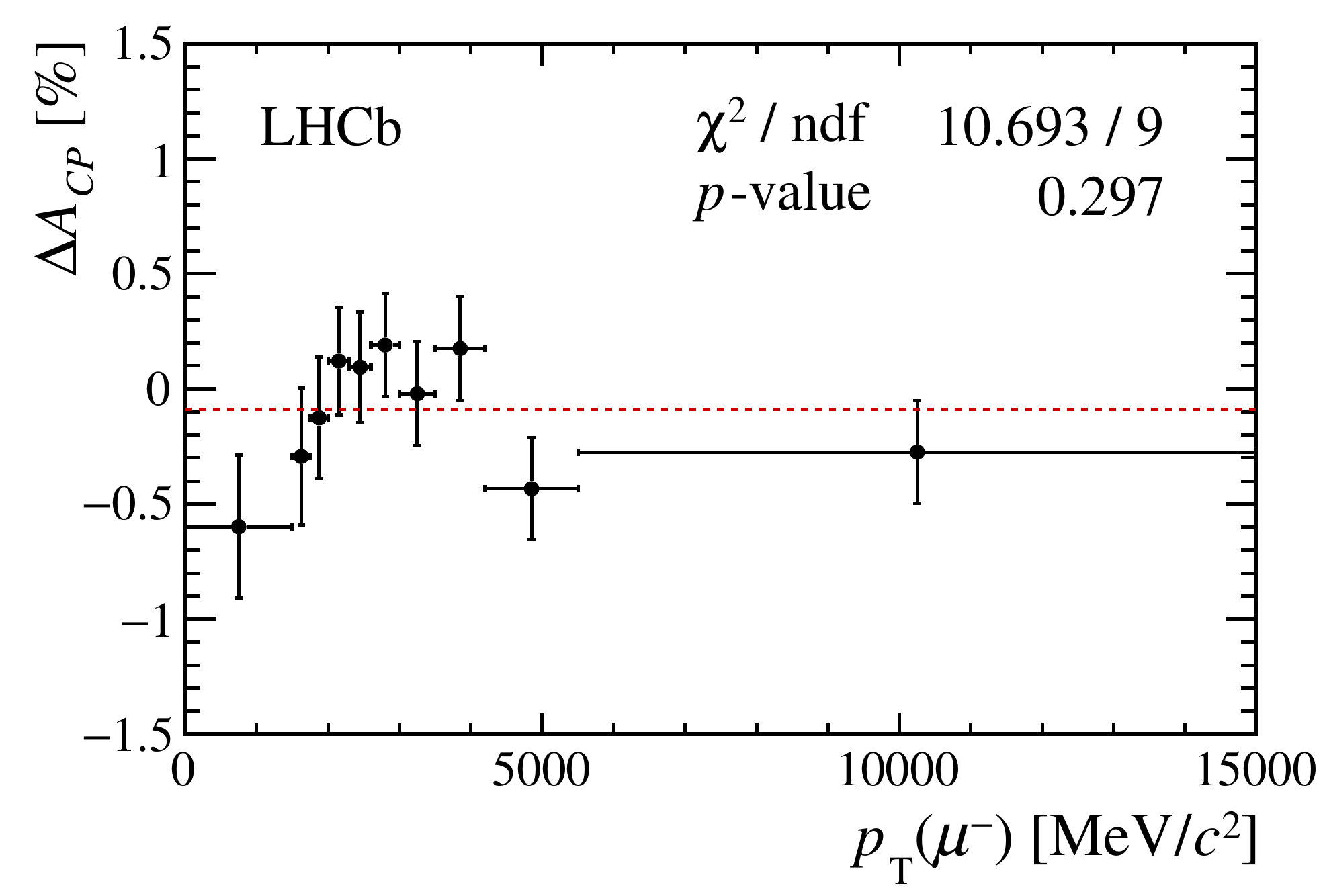}
\includegraphics[width=0.48\textwidth]{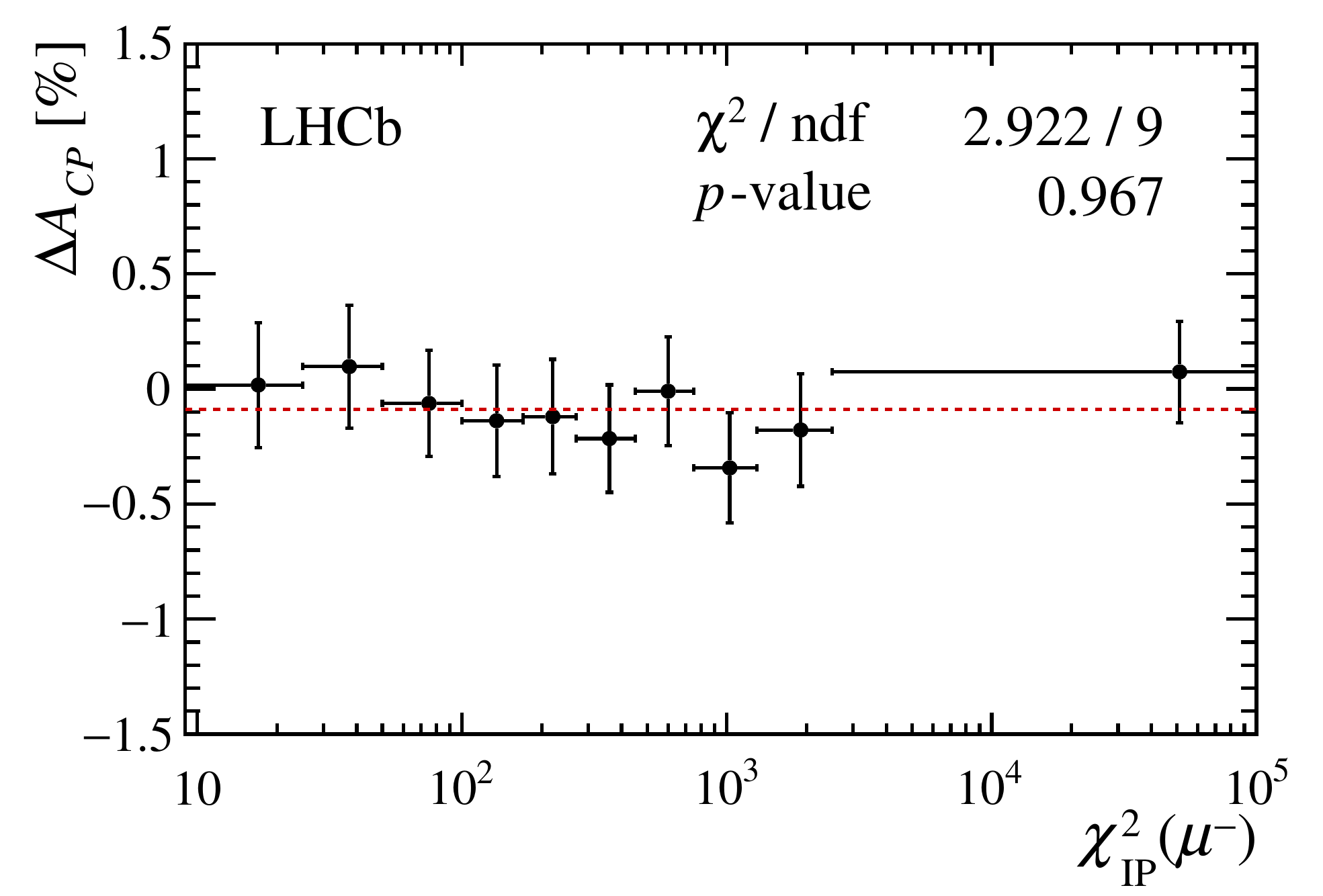}
\caption{Measurements of \DACP in bins of (top left)~pseudorapidity, (top right)~azimuthal angle, (bottom left)~transverse momentum and (bottom right)~\chisqip of tagging muons for the semileptonic sample. In each plot but that of the azimuthal angle, the last bin on the right also includes a few overflow candidates. The uncertainties are statistical only. The horizontal red-dashed lines show the central value of the nominal result.}
\label{fig:tagginmuon}
\end{figure}

\begin{figure}[htb]
\centering
\includegraphics[width=0.48\textwidth]{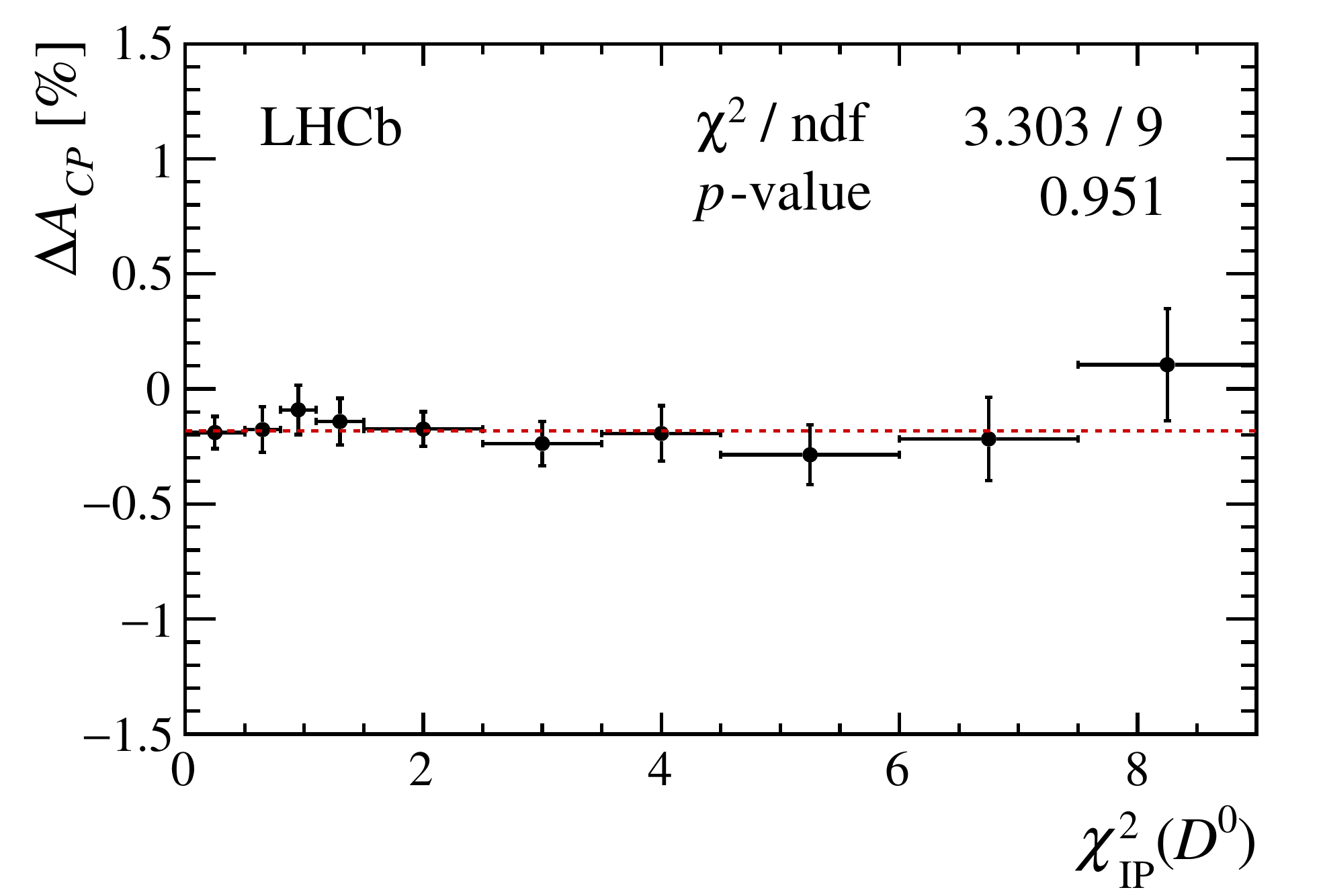}
\includegraphics[width=0.48\textwidth]{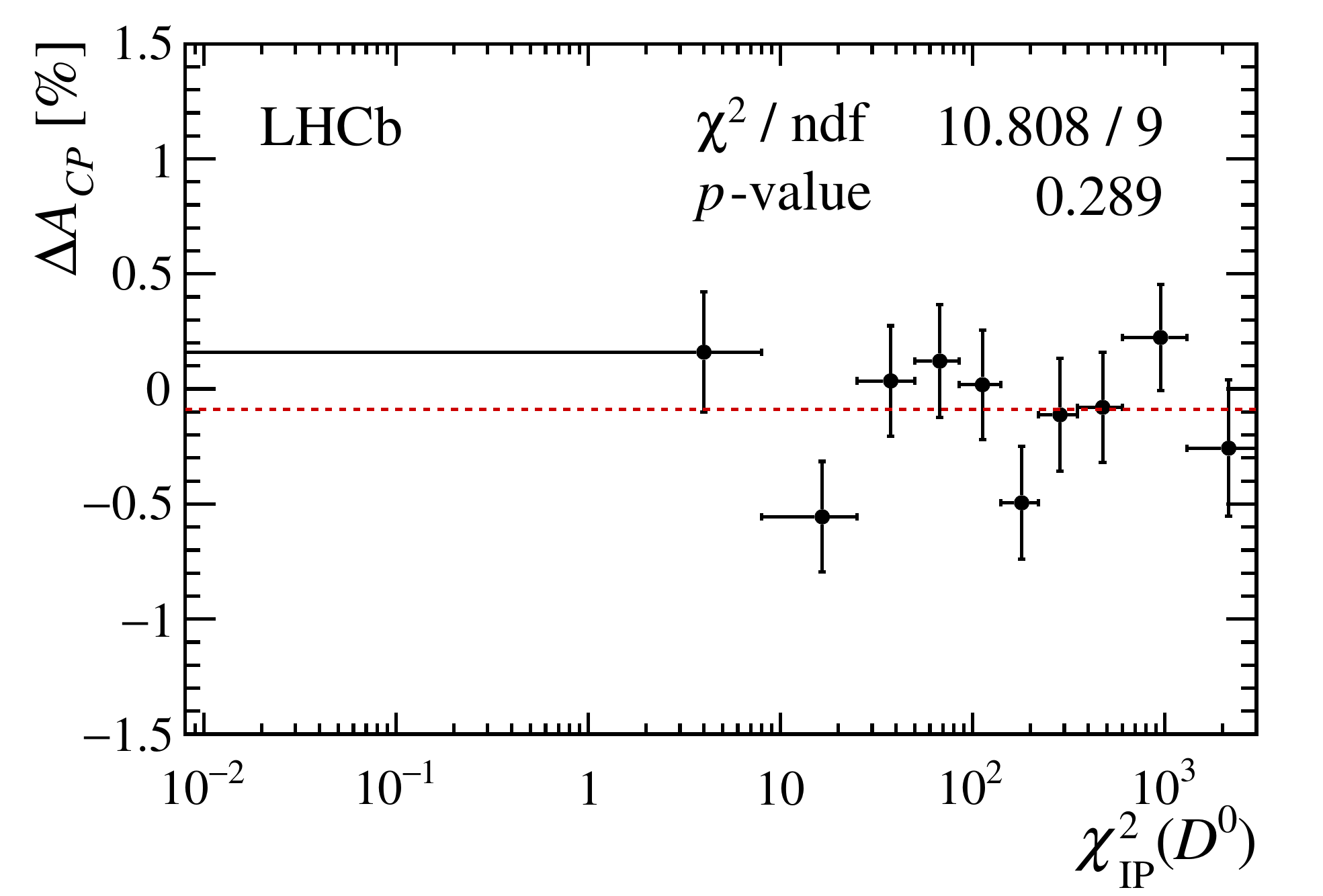}
\caption{Measurements of \DACP in bins of \Dz \chisqip for (left)~prompt and (right)~semileptonic samples. In each plot, the last bin on the right also includes a few overflow candidates. The uncertainties are statistical only. The horizontal red-dashed lines show the central values of the nominal results.}
\label{fig:ipchi2}
\end{figure}

\end{document}